# Operator-Controlled 6G: From Connectivity Infrastructure to Guaranteed Digital Services

**DAVID SOLDANI[1] (Senior Member, IEEE)**

Rakuten Mobile Inc., Tokyo 158-0094, Japan

CORRESPONDING AUTHOR: D. SOLDANI (e-mail: david.soldani@ rakuten.com).

This work received no external funding. The author was supported by Rakuten Mobile Inc. in his employed capacity.

[1] https://orcid.org/0000-0001-9677-5545

# Table of Contents

## Abstract

*Sixth-generation mobile networks (6G) are approaching a structural inflection point. Five generations of vendor-led architecture have left operators proficient at procuring, integrating, and operating networks they do not own, on platforms they cannot modify, with AI inference layers they cannot audit. This paper argues that 6G must reverse this trajectory through a deliberate reordering of five operator priorities:* ***Control First*** *(owning the software-defined control plane);* ***Customer First*** *(delivering verifiable outcomes, not peak data rates);* ***Business First*** *(pricing guarantees, not megabytes);* ***Operations First*** *(running networks as software with agentic AI); and* ***Technology Last*** *(placing architecture and technologies in service of the preceding four priorities), not the reverse. Two paired contributions operationalize this thesis: the* ***6G Control Compact***[2]*, a three-layer ownership taxonomy, namely own, federate, and consume, that allocates architectural sovereignty according to strategic value; and the* ***Guarantee Economy****, a five-tier, outcome-priced commercial model that translates operator control into enforceable service-level objectives. The framework is grounded in operational evidence from Rakuten Mobile, the world's first national-scale, fully cloud-native, fully Open RAN deployment, which reached full-year EBITDA profitability in FY2025. It is also aligned with the ITU-R IMT-2030 framework; 3GPP 6G use cases and service requirements; NGMN views and recommendations; ETSI standards; O-RAN Alliance and AI-RAN Alliance technical specifications; IOWN Global Forum sustainability metrics; Linux Foundation initiatives; and the work of leading industry and academic programs and organizations. A three-phase deployment roadmap, covering 2025-2027, 2027-2029, and 2029-2032 and beyond, together with seven stakeholder-specific calls to action, translates the architecture into actionable industry commitments. The central claim is that Rakuten Mobile's national-scale deployment demonstrates the technical and commercial feasibility of this framework. The architectural decisions made during the 2026-2028 window will determine whether the industry's default 6G model is operator-controlled or vendor-dependent.*

***Index Terms*** *– 6G, 5G Advanced, IMT-2030, Open RAN, agentic AI, AI-RAN, network sovereignty, Guarantee Economy, Control Compact, zero-touch service management.*

## Introduction

Mobile communication has achieved what no other industry has managed in the same timeframe: it has connected more than five billion people, built continent-spanning infrastructure, and repeatedly reinvented its own technical foundations roughly every decade. Yet beneath this record of technical accomplishment lies a structural paradox. Each successive generation – from the analogue circuits of first generation (1G) to the cloud-native radio of fifth generation (5G) – has delivered new capabilities while simultaneously ceding more architectural control to a small set of equipment vendors. Operators have become expert at procuring, integrating, and operating technology they do not own, using interfaces they did not design, and providing services on top of platforms they cannot modify. The result is an industry that has progressively outsourced its own strategic autonomy. As the world prepares for sixth generation networks (6G) –with the International Telecommunication Union's Radiocommunication Sector (ITU-R) framework IMT-2030 setting the technical horizon [1][2][3], – the central question is not whether 6G will be technically superior to 5G. It will be! The question is whether operators will design and control the platform or merely consume it.

[2] The term draws its rhetorical weight from the political/philosophical notion of a "***social compact***" (Rousseau, Locke): the idea of a binding agreement that defines rights, obligations, and sovereignty. Here it is repurposed to describe the architectural sovereignty question operators must resolve for 6G: what must they own, what can they federate, and what should they consume as commodity.

This paper argues that 6G represents a decisive inflection point requiring operators to reorder five fundamental priorities – not incrementally, but structurally. As illustrated in Figure 1, these five "reorderings" are:

- ***Control First***: Operators must own the software-defined control plane.
- ***Customer First***: Networks must deliver verifiable outcomes, beyond peak data rates.
- ***Business First***: The emerging Guarantee Economy demands outcome-based contracts, not connectivity pricing.
- ***Operations First***: Networks must operate as software, with agentic artificial intelligence (AI) at the helm.
- ***Technology Last***: Architecture and enabling technologies must serve the four priorities above, not drive them.

***Control First*** addresses the most fundamental failure mode of the vendor-driven era: operators do not own the software that governs their own networks. Across 2G through 5G, the operations support system (OSS), business support system (BSS), RAN, and core have been delivered as vertically integrated stacks in which the vendor controls the data model, the upgrade cycle, the integration interface, and increasingly the AI inference layer. An operator wishing to change a network behavior must raise a change request with its vendor and wait – sometimes years – for a software release.

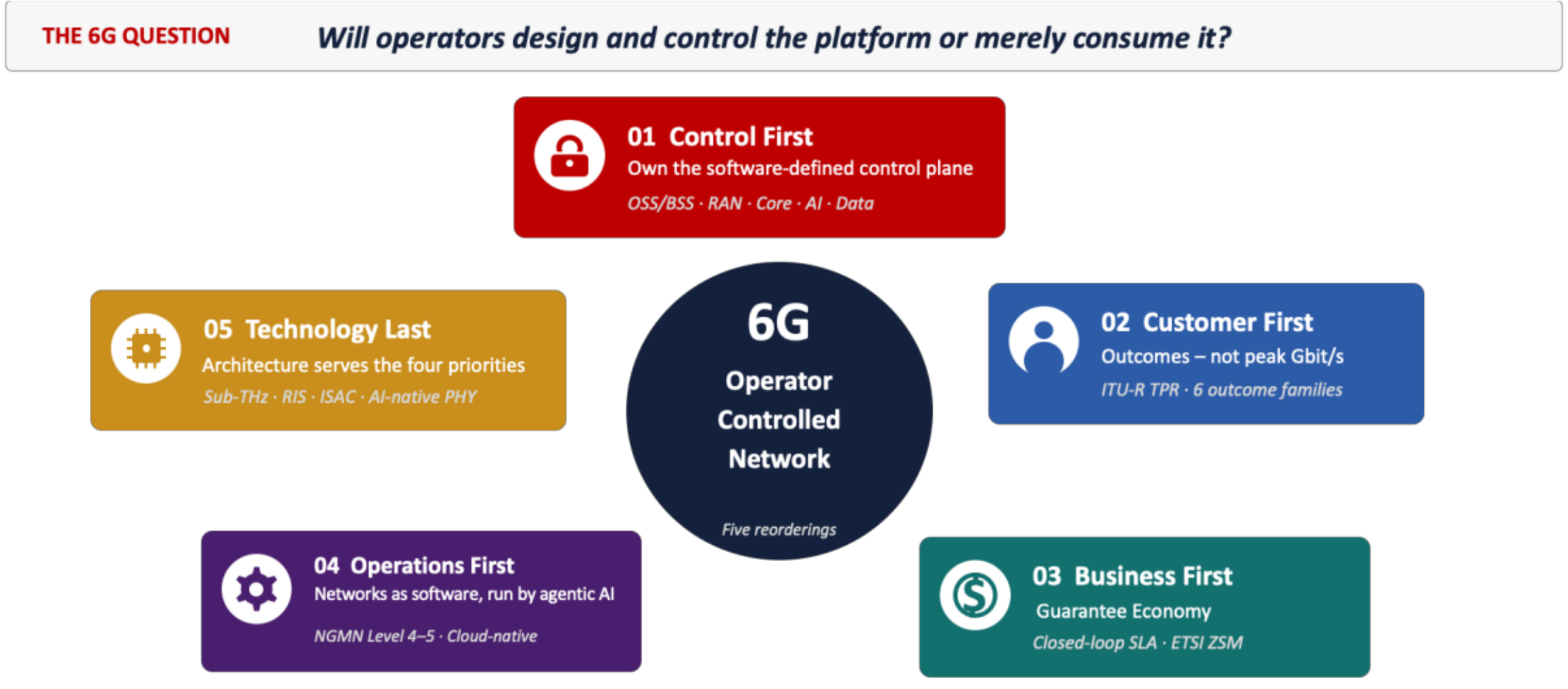

*Figure 1. Five-axis priority reordering; Technology Last is a sequencing principle, not a downgrade.*

This is not a procurement inconvenience; *it is a structural loss of sovereignty*. For 6G, the Control First principle demands that operators define and own the control plane: the software layer that expresses policy, enforces service guarantees, manages spectrum, and directs AI-driven automation. Rakuten Mobile demonstrated that this is achievable at scale: its network runs on a fully disaggregated, software-defined architecture in which Rakuten Symphony controls the complete software stack, from the distributed unit (DU) to the core and orchestration layer. The 6G Control Compact must encode this model as the default, not as an exception.

***Customer First*** reorders what networks are optimized for. Since 3G, mobile network generations have been marketed around peak data-rate specifications: 3G promised megabits per second, 4G Long-Term Evolution (LTE) delivered tens of megabits, and 5G New Radio (NR) reaches multiple gigabits per second under controlled conditions. The ITU-R IMT-2030 technical performance requirements (TPR) [3] for 6G extend this further, specifying a peak downlink of 36 Gbit/s and uplink of 18 Gbit/s, with user-plane latency of 4 milliseconds for immersive communications (IC) and 1 millisecond for hyper-reliable low-latency communications (HRLLC), and a connection density of $10^6$ devices per square kilometer. These are necessary engineering targets. However, they are not what enterprise customers and consumers purchase. Customers purchase *outcomes*: a video call that does not drop, a surgical robot that does not lag, a private campus network that guarantees throughput to every machine tool, etc. The Customer First reordering demands that 6G networks be designed and measured against outcome delivery, such as, e.g., latency experienced, reliability realized, and coverage provided, rather than peak specifications achieved under optimal conditions. This shift requires verifiable service-level agreement (SLA) enforcement in real time, which in turn requires operator control of the AI substrate that monitors, predicts, and adapts network behavior.

***Business First*** recognizes that the connectivity pricing model, e.g., charging for megabytes transferred or subscriptions held, has reached its structural ceiling. Average revenue per user (ARPU) in most mature markets has been flat or declining for a decade, even as network investment has grown. The emerging Guarantee Economy offers a different model: *operators contract to deliver specific, quantifiable, measurable outcomes and are compensated accordingly*. A manufacturing enterprise does not want to pay for bandwidth; it wants to pay for zero-defect automated quality inspection enabled by a guaranteed 1 ms latency link. A media company does not want to pay for data volume; it wants to pay for immersive 8K volumetric streaming delivered without interruption to a stadium of 80,000 concurrent users. These outcome-based contracts require operators to instrument their networks deeply, model the relationship between network state and application performance, and provide contractual guarantees backed by real-time AI-driven assurance. The Business First reordering is therefore inseparable from Control First and Customer First: *without owning the control plane and measuring outcomes, an operator cannot price on outcomes.* AI traffic today represents approximately 0.06% of total mobile data volume, but AI-native sessions already exhibit a fundamentally different per-session traffic profile: an uplink-to-session-traffic ratio of approximately 26%, compared with around 10% for conventional broadband sessions [4] – driven by the continuous upload of context, sensor data, and intermediate reasoning artefacts that agentic AI requires. As AI session shares grow, this per-session characteristic will structurally reshape how networks must be dimensioned and priced.

***Operations First*** confronts the unsustainable complexity of running a 6G network with human-scale tooling. A fully instrumented 6G network will generate data volumes, event rates, and configuration decisions that no operations team can process manually. The Next Generation Mobile Networks Alliance (NGMN) has identified network simplification as a prerequisite for viable 6G operations [5]. The Operations First reordering goes further: it requires that the network be designed as a software system from the outset, with agentic AI: AI modules that can observe, reason, plan, and act across the full network stack without per-action human approval, as the primary operational agent. This does not eliminate human operators; it elevates them from reactive fault responders to strategic supervisors of an AI-driven system. Rakuten Mobile's deployment has already demonstrated early elements of this model, with closed-loop automation handling fault detection, traffic steering, and capacity management at a scale and speed impossible with manual processes. The 6G target is a zero-touch network in which agentic AI manages the full lifecycle, namely *provisioning, assurance, optimization, and decommissioning*, while human oversight is reserved for policy and exception handling.

***Technology Last*** is the most counterintuitive of the five reorderings, and the most important. It does not mean technology is unimportant, conversely, the innovations embedded in 6G physical layer, from sub-terahertz (sub-THz) spectrum to reconfigurable intelligent surfaces (RIS) and integrated sensing and communication (ISAC), are genuinely transformative. What it means is that technology selection must follow from the four preceding priorities, not precede them. NTT DOCOMO's three-wave model of mobile service evolution (connectivity, programmability, and cyber-physical fusion [6]) captures the direction of travel: 6G is the network in which the digital and physical worlds merge, enabling applications from digital twins to AI-native services that sense, reason about, and act upon the physical environment. These applications define the requirements; technology must serve those requirements. When operators allow technology roadmaps, driven by vendor research agendas and standards committee momentum, to define the network, they reproduce the dependency cycle that has characterized 2G through 5G. *Technology Last means operators set the priorities, define the use cases, specify the outcomes, and then procure or develop the technology that serves them*.

The priority reordering is a direct response to a structural failure in the current model, and each assertion is grounded not in aspiration but in operational evidence drawn from the world's first fully cloud-native, Open Radio Access Network (Open RAN) greenfield operator: Rakuten Mobile, in Japan.

Rakuten Mobile occupies a unique position in this debate. It is the only operator that has already built, deployed, and operated a fully cloud-native, Open RAN network at national scale, exceeding 10 million subscribers in 2025 and reaching positive EBITDA in FY2025 – its first full-year EBITDA profitability since launch [7]. These are not pilot metrics; they are the financial outcomes of a national commercial deployment, proving that the Control First model is not a future aspiration but a present reality. Through Rakuten Symphony, this architecture has been commercialized as a platform engaged with operator partners across multiple markets, with the segment achieving operating-income breakeven in FY2025. The lessons encoded in that deployment, about what operators must own, what they can federate, and what they should consume as commodity, form the empirical backbone of this paper.

The five reorderings are not a wish list; they are a design specification derived from operational experience. The industry now faces a choice. It can approach 6G as it has approached every previous generation – as a technology upgrade cycle led by vendors and ratified by operators – and arrive at the same structural dependency, compounded by AI systems it does not control. Or it can treat 6G as the once-in-a-generation architectural reset it is and build the Operator-Controlled Network from the outset. This paper makes the case for the latter and provides the blueprint.

This paper makes four original contributions. First, the *6G Control Compact*: a three-tier ownership taxonomy (own / federate / consume) that provides explicit, standards-grounded criteria for classifying each layer of the 6G stack by strategic operator value – distinct from prior SDN/NFV disaggregation frameworks in that it integrates AI substrate ownership and data sovereignty as first-class design dimensions. Second, the *Guarantee Economy*: a five-tier outcome-priced commercial model with specific SLOs, breach consequences, and billing mechanisms that translates network control into contractually enforceable service products – extending ETSI ZSM and TM Forum IG1252 into a complete commercial architecture. Third, *Rakuten Mobile's FY2025 operational evidence* as empirical validation that a fully cloud-native, Open RAN operator can achieve national-scale deployment and commercial profitability, providing the first publicly documented proof point for the Control Compact's technical and business feasibility [7]. Fourth, *a three-phase deployment roadmap* (2025–2027 / 2027–2029 / 2029–2032+) with explicit standards milestones, risk scenarios, and seven stakeholder-specific calls to action that operationalize the framework for immediate industry adoption.

The remainder of this paper is organized as follows: Section I establishes the Control Compact and the three phases of telecom evolution; Section II defines the six outcome families; Section III presents the Guarantee Economy and three revenue engines; Section IV describes the agentic AI operating model; Section V details the end-to-end 6G architecture; Section VI addresses zero-trust security, data sovereignty, and sustainability; Section VII presents the phased deployment roadmap and calls to action. Annexes A–I provide supporting reference material.

# Section I – Control First: Why 6G Must Be Operator-Owned

## Related Work

The operator sovereignty argument advanced in this paper is positioned against four bodies of prior work. On 6G vision, requirements, and AI integration: the ITU-R IMT-2030 framework [1]-[3] and 3GPP study on 6G use cases and service requirements [8] establish the standards baseline;

NTT DOCOMO's 6G vision [6] provides the cyber-physical fusion context; Chatzieleftheriou and Liotou [9] survey the landscape of AI techniques applicable to 6G, mapping challenges across PHY, RAN, and network management that inform the AI substrate requirements of the Control Compact; Zheng et al. [10] demonstrate AI-native physical layer design with cross-module optimization, grounding Section V's Stage 2–3 air interface evolution in published prototype evidence; and the AI-RAN Alliance architecture and working group reports [11][12][13] provide the deployment framework for AI-for-RAN and AI-on-RAN that this paper's Guarantee Economy commercial model depends on. On Open RAN and RAN-core convergence: Salmi et al. [14] address real-time adaptation and conflict resolution in AI-native O-RAN architectures, identifying the multi-vendor integration challenges that Section V acknowledges; the O-RAN Alliance nGRG reports on Near-RT RIC [15], dApps [16], and scalable RAN [17] provide the specification baseline for Section V's three-tier intelligence architecture; Harkous et al. [18] examine RAN-core convergence for the 6G user plane, which informs the distributed UPF placement and E3 interface design discussed in Section V. On autonomous network management and agentic AI: the ETSI ZSM reference architecture [19] and NaaS/Agents group reports [20][21] define the closed-loop automation framework against which this paper's Level 4–5 autonomy targets are stated; TM Forum IG1252 [22] provides the Autonomous Networks Levels scale; the NGMN agentic AI framework [23] defines the Level 3–5 taxonomy; Sun et al. [24] survey explainability requirements across 6G communications and network slicing – covering XAI techniques from PHY to service layer – that reinforce Section IV's argument for operator-auditable AI substrate; and Provvedi et al. [25] present an intent-driven LLM framework for automated network configuration that directly instantiates the Intent → Observe → Decide → Act loop described in Section IV. On security for autonomous and AI-native networks: NIST SP 800-207 [26] establishes zero-trust architectural principles; ETSI GR ZSM 017 [27] applies them to closed-loop automation; Choudhary et al. [28] address post-quantum transition for 5G-AKA, grounding Section VI's PQC migration discussion in published protocol-level analysis; and Altintaş et al. [29] survey adversarial threats specific to AI-native 6G systems – including prompt injection and goal misgeneralization – that Section VI's LLM-security paragraph addresses. *The gap this paper fills is the absence of an integrated framework connecting operator control architecture, outcome-based commercial models, autonomous operations, zero-trust security, and a phased deployment roadmap in a single standards-grounded treatment validated by operational evidence at national commercial scale.*

Over two decades of network evolution, mobile operators have ceded architectural control to equipment vendors through proprietary stacks, monolithic platforms, and standards processes shaped by suppliers rather than service providers. The consequence is a structural dependency that erodes operator margins, constrains innovation velocity, and ultimately threatens the operator's role as a sovereign infrastructure provider. 6G offers a once-in-a-generation opportunity to reverse this trajectory, but only if operators design the *Control Compact* from the outset, before standards are frozen, before ecosystems calcify, and before procurement decisions lock in the next decade of dependency.

## The Control Illusion: Two Decades of Outsourced Architecture

The history of mobile network generations is conventionally told as a story of technical progress: each generation added speed, capacity, and new service classes, broadening the reach and utility of mobile connectivity. This narrative is accurate as far as it goes. What it omits is the corresponding trajectory of operator agency, i.e., the degree to which operators have been architects of their own networks rather than consumers of pre-packaged infrastructure. Measured against this dimension, the history of mobile looks quite different. With each generation, the complexity of the network increased, the integration surface between subsystems deepened, and the practical cost of switching vendors or modifying core behaviors rose. The 2G Global System for Mobile Communications (GSM) era established the pattern: a standards-defined air interface combined with proprietary vendor implementations of the core and OSS/BSS, creating a de-facto lock-in even within a formally open standard.

The 3G Universal Mobile Telecommunications System (UMTS) and 4G LTE generations intensified this dynamic. The shift to all-IP architecture in 4G was genuinely transformative, but the Evolved Packet Core (EPC) was delivered almost universally as a monolithic appliance from a handful of vendors, with proprietary northbound interfaces into OSS/BSS stacks that were themselves vendor specific. An operator running a four-vendor network, a common situation in large national deployments, would typically operate four separate OSS instances, with no common data model, no shared automation layer, and no ability to implement cross-domain AI optimization without bespoke integration work costing tens of millions of dollars. This is not a failure of engineering skill on the operator side; it is the predictable outcome of a supply chain structured around vendor-controlled integration.

5G's Service-Based Architecture (SBA) and the Network Functions Virtualization (NFV) paradigm that preceded it offered genuine promise. By decomposing the core into software network functions (NFs) running on commercial off-the-shelf (COTS) hardware, NFV appeared to commoditize the infrastructure layer and restore operator control at the software level. In practice, the transition was slower and more constrained than anticipated. Virtual network functions (VNFs) were frequently ports of existing appliance software rather than genuinely cloud-native designs; the management and orchestration (MANO) layer, intended to provide vendor-agnostic lifecycle management, became another integration battleground; and the AI/ML capabilities that vendors embedded in their 5G stacks –for predictive maintenance, traffic steering, and energy optimization – were deliberately opaque, locking the data and the inference layer inside the vendor's system. The NGMN Framework for Network Simplification [5] identifies this fragmentation and opacity as a primary driver of operational cost and a barrier to realizing the automation promise of 5G.

The Open RAN movement, formalized through the O-RAN Alliance's interface specifications, represents the most significant attempt to break this cycle in the RAN domain. By defining open fronthaul, midhaul, and management interfaces, and by separating the RAN Intelligent Controller (RIC) from the distributed and centralized units (DU/CU), O-RAN created the architectural preconditions for multi-vendor RAN deployments and third-party xApps running on the Near-RT RIC. This is an important and necessary step. It is not sufficient. Open interfaces at the RAN layer do not, by themselves, give operators control of the full network stack. An operator can deploy a multi-vendor O-RAN while still running a proprietary core, a vendor-locked OSS, and AI models it cannot inspect or modify. The vendor lock-in spiral operates at every layer simultaneously; addressing one layer without a systemic model for operator ownership reproduces the problem elsewhere. NTT DOCOMO's analysis of service evolution [6] reinforces this point: the shift to cyber-physical fusion in 6G will require operators to integrate AI, sensing, and compute in ways that no single-layer openness initiative can address.

Data and AI access problems deserve particular emphasis because it is the dimension most likely to define competitive outcomes in the 6G era. AI-driven network optimization depends on access to high-fidelity, real-time telemetry from every layer of the network stack (physical layer measurements, protocol state, user-plane performance, and application quality signals). In current vendor-operated stacks, these data either do not exist in accessible form, are aggregated beyond utility, or are available only through proprietary APIs under vendor-controlled licensing terms. An operator that cannot access its own network's data cannot train its own AI models, cannot audit vendor AI claims, and cannot build differentiated automation capabilities. It is, in effect, renting intelligence it does not own from a vendor whose incentives are not aligned with the operator's. This is the Control Illusion at its most consequential: operators believe they are operating their networks, when in practice the networks are operating themselves according to logic the operator cannot see, modify, or own.

## Three Phases of Telecom Evolution: From Pipe to Platform

Understanding what 6G must become requires understanding what telecoms have been. As illustrated in Figure 2, the industry's trajectory can be characterized as three distinct phases of evolution, each defined by what operators do with their infrastructure and what value they create from it. These phases are not strictly sequential network, but they represent progressively more sophisticated relationships between network capability and economic value.

***Telecom Evolution Stage 1 – Connectivity:*** defines the operator as a dumb bit-transporter. In this phase, which dominated from 1G through the early 4G era, the operator's primary function is to move information from one point to another reliably and at acceptable cost. The network is a utility: customers pay for access, measured in minutes of voice or megabytes of data, and the operator's competitive differentiation is primarily geographic coverage, price, and raw throughput. This model generated enormous revenues during the period of mobile adoption, when simply having a mobile connection was itself the value proposition, but it contains the seeds of its own commoditization. As connectivity became ubiquitous, the differentiation collapsed, and operators found themselves competing primarily on price in markets with high capital intensity and regulatory constraints on spectrum and infrastructure sharing. The operator as a pure connectivity provider is structurally vulnerable to anyone who can deliver bits more cheaply.

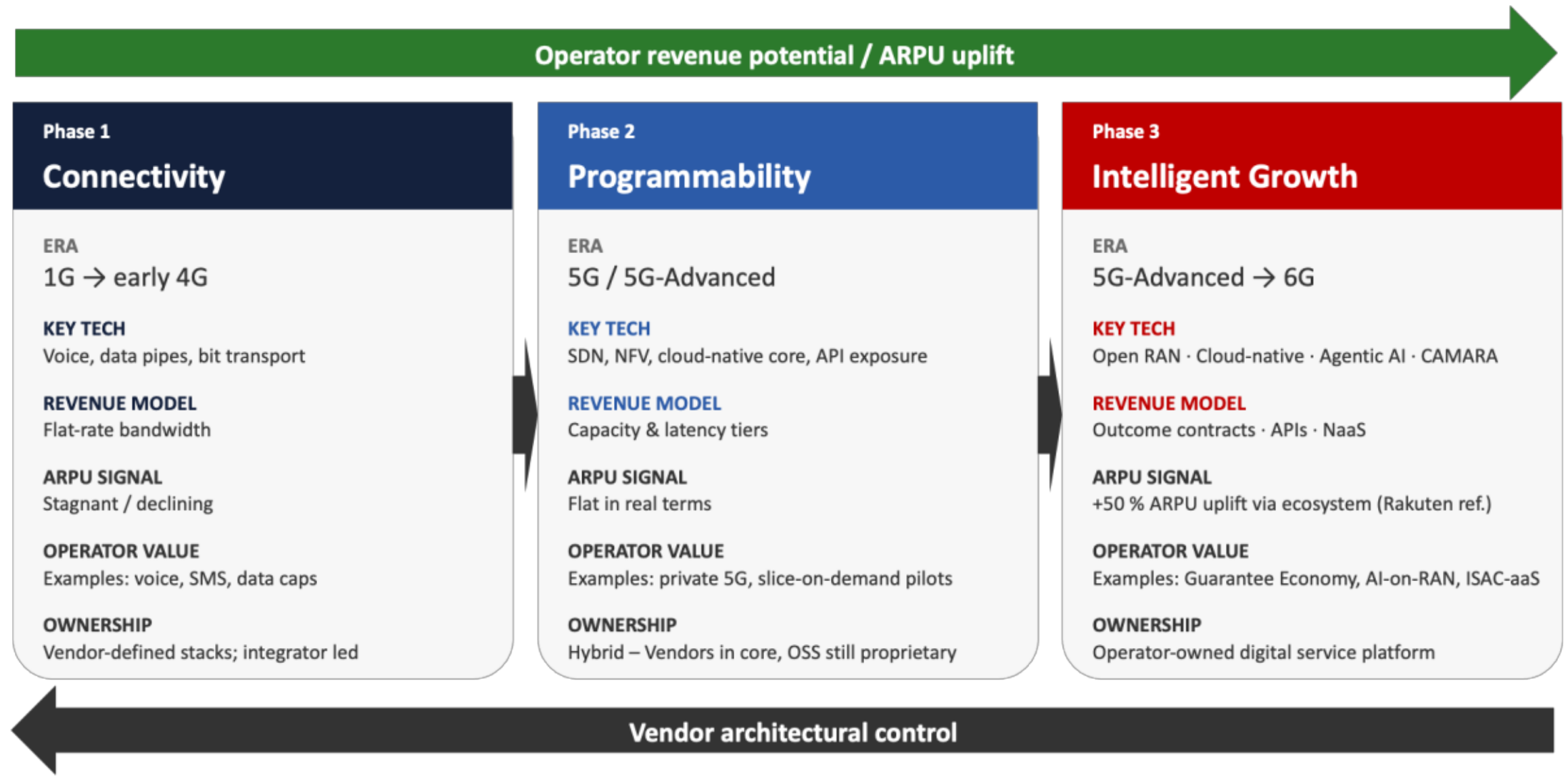

*Figure 2. Operator revenue potential rises as vendor control falls; Phase 3 reverses two decades of disaggregation drift.*

***Telecom Evolution Stage 2 – Programmability:*** marks the operator's first attempt to reclaim the software layer. Driven by Software-Defined Networking (SDN) and NFV, and later by the cloud-native 5G core architecture, Phase 2 replaces purpose-built hardware appliances with software running on commodity compute. The theoretical promise of programmability is that operators can define their own network behaviors, create differentiated service slices, and build APIs that expose network capabilities to third-party developers, transforming the network from a closed pipe into an *open platform*. In practice, the transition to Phase 2 has been partial and uneven. Most operators have virtualized their infrastructure without achieving genuine programmability; they run vendor software on commodity hardware, but the software remains as closed, and vendor controlled as the appliances it replaced. The 5G network slice promise that operators could offer guaranteed, isolated virtual networks to enterprise customers has been realized in proof-of-concept deployments but rarely at commercial scale, largely because the orchestration and assurance capabilities required to deliver and guarantee slices across a multi-vendor network remain immature.

***Telecom Evolution Stage 3 – Intelligent Growth:*** is the phase that 6G must fully realize: the operator as a digital service platform owner. In this phase, the network is not merely a programmable infrastructure but an intelligent service delivery system that can sense application requirements, guarantee outcomes, expose capabilities through well-defined APIs, and price based on value delivered rather than resources consumed. Phase 3 requires everything that Phases 1 and 2 built (ubiquitous coverage, cloud-native infrastructure), but adds three capabilities that previous generations have not delivered at scale:

- Real-time AI-driven assurance of guaranteed SLAs.
- A data layer that the operator owns and can monetize.

- A business model that maps network outcomes to customer outcomes and prices accordingly [30].

The cyber-physical fusion applications that NTT DOCOMO identifies as the defining use cases of 6G [6] – digital twins, AI-native services, immersive extended reality (XR), and autonomous system coordination –all require Phase 3 capabilities. They cannot be delivered from a Phase 1 or Phase 2 network.

The practical implication of this three-phase model is that 6G network design must explicitly target Phase 3 from the outset. This means making architectural choices (e.g., about where AI runs, who owns the data model, how SLAs are instrumented and enforced, and how capabilities are exposed to applications) that are informed by the monetization model, not just the radio performance specifications. *An operator that designs a 6G network optimized for Phase 1 throughput and Phase 2 virtualization, and then attempts to retrofit Phase 3 monetization, will reproduce the integration debt and vendor dependency that has characterized every previous generation.* The transition from connectivity to digital service platform is not a feature that can be added later; it is an architectural posture that must be established first.

## The 6G Control Compact: What Operators Must Own, Federate, and Consume

The central design question for 6G operator strategy is not “how much should we build versus buy?” It is “what must we own to remain sovereign?” The answer requires a three-layer model, illustrated in Figure 3, that distinguishes between what operators MUST OWN (because ceding it creates structural dependency), what they can FEDERATE (because shared ownership creates mutual value without dependency), and what they should CONSUME AS COMMODITY (because ownership creates cost without strategic benefit). This 6G Control Compact provides a framework for every architectural and procurement decision in the 6G era.

Operators MUST OWN eight core elements (mapped in full in Annex D) the most strategically critical of which are examined here.

- First, the ***control plane***: the software layer that expresses network policy, enforces service guarantees, manages spectrum allocation, and coordinates cross-domain behavior. This is the operator's fundamental act of sovereignty. A vendor-controlled control plane is structurally equivalent to a national government that has outsourced its legislative process.
- Second, the ***AI substrate***: the data pipelines, model training infrastructure, inference engines, and feedback loops that drive automated network behavior. AI that the operator does not own cannot be audited, cannot be modified to serve operator-specific objectives, and creates a more profound dependency than any previous form of vendor lock-in [23][30].
- Third, the ***data layer***: raw telemetry, processed analytics, customer behavior data, and network performance records. These data are the raw material for AI training, SLA verification, and business model innovation; an operator that cannot access its own data are permanently dependent on vendor interpretations of network state.

- Fourth, the ***OSS/BSS***: the operational and business support systems that translate network state into customer experience and business outcomes.
- Fifth, ***spectrum policy and licensing***: the physical layer asset that underpins all network value and must remain under direct operator control.

Operators can FEDERATE three categories of capability:

- ***Roaming and interconnect agreements*** –the bilateral and multilateral arrangements that allow subscriber mobility across operator boundaries –benefit from shared governance without requiring single-operator control.
- ***API platforms that expose network capabilities to third-party developers*** –such as the GSMA Open Gateway initiative [31] – can be operated as industry utilities, distributing the cost of developer ecosystem development while allowing individual operators to differentiate on top.
- ***Shared infrastructure in rural or low-density environments*** – towers, fiber backhaul, spectrum sharing arrangements –reduces capital expenditure without compromising competitive differentiation in urban high-value markets.

Federation is appropriate where the shared governance does not compromise the operator's ability to control its own service quality, data access, or pricing.

Operators should CONSUME AS COMMODITY those elements where ownership creates cost without strategic differentiation:

- ***Cloud infrastructure-as-a-service (IaaS)*** – the compute, storage, and networking fabric on which cloud-native network functions (CNFs) run –is a commodity that hyperscale cloud providers can deliver more efficiently than any individual operator.
- ***Open-source CNFs*** –where the source code is publicly available and the community of contributors is broad enough to prevent any single-vendor capture – can be consumed without creating strategic dependency, provided the operator retains the ability to modify and contribute.
- ***Hardware –radio units (RUs)***, server hardware, optical transport –follows the same logic: *as long as open interfaces prevent vendor lock-in at the hardware layer, commodity procurement drives down costs without sacrificing control*. The critical discipline is to maintain the boundary: commodity consumption must not drift upward into the control plane, the AI substrate, or the data layer.

An example of compelling available proof point for the Control Compact in practice is at Rakuten Mobile, in Japan. The Rakuten Mobile deployment is built on Symphony's software platform, which owns the complete software stack from the RAN layer –including the Central Unit (CU), Distributed Unit (DU), and RU management –through the 5G core, orchestration, OSS/BSS, and AI/ML automation layer. The platform is cloud-native from inception, not by retrofit: every network function is a containerized microservice, managed by Kubernetes, with open APIs exposed at every layer. This architecture gives Rakuten Mobile what no other operator has published equivalent operational evidence at national scale: the ability to modify any network behavior, retrain any AI model, access any network data point, and deploy any new capability, independently of any equipment vendor's roadmap or release cycle. Cloud-native, open-source disaggregated architecture is the necessary – yet not merely sufficient – condition for this control: a proprietary cloud-native platform would recreate vendor lock-in at the software layer, and a non-cloud-native open platform would reproduce the operational inflexibility of legacy RAN. It is the specific combination of open-source foundations, cloud-native execution, and operator-owned control plane that makes the Control Compact architecturally achievable and commercially durable. The commercial outcomes of this architectural choice are now quantified [7]: more than 10 million subscribers served, full-year EBITDA profitability achieved in FY2025 for the first time, operating-income breakeven, and a continuing year-on-year improvement in network quality satisfaction – driven by continuous cloud-native software updates rather than hardware replacement cycles. It is the Control Compact in operation, and it demonstrates that the model is not theoretically attractive but practically feasible at national scale.

The Control Compact is achievable for *greenfield operators* by design, but the industry majority consists of brownfield incumbents carrying existing vendor agreements, embedded OSS/BSS systems, and RAN hardware with remaining years of depreciated value. For these operators, migration to an operator-controlled architecture is a multi-year programme, not a single procurement decision. The practical path is *sequencing*: new spectrum deployments and new market entries should adopt Control Compact architecture from the outset, while existing infrastructure migrates as vendor contracts expire and equipment reaches end-of-life.

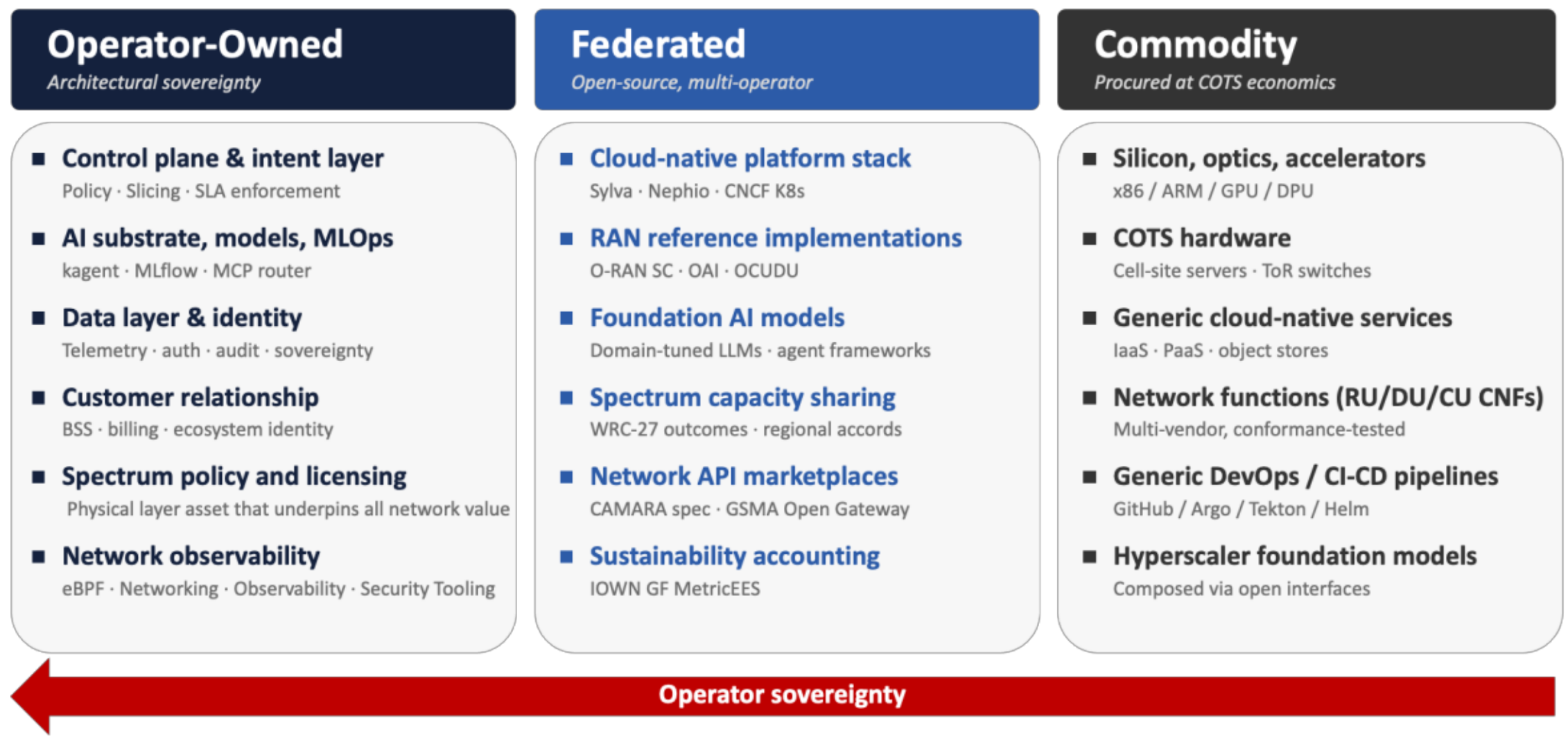

***Figure 3. Operator sovereignty is allocated by layer: control plane is owned, ecosystems are federated, hardware and IaaS are commoditized.***

*Operators that defer even this sequencing decision, defaulting to like-for-like vendor replacements at each refresh cycle, will find themselves carrying the same structural dependency into the 6G era.* The NGMN Network Simplification initiative [5] documents the specific brownfield barriers that must be sequenced through proprietary appliance-based RAN architectures, vendor-specific OSS/BSS integration, and the engineering skills transition from appliance management to cloud-native operations. The phased roadmap in Section VII provides a concrete sequencing framework for operators beginning this transition from existing deployments. The minimum viable first step for a brownfield incumbent that cannot immediately restructure its core vendor relationships is threefold: require CNTi CNF conformance [32] and open API documentation as contractual conditions for all new network function procurement from 2026 forward; deploy CNCF-conformant MLOps tooling, such as, but it is not limited to, MLflow, Prometheus, open-source observability, alongside existing network functions on a parallel data plane; and insert data access rights as a non-negotiable clause in all vendor contract renewals, ensuring the operator can extract its own telemetry and model artefacts independently of vendor consent. None of these steps require replacing existing RAN hardware; all three begin accumulating the operational capability the Control Compact requires.

The Rakuten evidence is most precisely interpreted as proof of technical feasibility for the Control Compact architecture, not necessarily as a directly replicable commercial roadmap for brownfield incumbents. Rakuten's advantages were structural: a greenfield license awarded in 2019 with no legacy infrastructure to migrate, a cloud-native toolchain that had matured by 2020, and access to a Group ecosystem of more than 100 million users as a platform multiplier requiring decades of parallel investment to construct. Published operator journeys toward cloud-native Open RAN – including Vodafone's Open RAN deployments in the United Kingdom and Germany, DISH Network's greenfield US deployment, and NTT's IOWN program – confirm that the technology stack is replicable but that migration timescales of five to ten years are realistic for incumbents managing live customer traffic, existing vendor contracts, and accumulated technical debt. The conclusion this paper draws is that the Control Compact represents the direction all operators must move toward, and that the technical barriers to that movement are lower than the organizational and commercial ones.

The Control Compact also confronts a structural competitive threat that must be named directly: *hyperscale cloud providers* (AWS, Microsoft Azure, and Google Cloud) are actively building toward the same control-plane value the Control Compact reserves for operators. AWS Wavelength zones embed computing at carrier infrastructure. Azure Private MEC extends operator-deployed edge under Azure management. Google's network partnerships position its AI platform inside operator infrastructure. These moves are not merely complementary; they are *structurally competitive*, advancing with engineering capability and capital that no individual MNO can match unilaterally. The operator's defensible advantage is not technology in isolation: it is the *combination of licensed spectrum, physical presence at the radio access edge, carrier-grade SLA accountability under national regulatory licensing, and existing subscriber relationships that no hyperscaler can replicate from a central cloud region*. The Control Compact argument is precisely that *this* *advantage is durable only if operators own the software layer that sits above the spectrum and below the application*, because that is the layer hyperscalers are building toward from above. *Operators that cede this layer under the logic that hyperscaler infrastructure is more cost-efficient will find that the apparent savings come at the cost of the strategic position the layer represents*. Section VII addresses how operators should structure hyperscaler engagement to preserve rather than dissolve this boundary.

The 6G Control Compact has direct implications for how operators engage with standardization bodies, vendor roadmaps, and enterprise customers. In standardization, operators must advocate for interface specifications that preserve owner control of the *AI substrate and data layer*, not merely open-air interfaces. In vendor engagement, operators must require contractual guarantees of data access, model transparency, and API stability as conditions of procurement. In enterprise sales, *operators can only offer guaranteed outcomes if they own the assurance layer that verifies and enforces those outcomes in real time*. The Control Compact is not simply an internal architecture choice; it is the prerequisite for every other strategic objective in the 6G era. The commercial architecture that the Control Compact makes possible (the Guarantee Economy, Programmable Network APIs, and Network-as-a-Service) is examined in Section III.

## What 6G Must Not Repeat: Failure Modes and Forward Commitments

The most important design inputs for 6G are not the use-case visions or the radio performance targets. They are the *failure modes* of every previous generation, i.e., the specific architectural decisions that created the structural dependencies operators now carry. Five failure modes are particularly consequential and must be explicitly designed against.

- ***The first failure mode is AI and data locked in vendor black boxes***. As AI becomes the primary mechanism by which 6G networks are configured, optimized, and assured, the question of who owns the AI layer becomes existential. 3GPP TR 38.914 [33] identifies AI/ML as a fundamental component of 6G RAN design, with native AI-based air interface adaptation, beam management, and channel estimation. If these AI components are delivered as opaque, vendor-proprietary systems, as has been the case with vendor-embedded AI in 4G and 5G, operators will be in the position of operating networks whose behavior they cannot explain, audit, or modify. The Zero-touch Service Management (ZSM) framework developed by ETSI [19] provides a reference architecture for end-to-end network automation, but its value depends entirely on operators having access to the data and control interfaces that feed the automation engine. ZSM without data access is an automation framework with no fuel.
- ***The second failure mode is standards written by vendors for vendors***. The 3GPP technical specification process has delivered the technical foundations of every mobile generation and deserves frank credit for enabling global interoperability. It has also, systematically, produced specifications that reflect the capabilities of the dominant equipment vendors more than the operational requirements of service providers. Feature complexity has grown with every release; the option space has expanded; and the practical cost of full standards compliance has risen to the point where only a handful of global vendors can implement a complete stack. This dynamic must change for 6G. Operators must engage at the level of requirements definition, not merely feature review, and must be willing to oppose features that serve vendor differentiation rather than operator or customer needs. The NGMN Framework for Network Simplification [5] is a direct response to this failure mode, advocating for leaner specifications, reduced optionality, and explicit consideration of operational cost in the standards process.

- ***The third failure mode is RAN integration lock-in through systems integration dependency***. Open RAN has established the principle of open interfaces, but integration complexity remains a significant barrier. In practice, multi-vendor Open RAN deployments require extensive testing, tuning, and ongoing management of inter-vendor interactions: work that is currently performed either by the operator's own engineering teams (at high cost) or by a systems integrator (creating a new form of dependency). For 6G, the integration layer itself must be designed as an operator-owned capability: a software-defined integration fabric that can incorporate new vendors and new network functions without bespoke integration work for each combination. This is achievable through consistent cloud-native design patterns, open data models, and AI-assisted integration tooling, e.g., capabilities that Rakuten Symphony has developed and deployed as part of its commercial platform.
- ***The fourth failure mode is open-source platform capture by dominant contributors***. For instance, Nephio [34], Sylva [35], Cilium [36], O-RAN SC [37], and kagent (and kmcp) [38] are examples of open-source foundations that prevent vendor lock-in, but open-source communities can be captured, forked, or marginalized when a small number of dominant commercial contributors align a project's direction with their proprietary interests. The pattern is documented: Kubernetes is governed by CNCF [39] with formal neutrality requirements, yet the practical enterprise implementations –AWS EKS, Azure AKS, Google GKE – are instances of vendor-specific managed services that reintroduce lock-in at a higher abstraction layer. For 6G, operators must exercise their upstream contributor role actively enough to constitute a countervailing influence against any single vendor's project dominance. The safeguards are CNCF's graduated project neutrality requirements, O-RAN Alliance's contributor licensing framework [40], and the operator community's collective obligation to maintain engineering presence in these communities at a level that makes strategic redirection by any single contributor detectable and contestable. A governance failure mode distinct from technical capture is standardization drift, where open-source reference implementations diverge from their specification baselines, creating de facto ecosystem fragmentation that benefits the vendor with the largest proprietary deviation. Operators must collectively fund the conformance testing infrastructure – 3GPP's testing event program, the O-RAN Alliance plugfest series, and CNCF's cloud-native network function conformance suite – that keeps open-source implementations standards-aligned, rather than leaving conformance testing funded exclusively by the vendors whose interests include selective standards interpretation.

The forward commitments implied by these failure modes are clear. Standards and technical specification bodies, such as 3GPP, ITU-R, IETF, and the O-RAN Alliance, must explicitly incorporate *operator sovereignty requirements* into their work programs: mandatory open data interfaces, AI transparency requirements, and complexity reduction as a primary design criterion. The NGMN Alliance provides the most effective organizational mechanism for this: as the principal operator requirements bloc in 3GPP and ITU-R, NGMN can translate collective operator intent into structured requirements documents that standardization bodies must respond to, as its Network Simplification initiative [5] and Agentic AI Operating Models framework [23][30] have already demonstrated. Operators that are not active NGMN participants are, in effect, delegating their standards representation to vendors. Operators must invest in the engineering and commercial capabilities required to exercise ownership, not merely to procure open-interface products, but to design, integrate, and operate architectures they genuinely control. And the industry must develop shared frameworks for what operator ownership means in a cloud-native, AI-driven network – frameworks like the 6G Control Compact proposed in this paper – so that the diversity of operator approaches does not fragment the ecosystem that 6G's scale benefits depend on. 6G will either be the generation in which operators reclaim their architecture, or the generation in which they permanently concede it. There is no neutral outcome.

# Section II – Customer First: The Outcomes 6G Must Deliver

## The Specification Trap: Why Technical KPIs Fail the Market

The mobile industry has an uncomfortable habit of celebrating the metrics it can measure rather than the outcomes that matter to the people and organizations it serves. Fifth-generation networks were marketed to the world on headline peak data-rate targets of up to 36 Gbit/s downlink [3], figures that captured boardroom imagination but reflected laboratory conditions achievable only with a single user in full line-of-sight of a millimeter-wave base station. No commercially deployed application, no enterprise workflow, and no consumer device has ever required or sustained that rate in operational conditions. ITU-R M.2160 [2] articulates the IMT-2030 framework with similarly ambitious aggregate targets: a downlink peak of 36 Gbit/s and an uplink peak of 18 Gbit/s per cell according to the ITU-R Technical Performance Requirements [3], yet the gap between these standardized maxima and the performance experienced at the cell edge, during peak-hour congestion, or inside a dense industrial building remains a structural feature of every generation of mobile technology, not an engineering anomaly to be corrected in the next release. The implication for 6G is profound: if the industry once again leads with gigabit per second throughput as its primary commercial narrative, it will repeat the disappointment cycle that has undermined enterprise confidence in 5G.

Enterprises do not purchase gigabit per second, milliseconds, or antenna elements in isolation; they purchase outcomes, service-level agreements, and, critically, verifiable guarantees that those SLAs will be enforced. A hospital system deploying remote surgical assistance requires a contractual commitment to millisecond latency and five-nines availability, not a theoretical note that the underlying standard supports such performance under optimal conditions. A logistics operator deploying hundreds of autonomous guided vehicles across a port terminal requires deterministic packet delivery, not a best-effort promise qualified by 'up to' language. 3GPP TR 22.870 [8] begins to acknowledge this commercial reality by framing requirements around use-case families rather than raw physical-layer parameters, yet the translation from standardized requirement to deployable, monetizable service tier remains the industry's most pressing unsolved problem. *The central argument of this part of the white paper is that 6G can break the specification trap only if operators, vendors, and standards bodies commit, from the outset, to defining success in the language of customer outcomes rather than the language of radio physics*.

## Six Outcome Families

***Immersive Experience*** represents the most visible consumer-facing promise of 6G. The convergence of extended reality, holographic communications, and spatial computing is generating application profiles that no existing generation of mobile technology can reliably satisfy at scale. Stereoscopic holographic video streams require not merely high throughput but low variance, consistently maintained throughput: the human visual and vestibular systems are acutely sensitive to rendering discontinuities, and a latency spike that would be imperceptible in a video call triggers motion sickness in an immersive environment. The ITU-R Technical Performance Requirements [3] establish an end-to-end interaction control latency target of less than 4 milliseconds, with fifth-percentile user-experienced downlink throughput of 300 Mbps and uplink throughput of 50 Mbps: figures that reflect not peak capability but the floor below which the experience degrades unacceptably. Achieving these guarantees at the cell edge and inside complex built environments requires a fundamental rethinking of radio resource management, backhaul dimensioning, and application-layer co-design that 6G must deliver as a system, not as a collection of independently optimized components.

***Mission-Critical Determinism*** addresses the evolution of what 3GPP has termed high-reliability, low-latency communications into a fully industrialized connectivity layer for autonomous systems, remote-controlled heavy machinery, power-grid management, and emergency-services command and control. The gap between the URLLC capabilities specified in 5G standards and the reality of industrial deployment is well documented: achieving one-millisecond user-plane latency with packet-error rates no greater than $10^{-5}$ requires a combination of ultra-lean frame structures, deterministic scheduling, and end-to-end network slicing that most commercially deployed 5G networks have not yet implemented in full. ITU-R TPR [3] and 3GPP TR 22.870 [8] both recognize an evolved HRLLC capability as a defining requirement for IMT-2030, targeting one-millisecond user-plane latency, reliability $\geq 1 - 10^{-5}$ (99.999 %) for the HRLLC baseline and $\geq 1 - 10^{-6}$ (99.9999 %) for the most demanding safety-critical use cases specified in TR 22.870 [8]. The commercial opportunity is substantial: industries from automotive to energy to mining are prepared to pay premium prices for connectivity that behaves like a utility, i.e., always available, always deterministic, and contractual accountable when it is not.

***Pervasive AI*** marks the point at which artificial intelligence transitions from a feature layered atop a connectivity pipe to a fundamental driver of network design. The rise of large language models, multimodal generative AI, and autonomous AI agents is already reshaping mobile traffic profiles in ways that challenge the downlink-optimized architecture of every previous generation. Ericsson's Network for AI Experiences report [4] documents that AI-native sessions generate an uplink-to-total-traffic ratio of approximately 26 percent, compared with around 10 percent for conventional broadband sessions: a shift driven by the continuous upload of context, sensor data, voice input, and intermediate reasoning artefacts that agentic AI systems require. The same report documents that generative AI applications grew to nearly one billion weekly active users within three years of mainstream availability, with most of those users accessing AI services from mobile devices. 6G must therefore be designed from the outset with symmetric or near-symmetric capacity, always-on low-variance uplink, and network-native AI inference offload capabilities that allow terminal devices to participate in distributed AI pipelines without exhausting their battery or computing resources.

***Sensing***, delivered through integrated sensing and communication (ISAC), elevates the 6G air interface from a pure communications medium to a shared sensing fabric capable of supporting environmental perception, asset tracking, autonomous navigation, and gesture recognition. The ITU-R TPR [3] defines quantitative targets that ground this capability in engineering reality: indoor factory environments require positioning accuracy of 0.75 meters and velocity estimation accuracy of 2 meters per second; urban macro deployments must achieve positioning accuracy of 6 meters with a detection probability of 95 percent for objects in the sensing zone. These targets enable a new class of network-embedded intelligence in which the radio access network itself becomes a distributed sensor layer, providing environmental awareness to logistics operators, smart-city platforms, and industrial automation systems without requiring the deployment of dedicated sensor infrastructure. The economic implication is significant: operators can monetize the sensing capability of their existing spectrum assets as a new revenue stream, unbundled from traditional connectivity services.

***Sustainable Connectivity*** recognizes that the climate crisis has transformed energy efficiency from a cost-optimization objective into a regulatory and reputational imperative. Healthcare alone illustrates the scale of the opportunity: analysis cited by the SNS eHealth consortium [41] projects that 6G-enabled digital health services could reduce the carbon footprint associated with clinical care episodes by 40 to 60 percent, through reduced patient travel and remote monitoring that prevents acute admissions. Across all verticals, the energy-per-bit target for 6G represents an order-of-magnitude improvement over 5G, achievable through a combination of near-zero-power communication modes for IoT endpoints, AI-driven cell sleep control that shuts down radio units during low-demand periods without service interruption, and hardware architectures that eliminate the analogue-to-digital conversion inefficiencies that dominate current radio access energy budgets. Sustainability, in this framing, is not a constraint on 6G design but a differentiating outcome that operators can sell to enterprises with ambitious net-zero commitments.

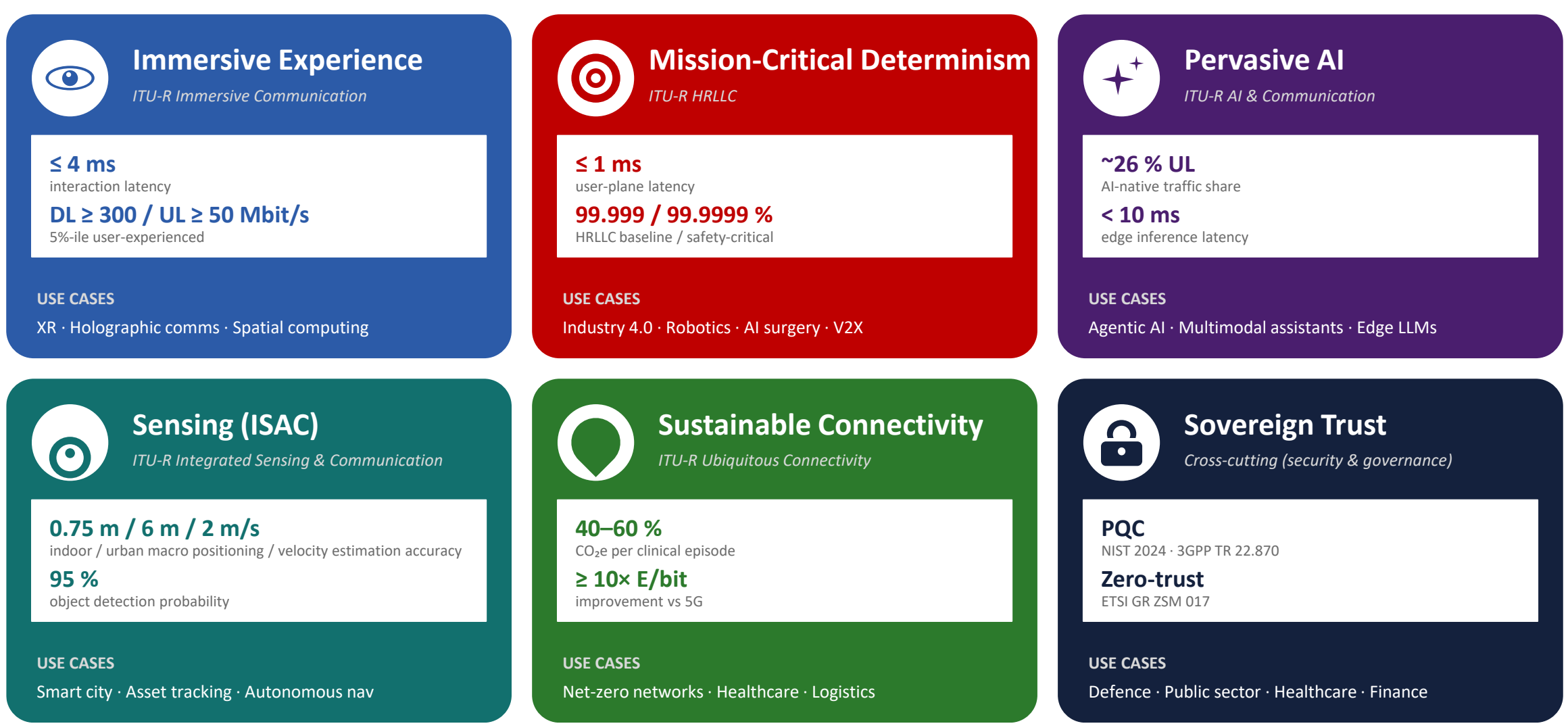

***Figure 4. Six families map to ITU-R IMT-2030 usage scenarios; Sovereign Trust is a cross-cutting governance dimension.***

***Sovereign** Trust* addresses the growing recognition – accelerated by geopolitical disruption, supply-chain scrutiny, and high-profile cyberattacks on critical national infrastructure – that connectivity infrastructure must be trustworthy by design, not by assertion. The migration to post-quantum cryptographic algorithms is no longer a speculative future requirement: the US National Institute of Standards and Technology finalized its first post-quantum cryptography standards in 2024, and 3GPP TR 22.870 [8] includes quantum-resistant security as an explicit use-case requirement for IMT-2030 systems. Zero-trust network architectures – in which no device, user, or workload is implicitly trusted regardless of its position within the network perimeter – are becoming the default expectation for enterprise and public-sector customers. Data sovereignty requirements, which mandate that data generated within a jurisdiction be processed and stored within that jurisdiction, add a further layer of architectural complexity that 6G must address through localized edge processing, policy-driven data routing, and verifiable compliance attestation mechanisms. As illustrated in Figure 4, these six families define the service design space for 6G, providing a comprehensive framework for translating technical capability into outcomes that customers across every segment can understand, evaluate, and purchase.

## Customer Segments and the Outcome Matrix

The six outcome families do not map uniformly across customer segments; rather, they form a configurable matrix in which different combinations of outcomes address the specific needs of consumers, enterprises, and public-sector organizations in ways that reflect their distinct operational contexts, risk tolerances, and regulatory environments.

- ***Consumer customers*** – particularly in high-density urban markets – prioritize immersive experience, pervasive AI, and sovereign trust, seeking networks that deliver seamless spatial computing, personalized AI services, and confidence that their data are protected under applicable privacy law.

- ***Enterprise customers***, especially in manufacturing, logistics, energy, and healthcare, weight mission-critical determinism, pervasive AI, and sensing most heavily, requiring the deterministic connectivity and environmental awareness that autonomous systems demand.
- ***Public-sector organizations*** – national defense, emergency services, critical infrastructure operators, and healthcare systems – place sovereign trust, resilience, and sustainable connectivity at the top of their priority hierarchy, reflecting their unique accountability to citizens and their exposure to geopolitical risk.

The operator's task is to build a service architecture flexible enough to deliver differentiated outcome packages to each segment without maintaining separate physical networks for each use case. Regional and cultural dimensions add further complexity to the outcome matrix. The Bharat 6G Vision [42] articulated by India's Department of Telecommunications places affordable, ubiquitous connectivity and digital inclusion at the center of its 6G ambition, reflecting a market context in which the primary connectivity challenge remains geographic coverage and price accessibility rather than peak performance. The SNS eHealth consortium's European research programme [41] foregrounds healthcare digitization and cross-border interoperability, operating within the European Union's stringent data sovereignty and AI governance framework. The NGMN AI Surge initiative [23] reflects the North American and global operator community's focus on AI-driven network automation and the revenue opportunities of AI-native services. These divergent national and regional priorities are not contradictions to be resolved but legitimate expressions of the different contexts in which 6G will be deployed; a genuinely global standard must therefore provide a flexible framework that accommodates all of them, rather than optimizing for the requirements of any single market.

As illustrated in Figure 5, the customer segment and regional dimensions together define a three-dimensional outcome space that operators must navigate when building their 6G service strategy.

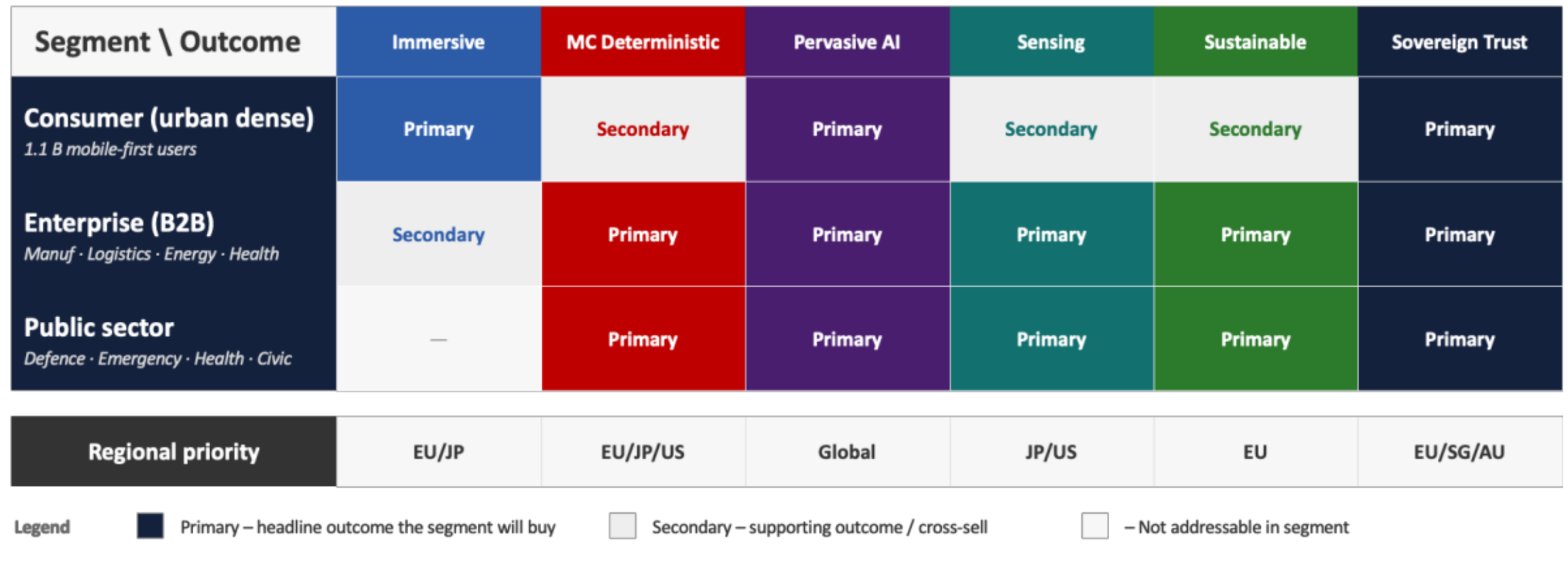

| Segment \ Outcome | Immersive | MC Deterministic | Pervasive AI | Sensing | Sustainable | Sovereign Trust |
|---|---|---|---|---|---|---|
| Consumer (urban dense)<br>*1.1 B mobile-first users* | Primary | Secondary | Primary | Secondary | Secondary | Primary |
| Enterprise (B2B)<br>*Manuf · Logistics · Energy · Health* | Secondary | Primary | Primary | Primary | Primary | Primary |
| Public sector<br>*Defence · Emergency · Health · Civic* | — | Primary | Primary | Primary | Primary | Primary |
| Regional priority | EU/JP | EU/JP/US | Global | JP/US | EU | EU/SG/AU |

Legend: ■ Primary – headline outcome the segment will buy; □ Secondary – supporting outcome / cross-sell; □ – Not addressable in segment

*Figure 5. Primary = headline outcome; Secondary = supporting outcome; — = not addressable in segment.*

A scoping note is warranted: the Guarantee Economy, Programmable Network APIs, and AI substrate investment model described in this paper are most immediately applicable to operators in high-ARPU markets with sufficient enterprise density to support outcome-contract revenue models. For developing-market operators serving subscribers at ARPU levels of two to four dollars per month – across South Asia, Sub-Saharan Africa, and Southeast Asia – the near-term 6G priority is connectivity-first: geographic coverage, affordable device ecosystems, and infrastructure investment to close digital access gaps. The Control Compact applies in these contexts in a simplified initial form – owning spectrum policy, controlling the data layer for nationally relevant AI applications, and ensuring that infrastructure investment does not recreate vendor dependency at a different price point. The specific elements that are economically viable for low-ARPU operators in the 2025–2032 timeframe are: spectrum ownership and O-RAN disaggregation (separates capital costs and prevents single-vendor lock-in); cloud-native infrastructure sharing through national or regional consortia (amortizes MLOps and platform engineering costs); CAMARA API [43] exposure at the basic tier (QoD, device location) as an incremental revenue stream; and participation in IOWN MetricEES [44] sustainability reporting as a precondition for accessing green infrastructure financing. The Premium Consumer Immersive and AI Inference Edge tiers of the Guarantee Economy are Phase 3 objectives in these markets; the Mission-Critical Determinism and Sensing-as-a-Service tiers may find earlier traction in verticals – healthcare, agriculture, mining – where productivity gains at sub-USD 10/month device economics are documented by the SNS JU eHealth programme [41]. The Bharat 6G Vision [42] explicitly frames this differentiated economics model as a national strategic priority; the 6G architecture must accommodate both contexts rather than prescribing uniform service tiers commercially viable only in high-ARPU markets.

## Mapping Outcomes to Standards

The six outcome families described above do not exist in isolation from the formal standards that will govern 6G implementation; rather, they provide a customer-centric lens through which the technical requirements embedded in those standards can be understood, prioritized, and translated into commercial deployments. ITU-R M.2160 [2], and 3GPP TR 38.914 [33], define the IMT-2030 framework around six usage scenarios – *Immersive Communication*, *Hyper-Reliable and Low-Latency Communication*, *Massive Communication*, *Ubiquitous Connectivity*, *AI and Communication*, and *Integrated Sensing and Communication* – which map with reasonable fidelity to five of the six outcome families identified in this paper, with *Sovereign Trust* representing a cross-cutting security and governance requirement rather than a discrete usage scenario. The ITU-R TPR [3], and 3GPP TR 38.914 [33], assign quantitative targets to each scenario, establishing the minimum technical floor that a compliant 6G system must achieve. 3GPP TR 22.870 [8] translates these scenarios into concrete service requirements, defining the use-case families – *remote multi-presence*, *high-precision manufacturing*, *AI-assisted surgery*, and *connected autonomous systems* among them – that will drive the detailed specification work in 3GPP Release 20 and beyond.

The operator's role in this mapping process is neither passive nor trivial. Standards define the performance envelope within which a compliant system must operate; they do not define the commercial architecture through which operators translate that envelope into deployable, differentiated service tiers. An operator must segment the standardized performance space into a manageable set of service categories – ranging from best-effort broadband through assured quality tiers to fully guaranteed mission-critical slices – and must build the measurement, enforcement, and billing infrastructure needed to make those categories credible to enterprise customers. This is precisely the challenge that Rakuten Mobile has confronted in deploying the world's first fully cloud-native, Open RAN-based network at commercial scale: the architectural choices made in that deployment – disaggregated RAN, Kubernetes-orchestrated core, intent-driven slice management – are not merely engineering preferences but commercial enablers that make programmable, verifiable service tiers operationally achievable. The following parts of this paper examine how that architecture is extended and generalized to meet the full ambition of the 6G outcome framework described here, anchored in the empirical evidence from Rakuten's deployment and informed by the standards trajectory defined by [2], [3], and [8].

## Section III – Business First: From Connectivity to Digital Service

### The Connectivity Trap: A Structural Revenue Problem

The mobile industry's financial trajectory over the past decade reveals a structural paradox: operators have invested hundreds of billions of dollars in successive generations of spectrum, equipment, and network transformation, while average revenue per user has stagnated or declined in most mature markets. ARPU in leading European and North American markets has remained flat in real terms since the commercial launch of 4G LTE, even as the volume of data traffic transported has grown by orders of magnitude [GSMA Mobile Economy, 2023–2025] – a structural paradox whose cost and complexity pressures are addressed by the NGMN Simplification initiative [5]. Rakuten Mobile's own trajectory illustrates both the challenge and the path beyond it: its Ecosystem ARPU is materially lifted by cross-group revenue contributions – Rakuten Points spend uplift, e-commerce referral, and affiliated financial services driven by the Rakuten Group's tens of millions of active ecosystem members, into which Rakuten Mobile subscribers are enrolled – rather than by data charges alone [7]. This ecosystem model demonstrates in commercial deployment the principle that the Business First reordering demands: revenue anchored in service value, not data volume. It must, however, be understood in context: Rakuten's ARPU uplift depends on a loyalty flywheel – Rakuten Points, an integrated portfolio of digital services, and a sizable cross-service ecosystem membership base – that required decades and substantial cross-industry investment to construct. Operators without a comparable pre-existing digital services ecosystem cannot replicate this trajectory through connectivity investment alone; in their context, the Business First reordering is realized through API monetization, enterprise Guarantee Economy contracts, or partnership-based service aggregation rather than consumer loyalty platform economics. The fundamental cause is structural rather than cyclical: the market has commoditized connectivity, pricing it as an undifferentiated utility while the economic value created by digital services – social media, streaming video, cloud computing, e-commerce, and now generative AI – has migrated entirely to over-the-top platforms that sit above the network layer. Operators have, in effect, invested in the pipes through which other companies deliver value, without capturing a commensurate share of the economic surplus their infrastructure enables.

The 5G business case was supposed to break this pattern through enterprise connectivity – *private networks*, *network slicing*, and *edge computing* were positioned as the mechanisms by which operators would capture value from Industry 4.0 digitalization [45]. The reality has been more constrained *without programmable, software-driven interfaces that allow enterprise customers to integrate network capabilities directly into their operational workflows*, the promise of *5G-as-a-platform* has remained largely theoretical. Enterprises that have deployed private 5G have frequently done so by purchasing spectrum licenses and network equipment directly, bypassing the public operator entirely. The lesson for 6G is unambiguous: connectivity alone is not a sustainable business; programmable connectivity, delivered through open APIs, governed by verifiable SLAs, and *integrated into the digital platforms that enterprises use*, is the only model that can generate the revenue growth the industry requires to fund the next cycle of infrastructure investment. The NGMN Simplification program [5] identifies *network API exposure*, *end-to-end slicing*, and *zero-touch operations* as the three structural enablers that must be in place before 6G can deliver on its commercial potential.

## Three Revenue Engines

*The Guarantee Economy* represents the most immediately actionable of the three revenue engines available to 6G operators. Market research consistently demonstrates that a significant portion of mobile users are willing to pay a premium for reliable connectivity when their current experience is demonstrably poor: as documented by the AI-RAN Alliance Working Group 3 [40], nearly half of global 5G users experience connectivity challenges during peak hours, and approximately half of those users indicate they would pay for guaranteed reliability – an industry estimate drawn from AI-RAN Alliance [12] member research rather than an independent consumer survey, but one that nonetheless points to a substantial, latent premium-tier market that current network architectures cannot serve because they lack the mechanisms to make and enforce reliable quality guarantees. The Guarantee Economy requires three architectural components that 6G must deliver: the ability to define service-level objectives in granular, measurable terms covering latency, throughput, reliability, and availability; the ability to enforce those SLAs in real time through closed-loop network automation; and the ability to verify compliance with those SLAs through transparent, auditable telemetry that both the operator and the customer can inspect. Without all three components, a guarantee is merely a marketing claim; with them, it becomes a contractually enforceable product that commands premium pricing across consumer, enterprise, and public-sector segments.

Realizing this contractual enforceability requires preconditions beyond the network's technical capability. Legal frameworks vary by jurisdiction in whether and how network performance penalties can be enforced in commercial contracts; operators entering the Guarantee Economy must map SLA terms to jurisdiction-specific contract law rather than applying uniform templates.

Commercial terms must survive vendor indemnity caps and force majeure clauses standard in operator supply agreements. And the telemetry demonstrating SLA compliance must be accepted by both parties as authoritative – an operator's own telemetry system is not, by itself, an independent audit. Building the Guarantee Economy therefore requires a parallel commercial and legal architecture alongside the technical one: standardized SLA term frameworks, independent performance verification mechanisms, and enterprise contracts that define outage in network-layer terms that map unambiguously to the application outcomes customers care about. Existing telecom regulatory frameworks – including the EU Electronic Communications Code's quality-of-service transparency obligations and Ofcom's general conditions – provide a floor for consumer SLA enforcement but do not govern enterprise network service contracts, which fall under general commercial law; SLA penalty enforcement therefore depends on arbitration clauses and choice-of-law provisions whose enforceability varies widely across jurisdictions. The industry must advocate for sector-specific SLA enforcement frameworks in major markets – analogous to the liability structures in banking's SWIFT messaging regime – as a precondition for the Guarantee Economy to scale from individual enterprise pilots to market-wide adoption. Concretely, TM Forum's Open Digital Architecture API standards (TMF630 SLA Management [46], TMF641 Service Order [47]) provide the data model for SLO term exchange, while ETSI ZSM closed-loop conformance testing provides the technical audit standard that both parties can accept as authoritative – a viable institutional infrastructure for standardized SLO verification available today. Operators should adopt these as contractual reference standards in enterprise service agreements from Phase 1 forward, building the industry precedent before seeking regulatory formalization.

The enterprise connectivity guarantee market has a precedent that demands honest assessment: the 5G enterprise SLA proposition – private networks, guaranteed network slicing, deterministic QoS – did not materialize at the scale initially projected [5], with most enterprise 5G deployments involving direct spectrum licenses and equipment purchases that bypass public operator SLA models entirely. Three structural differences make the 6G Guarantee Economy a distinct proposition. First, it is a pure service contract requiring no enterprise capital expenditure – a significantly lower adoption barrier than private network ownership. Second, the AI inference edge and Sensing-as-a-Service tiers create demand categories with no 5G precedent, where the service itself is novel. Third, the closed-loop enforcement mechanism of 6G – automated Service Level Objective (SLO) monitoring with contractual breach triggers and machine-readable compliance evidence – enables outcome contracts that 5G's manual operations model cannot provide. Phase 1 of the roadmap in Section VII is explicitly designed to generate the enterprise customer evidence that validates this proposition before Phase 2 commercial commitment at scale. The critical caveat is that enterprises that invested in private 5G networks did so partly for reasons of control, customization, and security that a managed service model does not fully address regardless of CAPEX elimination; the Guarantee Economy's initial total addressable market is most likely the segment of enterprises without the budget or scale for private network ownership, and operators should size commercial expectations accordingly.

***Programmable Network APIs*** constitute the second revenue engine, enabling operators to expose network capabilities as developer-accessible services rather than as opaque connectivity infrastructure. The CAMARA open-source project [43] and the GSMA Open Gateway [31] initiative have established the initial vocabulary of this approach in the 5G era, exposing Quality-on-Demand, device location, and fraud-detection APIs to enterprise application developers. 6G extends this vocabulary dramatically:

- ***AI inference offload APIs*** that allow computationally intensive workloads to be executed on network edge servers rather than on battery-constrained devices.
- ***Sensing-as-a-service APIs*** that expose the positioning and environmental detection capabilities of the 6G air interface to logistics and industrial applications.
- ***Network digital twin APIs*** that give enterprise customers a real-time, software-defined representation of the network conditions affecting their deployments.

ETSI ZSM [19]-[21] provides the management and orchestration framework within which these APIs are governed, ensuring that capability exposure does not compromise network integrity or create cross-customer interference in the underlying physical resources. The commercial model for network APIs mirrors the success of cloud platform APIs – consumption-based pricing, developer self-service, and a marketplace ecosystem in which third-party applications create value that flows back to the infrastructure provider through usage revenue.

***Network-as-a-Service (NaaS)*** represents the third and most transformative revenue engine – the delivery of complete, end-to-end network functionality as an on-demand, software-defined service consumed through intent-based interfaces rather than through the physical deployment and configuration of dedicated hardware. Under the NaaS model, an enterprise customer expresses its connectivity requirements – such as, but it is not limited to, *bandwidth*, *latency*, *coverage area*, *security posture*, and *availability* guarantee – through a high-level intent API, and the network autonomously instantiates, configures, and operates the required network slice without requiring any human engineering involvement. ETSI ZSM [19] provides the architectural framework for this capability through its zero-touch service management reference architecture, which decomposes network management into closed-loop automation domains that can operate independently or in coordination to deliver end-to-end service assurance. The economics of NaaS are compelling: by replacing labor-intensive, ticket-driven provisioning workflows with autonomous orchestration, operators simultaneously reduce operational expenditure and improve service velocity, enabling them to offer enterprise customers network capabilities on timescales – minutes or hours rather than weeks or months – that are competitive with the instant provisioning of public cloud services.

As illustrated in Figure 6, the three revenue engines are mutually reinforcing: network API exposure creates the developer ecosystem that generates demand for guaranteed service tiers, which in turn creates the commercial case for NaaS automation.

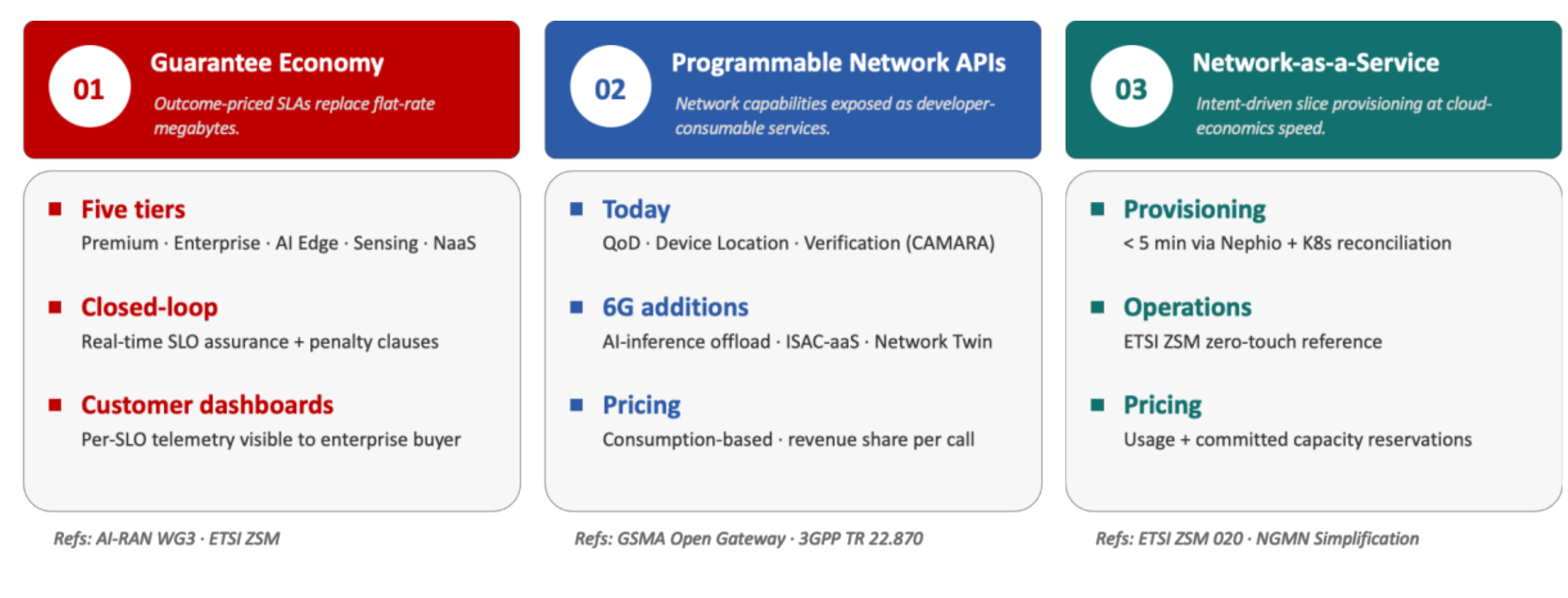

*Figure 6. The three engines reinforce each other; cumulative ARPU uplift accrues only when all three are operationally live.*

## The Guarantee Economy in Practice

Translating the concept of the Guarantee Economy into an operational commercial reality requires operators to build a service tier catalogue that maps customer outcome requirements to specific, enforceable technical commitments. The 6G Guarantee Economy comprises five named service tiers – defined in full in Annex E – each with guaranteed SLOs, target customer segments, and distinct billing models. The Premium Consumer Immersive tier targets gaming, XR, and live-event experiences, guaranteeing downlink ≥ 500 Mbit/s, latency ≤ 20 ms, and 99.99 % availability, billed as a monthly premium subscription [3][8]. The Enterprise Determinism tier serves manufacturing, surgery, robotics, and critical communications, guaranteeing latency ≤ 1 ms, reliability 99.9999 %, and jitter < 1 µs, contracted on per-SLO outcome terms with explicit penalty clauses [3][8]. The AI Inference Edge tier targets enterprise AI workloads and real-time computer vision, guaranteeing inference latency < 10 ms and model availability 99.99 %, billed per inference or as a committed capacity reservation [11]-[13]. The Sensing-as-a-Service tier covers smart-city, autonomous logistics, and environmental monitoring applications, guaranteeing positioning accuracy ≤ 0.75 m (indoor) / ≤ 6 m (urban macro) per ITU-R TPR [3] and sensing update ≤ 1 s; specialized ISAC deployments in controlled industrial environments have demonstrated sub-centimeter positioning in B5G/6G trials [41], representing an achievable premium-tier target beyond the standards minimum; priced per API call or per-area subscription [1]. The Network-as-a-Service (NaaS) tier enables hyperscalers and enterprise private networks to procure intent-driven slices provisioned in under five minutes with guaranteed bandwidth, latency, and isolation, billed on usage and committed capacity [19][21].

The SLO/KPI structure across all tiers covers five enforcement dimensions: *latency* (mean and variance), *reliability* (packet-error rate and session-failure rate), *throughput* (peak, sustained, and fifth-percentile), *availability* (uptime percentage and maximum outage duration), and *security posture* (encryption standard, authentication mechanism, and data-sovereignty compliance attestation) [20].

| Service Tier | Latency SLA | Reliability | Throughput / Accuracy | Pricing Model |
|---|---|---|---|---|
| Premium Consumer Immersive | ≤ 20 ms | 99.99 % | DL ≥ 500 Mbit/s | Monthly premium subscription |
| Enterprise Determinism | ≤ 1 ms | 99.9999 % | Jitter < 1 µs | Per-SLA outcome contract + penalties |
| AI Inference Edge | < 10 ms | 99.99 % model availability | Edge LLM / CV / NLP | Per-inference / committed capacity |
| Sensing-as-a-Service | ≤ 1 s update | — | ≤ 0.75 cm (indoor) ≤ 6m (outdoor) positioning | Per-API-call / per-area subscription |
| Network-as-a-Service | < 5 min provisioning | ≥ 99.9 % slice integrity | Intent-driven slice | Usage + committed capacity |

**SLA enforcement axes:** latency · reliability · throughput · availability · security & data sovereignty

*Figure 7. Five tiers anchor outcome-priced commercial models.*

The five-tier anchor outcome-priced commercial models are depicted in Figure 7. SLAs are enforced through closed-loop AI assurance. The commercial model for guaranteed service tiers must combine three pricing mechanisms to adequately capture the value delivered. An upfront capability premium reflects the network investment required to build and maintain the infrastructure needed to deliver guaranteed performance – including dedicated spectrum resources, low-latency transport, and the automation infrastructure needed for real-time SLO enforcement. SLA penalty mechanisms provide the credibility that makes the guarantee commercially meaningful: an operator that faces financial consequences for SLA violations has a strong incentive to build the resilience and automation capabilities needed to avoid them, and an enterprise customer that receives compensation for outages has a mechanism for quantifying the cost of connectivity failure that justifies the premium pricing. Usage-based metering enables flexible consumption of guaranteed tiers for applications that require high-assurance connectivity intermittently rather than continuously – an autonomous inspection robot that requires mission-critical connectivity during active operation but is satisfied with best-effort connectivity during idle periods. Enforcement of the commercial model across all three pricing components requires closed-loop automation that monitors SLO compliance in real time, triggers corrective actions when performance degrades, generates the telemetry evidence needed for billing and dispute resolution, and provides customers with transparent dashboards that make the value of their premium tier subscription continuously visible. As illustrated in Figure 6, this closed-loop commercial architecture is the operational foundation of the Guarantee Economy.

One structural constraint on Guarantee Economy scope must be stated directly: the SLA enforcement architecture described here applies to domestic network services under a single operator's management. When a customer travels internationally and uses roaming, the serving operator loses direct control of the network layer, making end-to-end SLA guarantees unenforceable under current inter-operator roaming architecture. For enterprise customers with globally mobile workforces – the primary target segment for the Enterprise Determinism and AI

Inference Edge tiers – this is not a fringe case. The Guarantee Economy is therefore initially a domestic proposition; international extension requires bilateral and multilateral inter-operator SLA agreements built on GSMA Open Gateway Quality-on-Demand extensions and inter-operator network performance monitoring frameworks [43]. Operators should not represent international SLA coverage to enterprise buyers until this inter-operator infrastructure is in place, positioning international Guarantee Economy coverage as a Phase 3 objective.

### The Rakuten Ecosystem Model: From Connectivity to Platform

Rakuten Group provides the most fully realized existing example of what a connectivity-anchored digital services platform can achieve on a commercial scale. The Group operates more than seventy distinct digital services – spanning e-commerce, financial services, insurance, travel, entertainment, sports, and digital content – accessible to a membership base exceeding one hundred million registered users in Japan and growing internationally. The commercial effect of this ecosystem integration is captured in Rakuten's disclosed ARPU premium: Rakuten Group's operating data show that ecosystem members who engage with multiple Rakuten services generate materially higher revenue per user than single-service subscribers, a premium driven by the compounding value of personalized cross-service recommendations, integrated loyalty rewards, and frictionless cross-platform transactions. Rakuten Mobile occupies a strategically central position in this ecosystem: the mobile network is the connective tissue that links ecosystem members to the full portfolio of digital services, providing the persistent, personalized connectivity layer that enables real-time recommendations, location-aware services, and seamless authentication across every touchpoint. The network is, in the most literal sense, the platform.

The 6G evolution of the Rakuten ecosystem model enables a qualitative deepening of this integration that 5G cannot support at the required scale or performance level. Ericsson's analysis [4] observes that generative AI grew to nearly a billion weekly active users in three years, with the majority accessing AI services from mobile phones – a trajectory that positions the mobile network as the primary delivery infrastructure for the AI economy. For an operator with Rakuten's ecosystem breadth, this creates the opportunity to deploy personalized AI services – shopping assistants, financial advisors, health monitors, travel planners – that operate continuously in the background, drawing on real-time network context, location data, and cross-service behavioral intelligence to deliver genuinely useful, proactive assistance rather than reactive query-response interactions. The 6G network enables deeper real-time logistics integration – connecting physical delivery operations with digital commerce workflows through ISAC-based asset tracking and deterministic low-latency control interfaces – and supports ambient commerce experiences in which product discovery, payment, and fulfilment occur seamlessly within spatial computing environments without requiring explicit user action. This vision of *network-as-platform* represents the commercial destination toward which Rakuten's cloud-native Open RAN architecture, Rakuten Symphony's software platform, and the 6G investments described throughout this white paper are all oriented.

# Section IV – Operations First: Intelligent Growth – Networks as Software

## The Operations Gap: Manual Processes Cannot Scale to 6G

The most consequential under-discussed problem in mobile network operations is not the complexity of the technology itself but the persistent gap between the automation ambition embedded in 5G standards and the operational reality of commercially deployed networks. Despite a decade of investment in network function virtualization, software-defined networking, and cloud-native architecture, the overwhelming majority of live 5G networks are still operated through a combination of command-line-driven configuration management, ticket-based change workflows, and reactive incident response that would be recognizable to an engineer who last worked on a 3G network. The NGMN Agentic AI Operating Models framework [23] defines a five-level AI adoption scale on which Level 3 – conditional autonomy, in which automated systems handle well-defined scenarios while escalating exceptions to human operators – represents the current industry ceiling for the most advanced deployed implementations. Levels 4 and 5, which represent proactive and optimized autonomy respectively, remain aspirational for most operators. The ETSI Zero-touch Service and Network Management (ZSM) architecture [19] provides the closed-loop automation reference model within which these autonomy levels operate. This gap is not merely an efficiency problem; it is a structural constraint on revenue growth, because the guaranteed service tiers and network-as-a-service capabilities described in Section III of this paper are operationally undeliverable without autonomous network management.

The NGMN Agentic AI initiative provides the most rigorous current framework for understanding the path from today's manual operations to the fully autonomous network management that 6G demands. The framework defines five AI adoption levels that map directly onto the trajectory operators must follow [23]:

- ***Level 1***, foundational rule-based automation, in which scripted workflows replace manual CLI operations for well-understood routine tasks.
- ***Level 2***, supervised machine learning, in which AI models provide recommendations that human operators review and approve.
- ***Level 3***, conditional autonomy, in which AI systems take independent action within defined boundaries and escalate out-of-boundary situations.
- ***Level 4***, fully autonomous proactive AI with closed-loop control, in which AI agents anticipate demand, detect and remediate faults, and optimize performance continuously without human intervention.
- ***Level 5***, optimized enterprise-scalable governed autonomy, in which multi-agent AI systems coordinate across network domains, business systems, and partner ecosystems to optimize outcomes at enterprise scale within a robust governance framework.

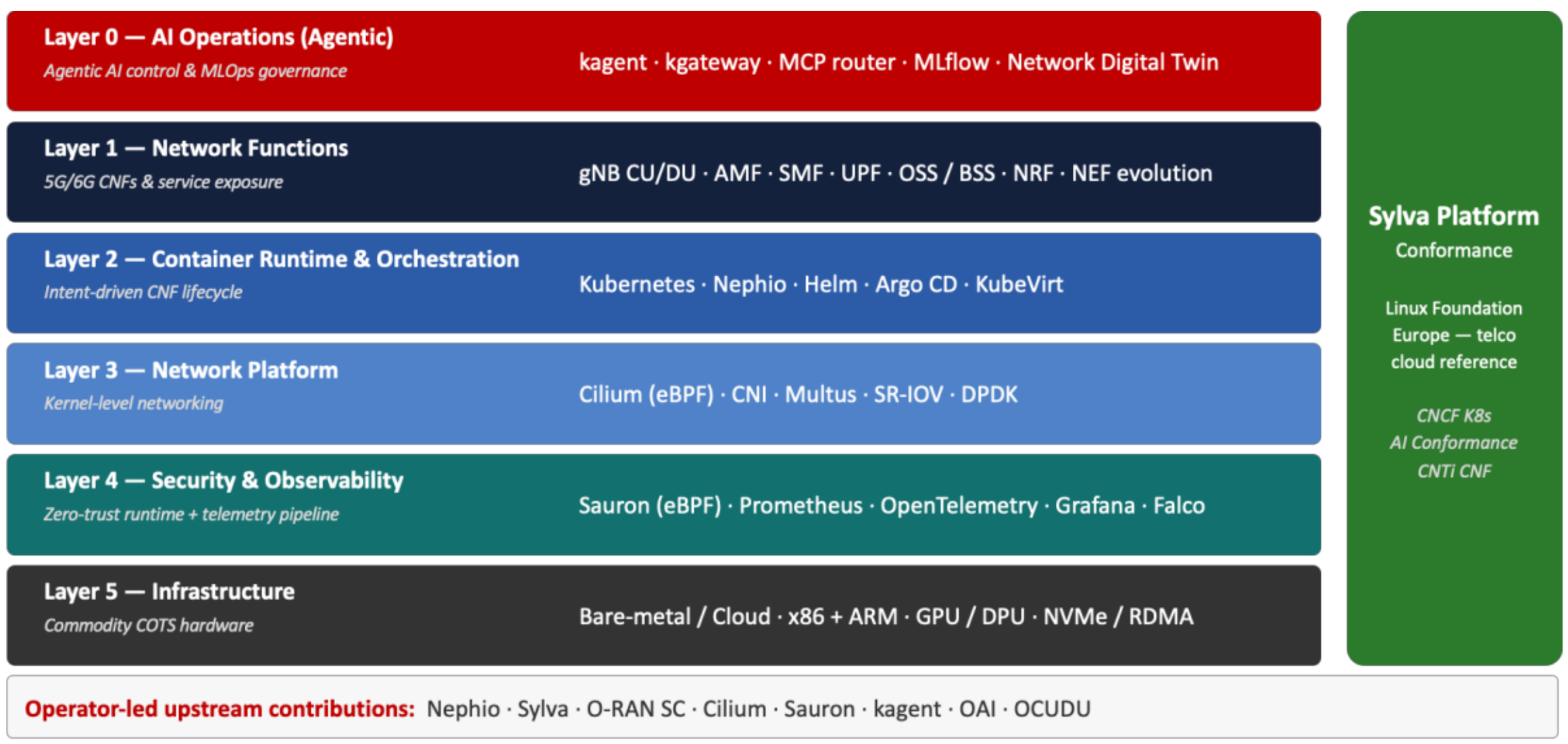

*Figure 8. Six layers of open, eBPF-instrumented, Kubernetes-orchestrated infrastructure replace proprietary middleware.*

As NGMN articulates [23], "*by mapping cloud-native maturity stages to AI readiness, operators can systematically plan their transition toward autonomous, intelligent network management*" – a roadmap that requires simultaneous investment in cloud-native infrastructure, MLOps tooling, and the organizational capability to govern AI-driven systems in production.

## Networks as Software: The Cloud-Native Operating Model

The architectural foundation of the autonomous 6G network is a cloud-native operating model in which every network function – from radio unit digital signal processing to core network session management to operational support systems, as illustrated in Figure 8 – executes as software on a shared, horizontally scalable compute infrastructure governed by a Kubernetes orchestration layer. Kubernetes functions, in this model, as the operating system of the network: it schedules workloads, manages resource allocation, enforces policies, and provides the lifecycle management primitives – rolling updates, health checks, auto-scaling, and self-healing – that make software-defined network functions as reliable and manageable as any cloud-native application. The networking and security layer of this infrastructure may be implemented using Cilium (Tetragon) [36], Sauron [48], or any other carrier grade extended Berkeley Packet Filter (eBPF) observability and security tooling platform, which provide network policy enforcement, traffic observability, and security monitoring at kernel level without requiring kernel module installation or vendor-proprietary hardware dependencies. Sauron [48] extends the eBPF security stack to runtime threat detection, providing kernel-level visibility into process execution, file access, and network activity that enables real-time detection and response to security incidents without interrupting normal network operations. Nephio [34], the cloud-native network automation project hosted by the Linux Foundation, provides the intent-driven lifecycle management layer for network functions, allowing operators to express desired network state in Kubernetes-native configuration objects and rely on Nephio's reconciliation controllers to drive the actual network configuration toward that desired state. As illustrated in Figure 8, this cloud-native stack provides the operational substrate on which agentic AI and autonomous network management are built.

Rakuten Mobile's deployment provides empirical evidence that this cloud-native operating model is not a theoretical construct but an operational reality achievable at commercial scale. Rakuten Mobile operates the world's first fully cloud-native, Open RAN-based mobile network, with every network function – including the radio access network – running as containerized software on commercial off-the-shelf hardware orchestrated by Kubernetes. Advanced network observability, active transport network layer monitoring, and security tooling are achieved with the deployment of Sauron eBPF Platform [49]. Rakuten Symphony, the technology platform and professional services business that emerged from Rakuten Mobile's deployment experience, owns the full software stack from OSS/BSS through RAN and core, providing an end-to-end, vendor-disaggregated platform that has been deployed and is being deployed by operators across multiple markets. This deployment experience has generated operational insights that are not available to any other operator or vendor in the industry: the specific failure modes of cloud-native network functions under production load, the automation capabilities required for zero-touch RAN configuration, the observability tooling needed to maintain service quality across a disaggregated multi-vendor stack, and the organizational processes required to operate a software-driven network with the rigor that telecommunications-grade availability demands – validated by Rakuten Mobile's FY2025 production data [7]. These insights directly inform the 6G operations architecture described throughout this part of the white paper.

The cloud-native movement in Telco extends beyond any single operator's deployment, and the industry-wide momentum behind open, standards-based cloud infrastructure is an important enabling condition for 6G's operational ambition. The Sylva project [35], hosted by the Linux Foundation Europe, provides a common, carrier-grade cloud platform specification that enables operators across Europe and beyond to build cloud-native network infrastructure on a shared, open-source foundation, reducing duplicated engineering effort and accelerating the development of the cloud-native ecosystem. The OCUDU Ecosystem Foundation [50] advances open, cloud-native disaggregation of the RAN baseband unit, providing the architectural separation between radio hardware and digital signal processing software that makes cloud-native RAN economically viable at scale. Together, these industry movements create the open, interoperable cloud-native platform ecosystem on which 6G's autonomous operations ambition depends – ensuring that the operational benefits of cloud-native architecture are accessible to operators of all sizes rather than being the exclusive domain of those with the scale and resources to build proprietary infrastructure.

## Agentic AI: The Operating Model for Autonomous Networks

The emergence of large language model-based AI agents represents a qualitative transformation in the capability of network automation systems – a shift from rule-based automation that executes predefined scripts to intent-driven agents that understand natural-language objectives, decompose them into executable tasks, select and invoke appropriate tools, and learn from the outcomes of their actions to improve future performance. The operational loop of an autonomous network agent proceeds through four stages:

- ***Intent***, in which the agent receives a high-level objective – 'optimize energy consumption in the northern cluster while maintaining service SLOs' – expressed in natural language or structured intent notation.
- ***Observe***, in which the agent collects and interprets relevant network telemetry, configuration state, and contextual information from connected data sources.
- ***Decide***, in which the agent applies its domain knowledge and reasoning capability to determine the optimal sequence of actions required to achieve the stated objective.
- ***Act***, in which the agent invokes network management APIs, configuration management tools, and orchestration interfaces to execute the determined action sequence, monitoring the effects of each action and adjusting its plan in response to unexpected outcomes.

The operational target is NGMN Level 4 – Fully Autonomous Proactive AI with Closed-Loop Control [23] – operating within a Level 5 governance framework in which human operators set policy objectives and exception thresholds rather than approving individual agent decisions. Level 4 autonomous execution within a Level 5 policy governance layer is precisely the combination required by EU AI Act [51] compliance and commercial accountability, as elaborated later in this section. This four-stage operational loop is illustrated in Figure 9.

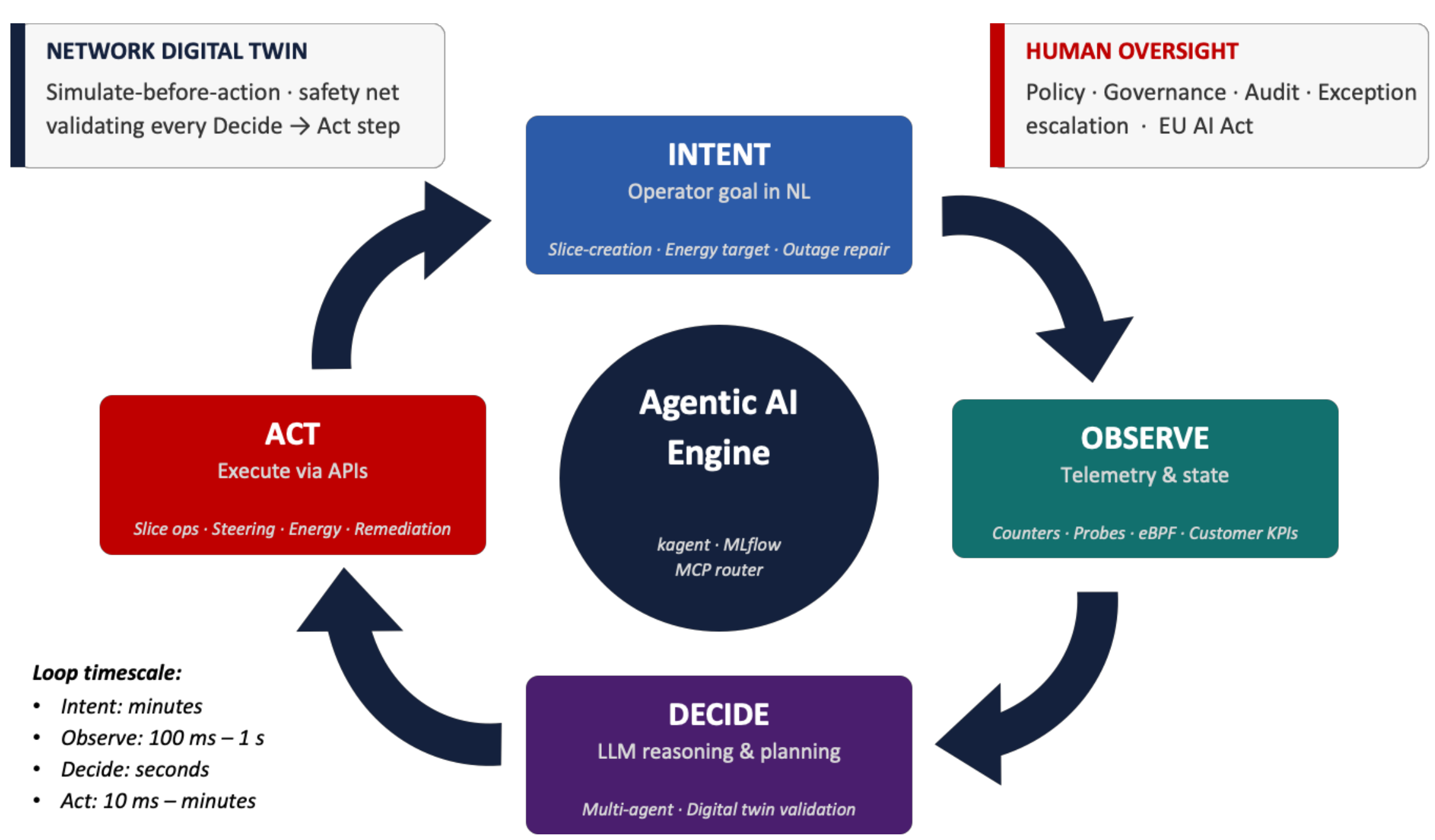

***Figure 9. Four-stage Intent → Observe → Decide → Act loop; Network Digital Twin validates actions before they are executed in the live network.***

The nGRG RR-2025-02 Generative AI report [52] documents the concrete operational capabilities that LLM-based AI agents can deliver in production network environments: "*AI Agents can autonomously perform network slice management tasks: creating, configuring, optimizing, and repairing network slices based on real-time requirements*." This capability directly addresses the operational bottleneck that prevents operators from delivering the Network-as-a-Service and Guarantee Economy propositions described in Section III – the inability to provision and manage network slices at the speed and scale that enterprise customers demand without proportional growth in engineering headcount. Retrieval-Augmented Generation (RAG) enhances the operational effectiveness of LLM-based agents by enabling them to draw on continuously updated knowledge bases – vendor technical bulletins, incident post-mortems, configuration best practices, and real-time network state information – rather than relying solely on the static knowledge embedded in their training data. This architecture keeps the agent's operational knowledge current with the evolving state of the network and the vendor ecosystem while improving the accuracy and relevance of its responses to novel operational situations, addressing one of the key reliability concerns that has historically limited the deployment of AI in network operations critical paths.

Multi-agent coordination extends the autonomous operations capability from single-domain tasks to the cross-domain, end-to-end service management that 6G's complexity demands. ETSI ZSM020 [21] defines a taxonomy of agent types that provides the architectural vocabulary for multi-agent network management: reactive agents that respond to stimuli with predefined action patterns, suitable for high-frequency, well-understood operational events; deliberative agents that maintain an internal model of the network and apply goal-directed reasoning to determine actions, appropriate for complex optimization and planning tasks; and hybrid agents that combine reactive and deliberative capabilities, enabling fast response to urgent events while maintaining long-term optimization goals. As ETSI ZSM020 [21] documents the value of multi-agent coordination for collaborative knowledge sharing and problem-solving across distributed network automation domains – capabilities that are essential when, for example, a major traffic anomaly in one network domain requires coordinated responses from RAN optimization, core network capacity management, and transport engineering agents operating in parallel. The Model Context Protocol (MCP) router provides the practical agent-to-tool interface in the telecom operations stack, enabling agents to discover and invoke the full catalogue of network management APIs, observability tools, and configuration interfaces available in the operator's environment through a standardized, secure protocol that maintains governance and auditability across all agent-initiated actions.

### AI/ML Infrastructure for Autonomous Operations

Deploying AI agents in production network operations is not merely a matter of selecting appropriate model architecture; it requires a comprehensive MLOps infrastructure that manages the full lifecycle of AI models from initial data collection through training, deployment, monitoring, and retraining. The data collection layer must aggregate telemetry from every network domain – RAN performance counters, core network session records, transport utilization metrics, security event logs, and customer experience indicators – into a unified, query able data platform that provides the training signal required to develop high-quality models and the real-time inference context required for autonomous agent decision-making. Model training requires the computational infrastructure and the engineering discipline to develop models that are not merely accurate in laboratory evaluation but robust under the distribution shifts and adversarial conditions that production network operations regularly present. The SNS JU AI/ML Landscape [53] identifies MLflow as a key enabler for implementing MLOps in dynamic and distributed environments, providing experiment tracking, model versioning, and deployment management capabilities that make it practical to maintain a portfolio of specialized network management models across a complex, multi-domain operational environment. The AI-RAN Multi-Agent Manager functions as the control tower for the autonomous agent ecosystem, providing centralized visibility into agent activity, resource allocation, policy enforcement, and performance measurement across the full fleet of deployed agents [11].

***Network Digital Twins*** provide the simulation-before-action capability that is essential for deploying autonomous agents in environments where incorrect decisions can cause service degradation or outages affecting millions of customers. A network digital twin maintains a real-time, high-fidelity software replica of the physical network – including its topology, configuration state, traffic load, and equipment characteristics – that allows autonomous agents to evaluate the expected impact of a proposed action before executing it in the live network. This capability transforms the risk profile of autonomous operations: rather than relying on the agent's training-derived confidence in a proposed action, operators can require that all significant configuration changes be validated against the digital twin before execution, providing a deterministic safety net that is independent of model quality. AI model versioning and governance aligned with the requirements of the EU AI Act (Regulation 2024/1689 [51]) must form the final layer of the governance framework. The EU AI Act classifies AI systems managing critical infrastructure – including mobile network operations – as high-risk under Article 6 and Annex III, triggering obligations including human oversight mechanisms (Article 14), technical robustness requirements (Article 15), and conformity assessment before deployment. This creates a direct tension with the Level 4 fully autonomous closed-loop target: Level 4 operates without per-action human approval, while the AI Act requires human oversight mechanisms for high-risk systems. The resolution lies in the governance layer above the agent: human oversight operates at the policy and exception level – operators set the objectives, constraints, and escalation thresholds within which the autonomous system acts, and the AI Act obligation is satisfied by this governance infrastructure around the system, not by inserting human approval into each individual agent decision. Concretely, every model deployed in the autonomous operations stack must be version-controlled, auditable, and accompanied by documentation of its training data, evaluation methodology, and known performance boundaries, enabling operators to demonstrate conformity assessment compliance and to rapidly roll back to a known-good model version when an updated model exhibits unexpected behavior in production.

A distinct challenge from adversarial security threats is the AI alignment risk: an autonomous system correctly optimizing for its defined objective function may, under specific load conditions or distributional shifts unseen during training, take actions that are technically correct but commercially or regulatorily unacceptable – for example, deprioritizing consumer traffic during SLA contention in ways that trigger regulatory complaints, or optimizing Telecommunications Energy Efficiency Ratio (TEER) metrics at the expense of geographic service equity. This is not a security failure but a calibration and governance challenge that the security architecture alone cannot address. The mitigation requires multi-objective agent design: reward functions must encode not only network performance metrics but also regulatory SLA floors below which no resource allocation is permitted, equity constraints across subscriber segments, and fairness criteria aligned with national regulatory obligations. The policy governance layer that distinguishes NGMN Level 5 from Level 4 – human operators defining the objectives and constraints within which the autonomous system operates – is precisely "*the mechanism that manages alignment risk alongside the security and compliance obligations the governance framework must satis*fy."

### Observability: Seeing the Network in Real Time

Autonomous network operations are only as effective as the observability infrastructure that provides agents with an accurate, timely, and comprehensive view of network state. eBPF-powered, kernel-level telemetry – e.g., on Rakuten Mobile network, deployed through the Sauron eBPF toolchain [49] – provides the deepest available layer of network observability without the instrumentation overhead of traditional monitoring approaches that require application modifications or the insertion of dedicated monitoring probes into the data path. AI-assisted observability, demonstrated at Cloud Native Telco Day EU 2026 [54], extends this infrastructure with LLM-based anomaly detection and root-cause analysis that can identify the signatures of emerging service degradation before they manifest as customer-visible failures, giving autonomous agents the advance warning needed to take preventive action rather than reactive remediation. The importance of this capability is underscored by the changing nature of the services that 6G networks must support: as the AI-RAN Alliance [11] observes, "*AI applications make network quality perceptible at the application layer in ways previous services did not – users notice when an assistant loses context mid-sentence*" – a qualitative shift in the user experience of network impairment that demands a correspondingly more granular and responsive observability and remediation capability than any previous generation of mobile services has required.

## Section V – Technology Last: Architecture in Service of Operator Priorities

Technology Last is the most counterintuitive of the five reorderings introduced in the previous sections, and the one most likely to provoke resistance from equipment vendors, infrastructure investors, and standards bodies whose primary orientation is technical possibility rather than commercial deployment. It does not argue that 6G technology is unimportant – the innovations described in this section are genuine and consequential. It argues that every technology choice – air interface design, RAN decomposition, spectrum strategy, ISAC integration, edge compute architecture – must be derived from the operator, customer, business, and operational requirements established in the preceding four sections, not from vendor research roadmaps or standards committee momentum. A network designed to enable the Control Compact, deliver the six outcome families, support the three revenue engines, and operate with Level 4 agentic AI autonomy [30] has specific, deterministic architectural requirements. This Section translates those requirements into a concrete blueprint, within the IMT-2030 technical performance framework [2][3], examining five technology domains: the end-to-end architecture, the AI-RAN platform, the AI-native air interface, the dApps and E3 control layer, and the spectrum strategy. In each domain, the analysis begins with the operator requirement and derives the technology specification from it – not the reverse. To make this derivation explicit: the Immersive Experience family's ≤20 ms end-to-end latency target mandates mmWave and dApp co-deployment; Mission-Critical Determinism's $10^{-6}$ reliability threshold mandates ISAC-based redundancy and AI-for-RAN predictive maintenance; the Pervasive AI family's <10 ms AI inference latency requirement mandates AI-on-RAN edge compute; and the Sensing-as-a-Service family's sub-meter positioning accuracy target mandates sub-THz or ultra-wideband sub-6 GHz signal processing. Each technology selection in this Section is thus a deterministic derivative of a Section II requirement, not an independent technical agenda.

A candid assessment of Open RAN's current maturity is essential context for this section. Published operator and academic studies have documented throughput gaps compared to integrated RAN in certain dense urban multi-vendor configurations, alongside higher integration complexity and operational overhead. These are engineering challenges being actively addressed: O-RAN Alliance plugfest testing, CNCF conformance requirements, and the xApp performance benchmarking work in nGRG RR-2025-04 [15] represent the industry's structured response. The argument for Open RAN is not that it is already performance-equivalent to integrated RAN in every scenario, but that AI-for-RAN optimization, improved integration tooling, and standards-driven interface stability are converging to close the performance gap on a trajectory that makes Open RAN the correct long-term architectural choice for operators prioritizing control. Section V describes the architecture with full acknowledgment that its commercial realization depends on continued execution against this convergence roadmap.

## The End-to-End 6G Architecture

The end-to-end 6G architecture, as depicted in Figure 10, extends the cloud-native disaggregation principles of Section IV to the full network stack: from the Radio Unit (RU) at the cell site through the Distributed Unit (DU) and Centralized Unit (CU), the 6G Service-Based Core, the edge compute layer, and the multi-domain management and orchestration plane. Three structural characteristics distinguish this architecture from its 5G predecessor and define its capacity to deliver the operator priorities established in this paper.

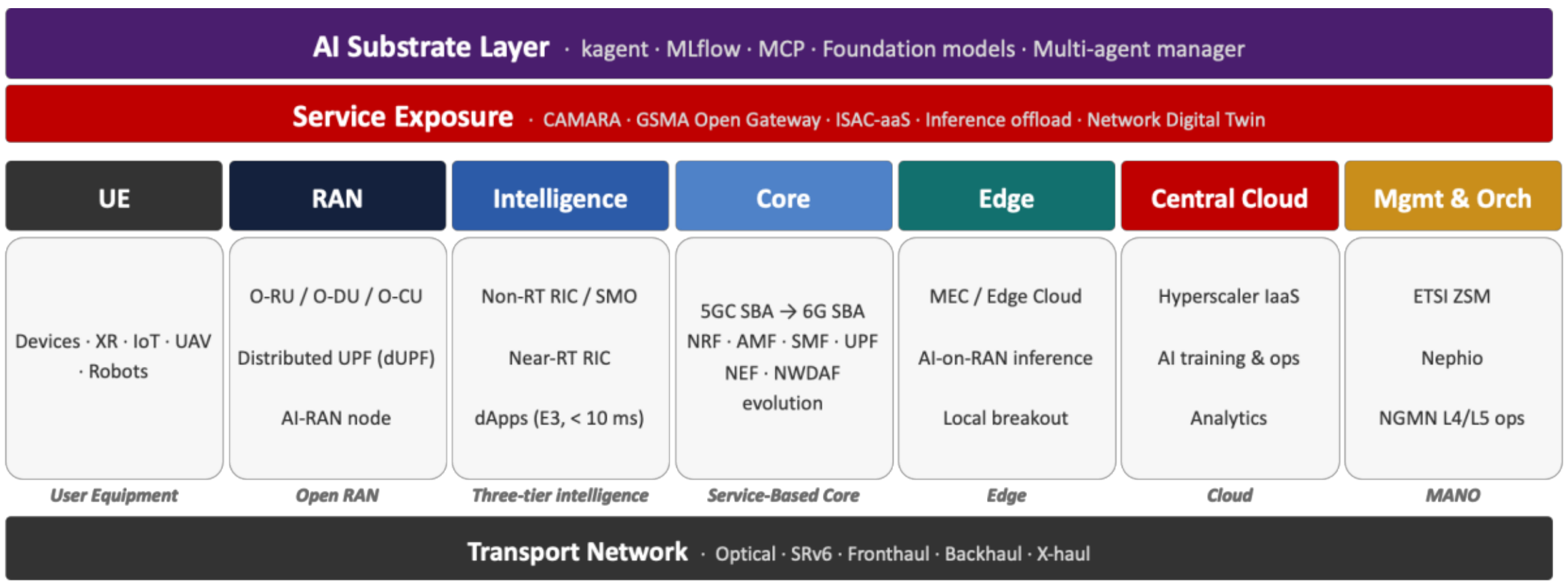

*Figure 10. Cloud-native disaggregation extended to the full stack; AI substrate on top, transport at the bottom, three-tier intelligence between.*

The first structural characteristic is the convergence of RAN and AI compute on shared, cloud-native infrastructure. In the AI-RAN architecture defined by the AI-RAN Alliance [11], the RAN functions – DU, CU, and RU – are co-hosted with AI inference workloads on the same accelerated hardware for computing, governed by a unified cloud-native platform layer comprising a real-time host operating system, a Kubernetes Container-as-a-Service layer, and an AI Runtime Platform optimized for the underlying accelerated hardware. This convergence eliminates the operational boundary between radio access and edge compute that has historically prevented operators from monetizing cell-site infrastructure for AI services. A distributed User Plane Function (dUPF) co-located with the RAN node provides local data breakout to AI-on-RAN workloads without traversing the transport network, enabling the sub-millisecond inference latency that AI-native services require. Rakuten Mobile's deployment – in which every network function from RU management to 5G core runs as a containerized microservice on commercial off-the-shelf hardware orchestrated by Kubernetes – demonstrates that this converged cloud-native model is not a theoretical aspiration but an operational reality deployable at national scale.

The second structural characteristic is hierarchical, multi-timescale intelligence architecture. The Non-Real-Time RAN Intelligent Controller (Non-RT RIC), operates with decision cycles of seconds to minutes, manages network-level policy, AI model training and lifecycle, and cross-domain resource optimization. The Near-Real-Time RIC (Near-RT RIC) [15], with decision cycles of 10 milliseconds to one second, performs per-UE and per-cell inference and control through xApps with access to real-time RAN telemetry. Distributed Applications (dApps) [16], a new architectural tier co-deployed on the DU/CU infrastructure, execute at sub-10 millisecond timescales, addressing real-time control loop requirements that neither RIC tier can satisfy. This three-tier intelligence hierarchy provides the full temporal resolution required by 6G use cases, from long-range energy optimization to microsecond-scale spectrum sharing decisions.

The third structural characteristic is a unified, standards-governed service exposure layer. GSMA Open Gateway [31] and CAMARA API [43] frameworks, extended with 6G-specific interfaces for ISAC-as-a-Service, AI inference offload, and network digital twin access, provide the developer-facing API surface through which the three revenue engines of Section III are operationally realized. The ETSI ZSM framework [19] governs automation and service assurance across the full architecture, while the nGRG Scalable RAN report [17] defines the elastic resource allocation and distributed intelligence capabilities that make user-centric, on-demand service delivery viable at commercial scale.

The architecture embodies the Control Compact of Section I: every management interface, data pipeline, AI model, and API is owned and operable by the operator, with no dependency on a vendor's proprietary management system to access, modify, or audit any network behavior.

## AI-RAN: The Intelligent Radio Platform

The AI-RAN architecture [11] redefines what a radio access network node is and what it can do. Every previous generation's base station was a purpose-built communications appliance: receive radio signals, process them, forward traffic to the core. The AI-RAN node is a general-purpose, accelerated compute platform that simultaneously hosts RAN functions and AI workloads, serving as both a *wireless access point* and *an edge AI inference engine* for the applications its connected users run. The AI-RAN architecture formalizes this dual role through two complementary function classes: AI-for-RAN and AI-on-RAN, as shown in Figure 11.

***AI-for-RAN*** applies machine learning across the full RAN stack to optimize the radio network own performance [12]. At the physical layer, neural-network channel estimators outperform legacy Minimum Mean Square Error (MMSE) algorithms in complex massive MIMO channels; learned CSI feedback encoders compress channel state information at higher fidelity and lower uplink overhead than conventional quantization schemes; transformer-based beam predictors reduce beam management overhead by anticipating UE mobility rather than reacting to it. At the MAC and RRM layers, deep reinforcement learning schedulers dynamically co-optimize throughput, fairness, and energy consumption simultaneously – a multi-objective problem that rule-based schedulers cannot solve without manual trade-off engineering.

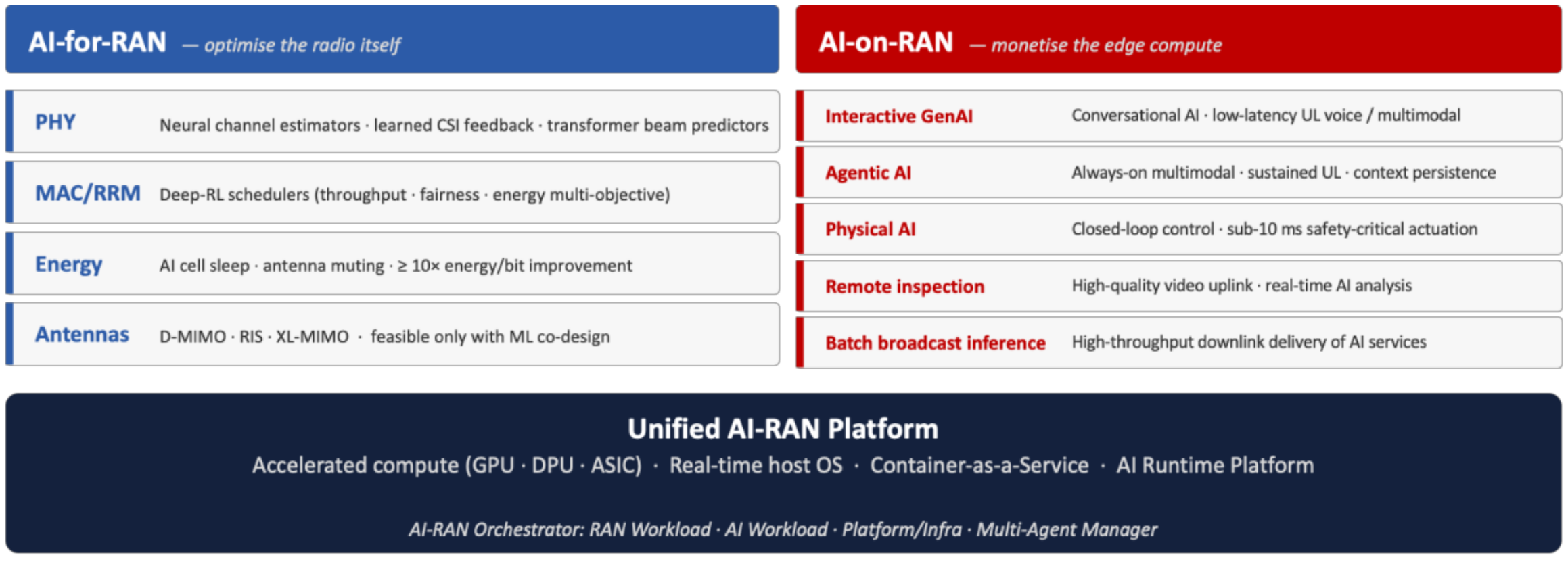

***Figure 11. Convergence of RAN and AI workloads on one accelerated cloud-native platform makes the cell site a profit center, not just a cost center.***

AI-driven cell sleep control and transmit antenna muting deliver the order-of-magnitude improvement in energy-per-bit that ITU-R TPR [3] establishes as a 6G sustainability requirement. Novel antenna architectures – Distributed MIMO (D-MIMO), Reconfigurable Intelligent Surfaces (RIS), and Extremely Large-scale MIMO (XL-MIMO) – depend on AI/ML for the channel estimation and coordination algorithms their scale makes analytically intractable. The standardization trajectory for AI-for-RAN follows the 3GPP Releases 18 through 20 roadmap [33], progressing from AI/ML-enhanced CSI and beam management in Release 18 toward AI-substituted physical layer functions and AI-native network management in Release 20. The critical prerequisite for realizing these gains is operator ownership of the underlying telemetry and model training infrastructure – precisely the AI substrate ownership requirement of the Control Compact.

AI-on-RAN uses the accelerated compute resources already present in the AI-RAN node – GPU, NPU, and FPGA capacity provisioned for RAN signal processing – to host AI, Generative AI, and XR applications at the network edge [11][13][52][53]. This capability directly enables the Pervasive AI and Immersive Experience outcome families of Section II. Drawing on the AI workload characteristics documented in the AI-RAN WG3 report [13], this paper identifies five representative deployment profiles that characterize the commercial demand for AI-on-RAN: the intermittent interactive generative AI assistant (conversational AI requiring low-latency uplink for voice and multimodal input); the always-on multimodal and agentic AI assistant (continuous sensing, reasoning, and action requiring sustained uplink quality and context persistence); physical AI for closed-loop autonomous systems control (sub-10ms round-trip for safety-critical actuation); remote inspection and field support (high-quality video uplink with real-time AI analysis); and batch inference for broadcast AI services (high-throughput downlink delivery). Each profile maps to a distinct SLA tier in the Guarantee Economy framework of Section III, creating a direct commercial architecture from AI workload requirement to network product. The AI-RAN WG3 finding [13] that "*AI applications make network quality perceptible at the application layer in ways previous services did not – users notice when an assistant loses context mid-sentence*" confirms that AI-on-RAN creates genuine, accountable demand for premium connectivity, not merely technical capability in search of a use case. As of mid-2026, no operator has yet commercially monetized RAN edge inference at scale – AI-on-RAN revenue models remain at proof-of-concept and early commercial trial stage; the commercial architecture presented here represents the projection from demonstrated technical capability and published demand research [11][13], with realization timelines contingent on accelerated compute cost trajectories, AI workload pricing norms, and enterprise AI adoption rates in each market.

The management and orchestration layer of the AI-RAN architecture [11] coordinates all RAN and AI workloads through a unified AI-RAN Orchestrator: the RAN Workload Orchestrator manages CNF lifecycle; the AI Workload Orchestrator manages AI model deployment and scaling; the Platform and Infrastructure Orchestrator manages hardware and platform resources across the AI-RAN estate; and the Multi-Agent Manager – described in Section IV – provides the autonomous coordination and governance capability that enables the Agentic AI operations model. This unified orchestration eliminates the management boundary between RAN operations and AI operations, treating both as workloads on a shared cloud-native platform and enabling the joint resource allocation decisions that AI-on-RAN monetization requires.

## The AI-Native Air Interface: A Three-Stage Evolution

The path from today's 5G NR air interface to a fully AI-native 6G physical layer follows a three-stage evolution documented in the AI-RAN WG1 report [12] and reflected in the 3GPP technical specification trajectory from Release 18 through Release 20 [33], as illustrated in Figure 12. Each stage represents a qualitative shift in the relationship between AI and the air interface: from enhancement of existing blocks, through replacement of conventional algorithms, to end-to-end joint learning of the complete transmit-receive chain.

***Stage 1 – AI-Assisted Enhancement –*** characterizes the current 5G-Advanced era and the opening phase of 6G deployment. AI augments specific blocks within a conventionally designed air interface without requiring changes to the underlying specification. Neural network channel estimators improve MIMO performance in complex propagation environments; learned CSI feedback compression reduces uplink overhead while maintaining reconstruction fidelity; AI-based beam predictors reduce measurement overhead by anticipating UE mobility trajectories using historical channel data and sensor fusion. 3GPP Release 18 and Release 19 define the initial standardized interfaces for AI/ML-based CSI management and beam management, establishing the interoperability framework that makes multi-vendor AI-assisted deployments commercially viable. The incremental nature of Stage 1 means it can be deployed on existing RAN hardware through software updates, providing near-term capacity and efficiency gains that fund the investment in Stage 2 and 3 infrastructures.

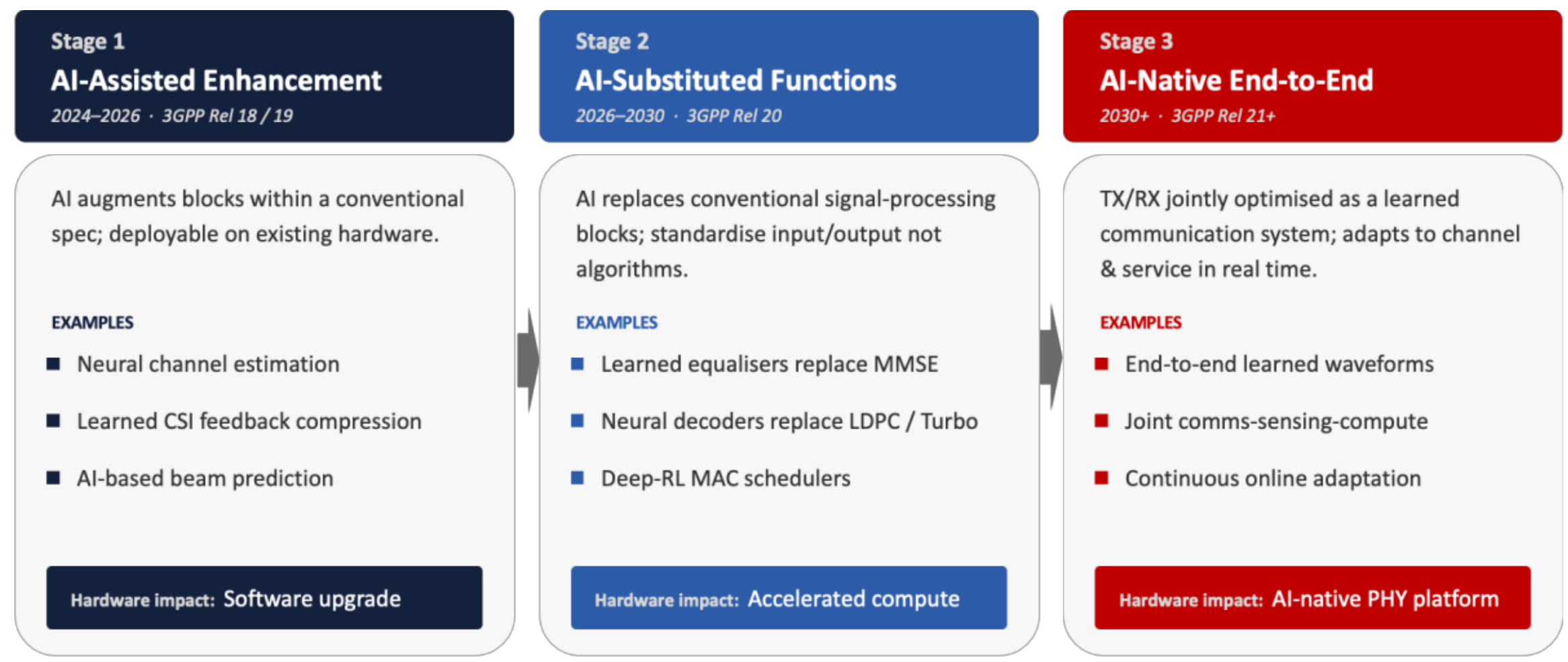

*Figure 12. From AI-augmented blocks to fully learned end-to-end transmit-receive chains over a decade-long standards trajectory.*

***Stage 2 – AI-Substituted Functions –*** begins in earnest with 3GPP Release 20 [33] AI models replace conventional signal processing blocks rather than augmenting them: learned equalizers replace MMSE receivers at the base station; neural network-based decoders replace LDPC or Turbo decoders; deep reinforcement learning schedulers replace rule-based MAC resource allocation. The AI-RAN WG1 report [12] documents demonstrated systems in which AI-based transceivers already deliver significant bit-error-rate performance gains in real-time environments, confirming that Stage 2 is not a research aspiration but an engineering reality advancing toward commercial deployment. The standardization challenge at Stage 2 is interoperability: *when AI models replace standardized algorithmic blocks, ensuring cross-vendor compatibility requires defining input-output interface specifications rather than algorithmic specifications* – a transition that will occupy 3GPP Release 20 and O-RAN specification work through the mid-2020s. Operators must actively drive these specifications to ensure that the AI-substituted blocks remain open, operator-auditable, and vendor-agnostic; allowing vendor-proprietary AI models to substitute standardized blocks would recreate, at the air interface, the same AI lock-in that the Control Compact is designed to prevent.

***Stage 3 – End-to-End AI-Native Air Interface –*** represents the full 6G vision for the mid-2030s deployment horizon. A complete physical layer in which transmitter and receiver are jointly optimized as a learned end-to-end communication system, with the air interface adapting the signaling strategy continuously to channel conditions, interference environment, and traffic profile. This self-adaptive air interface eliminates the engineering compromises embedded in hand-designed modulation and coding schemes – compromises that exist because conventional air interfaces must be fully specifiable before the deployment environment is known. NTT DOCOMO's three-wave model [6] positions the cyber-physical fusion applications of 6G as the demand driver for Stage 3 capability: digital twins, AI-native services, and immersive XR each require an air interface that can simultaneously optimize across communications, sensing, and computation dimensions in real time: an optimization space that only an end-to-end learned system can explore efficiently.

The most demanding challenge of Stage 3 is not performance but multi-vendor interoperability. A jointly learned transmitter-receiver system requires that the neural network model at the transmitter and the model at the receiver are jointly trained or, at minimum, trained to a shared set of input-output specifications. In a multi-vendor ecosystem – where UE chipsets (Qualcomm, MediaTek, Samsung, Apple, etc.) are designed independently of network infrastructure (Nokia, Ericsson, Samsung Networks, Rakuten Symphony, Red Hat, etc.) – there is no natural mechanism for joint training. The Stage 2 approach of specifying input-output interfaces rather than algorithmic specifications partially addresses modular substitution, but Stage 3 end-to-end learned architecture makes the compatibility challenge more acute. This remains an open research problem at the 3GPP Release 20 study phase; Stage 3 should be understood as a compelling long-term direction whose standardization pathway requires resolution of cross-vendor training compatibility before commercial deployment timelines can be committed with confidence.

Stage 2 and Stage 3 commercial timelines are also constrained by the device ecosystem, which is structurally independent of network infrastructure readiness. AI-based receiver functions required by Stage 2 must be implemented in UE baseband chipsets before they can be deployed at commercial scale. Each chipset vendor operates on approximately 18–24 month silicon design cycles with 12–18 months additional time to commercial device availability – meaning that a 3GPP Release 20 feature freeze in 2026–2027 will not produce widely available AI-capable handsets before 2029–2030 at the earliest. The 5G standalone adoption timeline provides a cautionary precedent: standardized in Release 15 (2019), 5G SA only became widely available in flagship handsets in 2021–2022. Operators planning Phase 2 AI-native air interface services should plan commercial launch timelines around device ecosystem availability rather than network infrastructure readiness alone.

## dApps and the E3 Interface: Sub-10ms Edge Control

The O-RAN Alliance existing Near-RT RIC provides inference and closed-loop control at timescales of 10 milliseconds to one second. This is sufficient for many RAN optimization tasks but insufficient for real-time spectrum sharing, ISAC-based positioning feedback, and deterministic low-latency control of autonomous systems. The dApps framework defined in nGRG RR-2025-05 [16] closes this gap by introducing a new control tier that executes directly on the same infrastructure as the DU and CU, eliminating the inter-node communication latency that limits Near-RT RIC response time.

A distributed Application (dApp) is a software entity deployed on the RAN node infrastructure with direct, low-latency access to user-plane data and real-time RAN state through the new E3 interface [16]. Three core E3 procedures define the dApp operational model. The dApp Registration procedure allows the dApp to declare its capabilities, resource requirements, and control scope to the RAN management system, enabling lifecycle management through the standard AI-RAN Orchestrator without requiring bespoke integration. The E3 Subscription procedure allows the dApp to subscribe to user-plane data streams and RAN telemetry at the sampling rates its real-time algorithms require, with data delivered via Unix domain socket-based exchange that minimizes communication overhead within the node. The E3 Control Message procedure allows the dApp to issue real-time commands to DU and CU functions – spectrum allocation updates, beam steering commands, slice admission decisions – within the sub-10ms cycle time that ISAC and deterministic latency applications demand. The nGRG prototype implementation on OpenAirInterface [16] validated this framework in over-the-air trials, demonstrating real-time spectrum sensing and sub-meter positioning using the dApp E3 architecture.

As depicted in Figure 13, the dApps tier integrates with the existing O-RAN control hierarchy rather than replacing it. dApps handle the fastest timescale loops – real-time interference management, microsecond-level beam steering, spectrum occupancy sensing – while the Near-RT RIC manages cell-level and UE-level optimization over tens of milliseconds, and the Non-RT RIC manages network-level policy and AI model lifecycle over seconds to minutes.

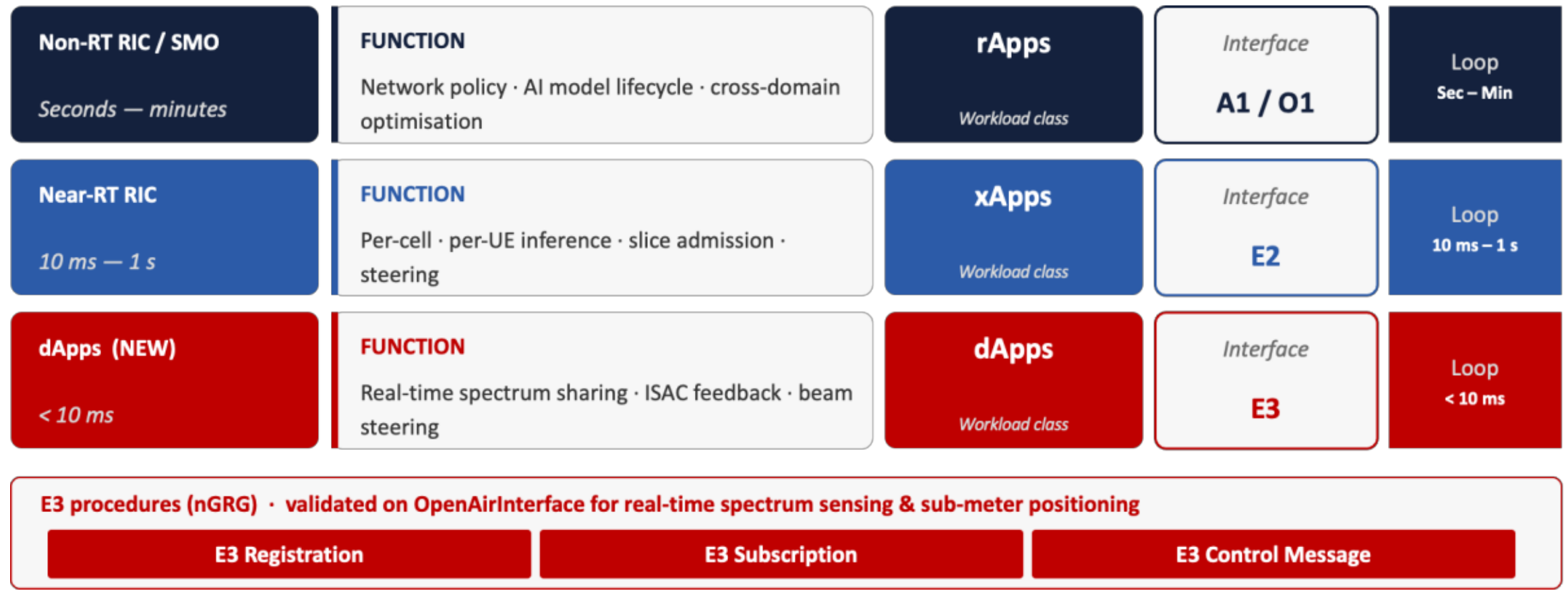

*Figure 13. Three-tier intelligence: Non-RT RIC (s–min) · Near-RT RIC (10 ms–1 s) · dApps (< 10 ms) co-located on DU/CU infrastructure.*

This hierarchical decomposition maps precisely onto the outcome requirements of Section II: the Mission-Critical Determinism outcome family requires the sub-10ms control that only dApps can deliver; the Sensing outcome family requires the direct user-plane data access that the E3 interface provides; and the Pervasive AI outcome family benefits from the in-node inference capability that dApps enable without routing data through the transport network. The dApps framework is a further expression of the Control Compact: *it is an operator-programmable, open-interface control layer that extends operator sovereignty down to the real-time control of individual radio resources.*

## Spectrum Strategy: Coverage, Capacity, and Sensing

The 6G spectrum strategy, illustrated in Figure 14, addresses three interdependent requirements derived from the outcome families of Section II: coverage for ubiquitous connectivity across the Pervasive AI and Sovereign Trust families; capacity for the Immersive Experience and Mission-Critical Determinism families; and sensing resolution for the Sensing outcome family. The ITU-R TPR [3] establishes formal requirements across all three dimensions – uniquely, this document is the first standard to include sensing-related requirements and AI-related requirements as formal technical performance criteria alongside the traditional communications parameters.

***Sub-6 GHz spectrum*** provides the coverage and ISAC foundation layer for 6G. The 3.5 GHz and 4.9 GHz bands proven in 5G, new mid-band allocations in the 7–24 GHz range under discussion in the ITU-R IMT-2030 spectrum identification process, and legacy sub-3 GHz bands collectively provide the indoor penetration, geographic reach, and connection density required for the $10^6$ devices per km² target [3]. Sub-6 GHz propagation characteristics – diffraction around obstacles, penetration through building materials, long coherence time in low-mobility scenarios – also make these bands optimal for ISAC-based positioning and environmental sensing.

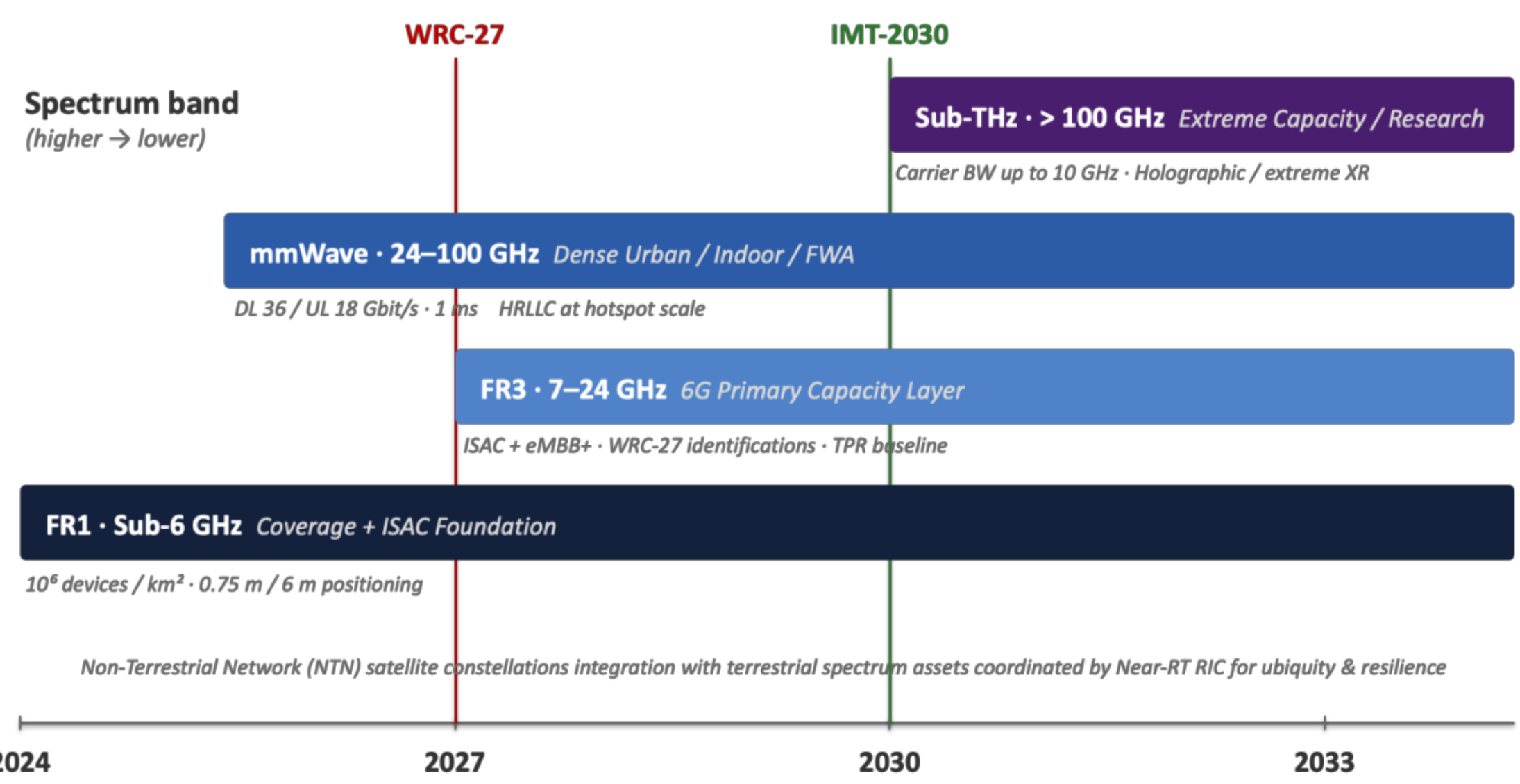

***Figure 14. Layered spectrum: sub-6 GHz coverage + ISAC, FR3 7–24 GHz upper-mid for 6G capacity, mmWave for hotspots, sub-THz for extreme research.***

The ITU-R TPR positioning accuracy targets of 0.75 meters in indoor factory environments and 6 meters in urban macro deployments are achievable through wideband signal processing in these bands. Rakuten Mobile's nationwide deployment in Japan, built primarily on sub-6 GHz spectrum, provides the empirical baseline for what coverage-first spectrum strategy can achieve at national scale when operated through a fully cloud-native, software-defined architecture.

***Millimeter-wave spectrum (26–40 GHz, FR2)*** delivers the capacity layer for dense urban environments, large-venue events, and enterprise campuses. The ITU-R TPR peak data rate targets – 36 Gbit/s downlink, 18 Gbit/s uplink [3] – and the 1 ms user-plane latency requirement for HRLLC are achievable in mmWave deployments using the ultra-wide channel bandwidths and lean frame structures that spectrum abundance in these bands enables. Achieving the 1 ms HRLLC latency target in a cloud-native deployment requires real-time kernel extensions – Linux PREEMPT_RT, CPU isolation, and DPDK-based user-plane I/O bypass – to eliminate the scheduling jitter that standard Kubernetes workload orchestration introduces; the AI-RAN architecture [11] specifies a real-time host OS layer for precisely this purpose. Operators targeting HRLLC service tiers must validate their cloud-native stack against deterministic latency requirements under production load, not merely average-case benchmarks. AI-driven energy management is essential for mmWave economic viability: the nGRG RS02 report [17] identifies AI/ML-driven O-DU consolidation – dynamically remapping O-RU traffic to fewer active O-DUs during low-demand periods – as a key mechanism for reducing mmWave infrastructure power consumption without compromising coverage or latency guarantees. The O-RAN management and orchestration framework, operated by Rakuten Symphony's platform in commercial deployments, provides the real-time orchestration capability that makes such dynamic consolidation operationally safe.

***Sub-terahertz spectrum (92–300 GHz)*** represents the frontier capacity layer for extreme short-range scenarios: holographic communications booths, immersive XR spaces, and high-density robotic manufacturing cells. The ITU-R TPR bandwidth requirements for the highest capacity scenarios [3] require carrier bandwidths of up to 10 GHz, achievable only in sub-THz allocations. Sub-THz deployment will remain complementary to sub-6 GHz and mmWave through the first deployment cycle of 6G, its role defined by the specific scenarios – *holographic communications, extreme capacity hotspots* – that lower frequency bands cannot support. The nGRG Near-RT RIC report [15] identifies Non-Terrestrial Network (NTN) integration – coordination between terrestrial spectrum assets and low-earth-orbit satellite constellations – as a further dimension of the 6G spectrum strategy, enabling the resilience and extended connectivity requirements of the ITU-R TPR [3] in remote, maritime, and airborne environments where terrestrial network economics do not support dedicated infrastructure.

### The Near-RT RIC Toward 6G: Intelligence at the Edge

The Near-RT RIC evolves significantly from its 5G O-RAN incarnation to serve as the primary edge intelligence hub for 6G RAN management. The nGRG RR-2025-04 report [15] defines the enhancements required. Enhanced AI/ML support – including federated learning, distributed model training, and data collection coordination across multiple local base stations – positions the Near-RT RIC as an AI training coordination point at the edge rather than a consumer of centrally trained models, reducing the latency and data-volume overhead of the model update cycle. Expanded service exposure enables direct xApp-to-xApp and xApp-to-external-system interactions without routing through centralized management layers, supporting the multi-agent coordination model of Section IV. Unified data management provides a consistent, query able view of all RAN, UE, and sensing telemetry, giving the AI substrate the comprehensive observability it requires for autonomous decision-making.

Two new coordination capabilities in the 6G Near-RT RIC are particularly consequential for the operator strategy described in this paper. First, interworking among multiple Near-RT RICs – enabling federated AI optimization across geographic clusters without centralizing all intelligence in the Non-RT RIC – supports the scalable, distributed architecture that national-scale 6G deployments require while respecting data locality and regulatory constraints on cross-domain data sharing. Second, Communication and Computing Integrated Network (CCIN) support [15] enables the Near-RT RIC to jointly optimize radio and compute resources as a unified pool. When connectivity and inference are treated as interdependent resources managed through a single intelligence layer, the end-to-end SLA enforcement that the Guarantee Economy demands becomes architecturally coherent: the Near-RT RIC can reserve both the radio resource and the edge compute capacity required to deliver a guaranteed inference SLA, and can enforce that reservation through the full intelligence hierarchy – from the sub-10ms dApp control layer to the minute-scale Non-RT RIC policy layer. Together, the AI-RAN platform, the AI-native air interface, the dApps E3 control tier, and the evolved Near-RT RIC constitute the complete technology architecture that the Operator-Controlled 6G Network requires open, software-defined, operator-owned, and designed in service of the commercial and operational priorities that precede it.

# Section VI – Trust, Sovereignty, and Sustainability

Three of the six outcome families presented in Section II – *Mission-Critical Determinism*, *Pervasive AI*, and *Sovereign Trust* – share a dependency that no amount of radio engineering or cloud-native automation can substitute: *the network must be trustworthy by design, sovereign by architecture, and sustainable by obligation*. Trust, sovereignty, and sustainability are not features to be retrofitted after the performance targets have been met; they are constraints that shape every architectural decision from the outset. An operator that cannot guarantee the security of its autonomous control loops, cannot demonstrate data sovereignty compliance to enterprise and public-sector customers, cannot quantify its carbon credentials with auditable precision, cannot access the premium market segments that the Guarantee Economy and the network API economy of Section III require.

This section examines how each dimension is realized technically, how they reinforce each other architecturally, and how together they constitute a durable competitive advantage rather than merely a compliance burden. The analysis draws on ETSI's closed-loop automation security framework [27], the IOWN Global Forum's energy efficiency metric architecture [44], and the SNS JU eHealth programme [41] validated sustainability evidence – three reference frameworks that together span the trust, sovereignty, and sustainability dimensions with rigor grounded in deployed and near-deployed systems.

## Zero-Trust Architecture for Autonomous 6G Networks

The security architecture of a 6G network managed by agentic AI is qualitatively different from that of any previous generation. In 4G and 5G, security was primarily a perimeter concept: authentication and authorization were performed at the network boundary, and entities that had cleared that boundary were implicitly trusted within it. Agentic AI demolishes this model. A 6G network operating at Level 4 autonomy, in which AI agents observe, reason, plan, and act across the full network stack without per-action human approval, as described in Section IV, introduces an attack surface that perimeter security cannot address. Compromising the telemetry that feeds an agent's Observe function, poisoning the model that governs its Decide function, or injecting unauthorized commands into its Act function does not require breaching the network perimeter; it requires only the ability to interact with the closed-loop automation infrastructure.

ETSI ZSM 017 [27] provides the most rigorous current analysis of this threat landscape, mapping security risks systematically across each stage of closed-loop automation and proposing a three-mechanism security architecture in response. It categorizes security threats across four stages of each closed-loop automation instance:

- At the ***Monitoring stage***, threats include telemetry data tampering – the insertion of fabricated or modified performance metrics designed to cause the AI agent to form an incorrect picture of network state – and unauthorized data collection that exfiltrates sensitive network and user data through the telemetry infrastructure.

- At the ***Analysis stage***, adversarial attacks on AI inference models can cause systematic misclassification of network conditions, and model poisoning during training or update cycles can embed persistent biases that cause chronic mis optimization under specific traffic or environmental conditions.
- At the ***Decision stage***, unauthorized command injection and privilege escalation can cause the AI agent to issue configuration commands outside its authorized scope, potentially degrading service for competing tenants or routing traffic to unauthorized destinations.
- At the ***Execution stage***, man-in-the-middle attacks on the configuration management interfaces translate the agent's decisions into altered, delayed, or suppressed network state changes.

The standard further identifies threats arising from the composition and coordination of multiple closed loops: a legitimately authorized loop can be manipulated into issuing commands that serve an adversarial loop's objectives, and cross-loop privilege escalation can enable a constrained automation domain to affect resources outside its intended scope – a threat that grows in severity as the number of coordinating agents increases with network scale.

ETSI ZSM 017 [27] proposes three interlocking security mechanisms that together constitute the foundation of a zero-trust architecture for autonomous network management. It addresses security within the closed-loop automation architecture as designed prior to LLM-based agentic AI becoming the primary implementation model for autonomous network management. It does not cover threats now prominent in AI security research: prompt injection, in which adversarial inputs manipulate an LLM agent's reasoning to produce unintended network configuration commands; tool poisoning, in which the API interfaces through which agents act on the network are compromised to redirect agent actions; and goal misgeneralization, in which an agent optimizing for a stated objective produces harmful network-level side-effects at scale [29]. These require defenses beyond ZSM 017's scope: adversarial testing pipelines for production agentic AI [29], output guardrails that validate network configuration commands against a policy whitelist before execution, sandboxed tool execution environments, and immutable audit logs enabling forensic reconstruction. Operators deploying agentic AI at Level 4 autonomy must treat LLM-specific threat modeling as a deployment prerequisite, not a post-deployment concern.

As shown in Figure 15, ETSI ZSM 017 [27] introduces three mechanisms layered (across the full zero-trust architecture) that extends the zero-trust principle (never trust, always verify, regardless of network position or claimed identity) from the physical network boundary into the AI inference and orchestration layer [26].

- ***Closed Loop Trust Management*** establishes a cryptographic trust hierarchy among automation domains: each closed loop is issued with a trust certificate that encodes its identity, authorization scope, and permitted interactions, enabling any loop to verify the credentials of any loop with which it coordinates without relying on implicit network-position trust. All inter-loop coordination events are logged with immutable timestamped evidence, providing the audit trail required for both operational forensics and regulatory compliance.

- ***Closed Loop Access Control*** enforces fine-grained, attribute-based policies governing which loops can read which telemetry data, issue which configuration commands, and modify which network parameters – implementing the principle of least privilege at the automation domain level rather than at the device or process level that conventional access control systems address.
- ***Closed Loop Security Exposure*** provides a standardized interface through which operators and regulators can inspect the real-time security posture of each closed loop, enabling automated anomaly detection within the security monitoring layer and human audit of AI decision-making chains under requirements imposed by the EU AI Act [51] for high-risk AI systems in critical infrastructure.

Zero-trust for 6G means that no AI agent, xApp, dApp, or orchestration component is implicitly trusted; every action is authenticated against a verified identity, every data access is authorized against a current policy, and every configuration change is logged with cryptographically signed, tamper-evident evidence. Post-quantum cryptographic (PQC) algorithms – whose first standardized set was finalized by the US National Institute of Standards and Technology in 2024 [55]–[57] and whose adoption is required for IMT-2030 systems under 3GPP TR 22.870 [8] - underpin the cryptographic integrity of the entire trust architecture. Quantum-resistant key exchange and digital signature schemes ensure that trust certificates, policy enforcement tokens, and audit records remain secure against adversaries equipped with cryptanalytically relevant quantum computers. The estimated arrival horizon for such systems – within the expected operational lifetime of 6G networks – makes PQC migration an urgent operational priority for operators today, not a deferred response to a speculative future threat. Operators that build PQC-ready key management infrastructure into their initial 6G deployments will avoid the costly, service-disrupting retrofit that late adoption will require.

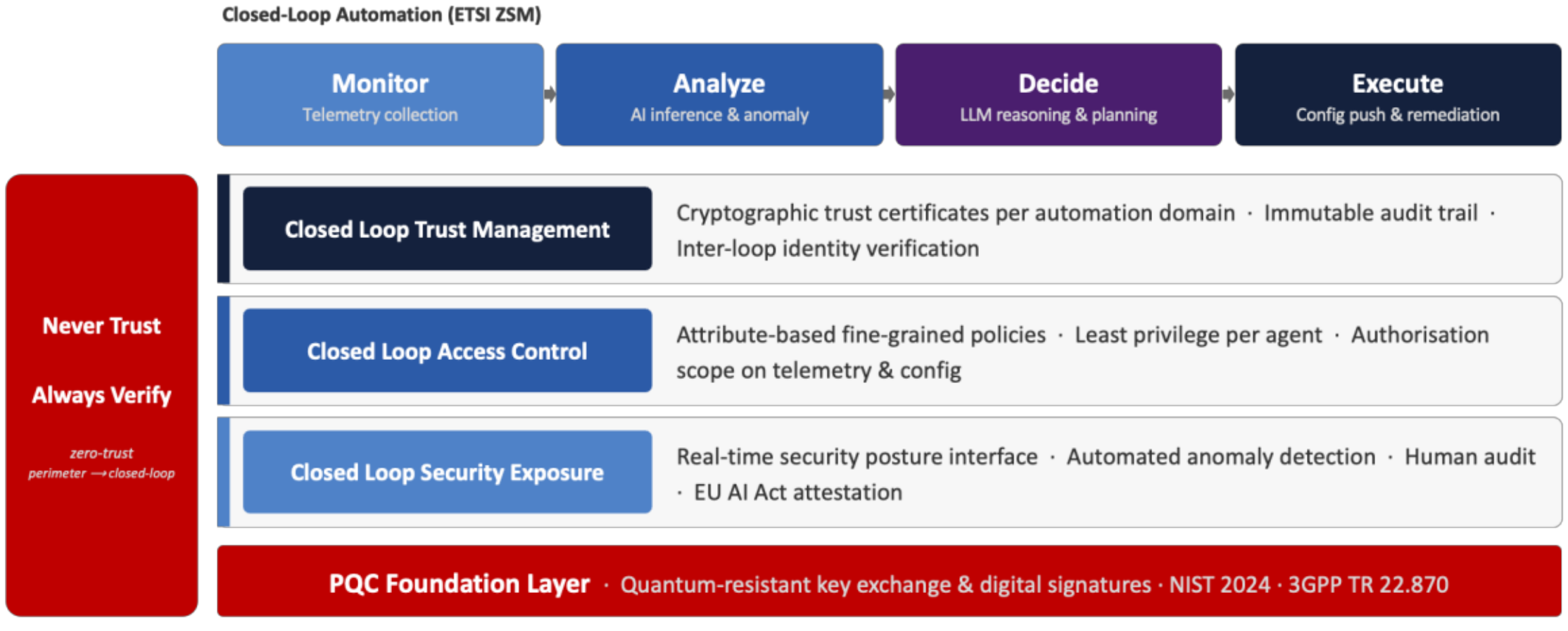

***Figure 15. Three ETSI ZSM 017 [27] mechanisms span every Monitor → Analyze → Decide → Execute stage; PQC underpins the trust foundation.***

## Data Sovereignty and Regulatory Compliance

*Data sovereignty* – the legal and architectural guarantee that data generated within a jurisdiction is processed, stored, and governed under that jurisdiction's legal framework – has emerged as a primary procurement criterion for the enterprise and public-sector customers who represent the Guarantee Economy's highest-value segments. Three regulatory frameworks now converge on this requirement for 6G deployments operating in the European market and across the jurisdictions that follow its data governance model. The EU General Data Protection Regulation (GDPR) [58] requires that personal data of EU residents be processed under explicit lawful basis, with technical measures ensuring encryption at rest and in transit, granular access control, and complete auditability of data access and processing decisions. The European Health Data Space (EHDS) [59] framework extends equivalent requirements specifically to health data, mandating interoperability across EU health systems through HL7 FHIR standards while requiring strict access governance that prevents health data from being repurposed without explicit consent – requirements that the SNS JU eHealth programme [41] validated as fully achievable in live B5G/6G deployments through network slicing for clinical workflow isolation and end-to-end encryption with granular per-patient access controls. The EU AI Act [51] provisions for high-risk AI systems in critical infrastructure – under which 6G network automation systems will be classified – require technical documentation, bias assessment methodologies, human oversight mechanisms, and complete audit trails for all AI-driven decisions that affect service delivery or resource allocation.

These regulatory requirements are not external constraints imposed on an otherwise unconstrained architecture; they are the technical specification for the data layer and AI governance architecture that the Control Compact of Section I requires operators to own. An operator that has built the data sovereignty infrastructure – edge processing nodes that keep jurisdiction-restricted data within regulatory boundaries, policy-driven routing that prevents data from transiting non-compliant paths, and cryptographically signed telemetry records that provide verifiable compliance attestation – has simultaneously built the competitive differentiator that separates its offering from public cloud alternatives. Public cloud hyperscalers process data in geographically distributed, jurisdictionally complex infrastructure that enterprise customers in healthcare, defense, financial services, and public administration cannot use for their most sensitive workloads. The 6G operator, operating a sovereign network with localized edge compute and policy-governed data flows, can provide the *data sovereignty guarantee* that those customers require as a contractual precondition for deployment. The SNS JU eHealth programme [41] documented this competitive dynamic in practice: healthcare system procurement decisions in the SNS JU portfolio were conditioned on GDPR and EHDS compliance architecture, and operators that demonstrated architectural compliance secured deployment contracts over competitors that could not. Rakuten Mobile's platform architecture – operator-owned data pipelines, Kubernetes-enforced access policies, and immutable audit logs native to the platform software stack – provides the operational template for data sovereignty architecture deployed at national network scale.

The SNS JU AI/ML Landscape report [53] identifies AI trustworthiness as a dimension of data sovereignty that is increasingly central to regulatory compliance and enterprise confidence. AI trustworthiness encompasses:

- ***Explainability***, i.e., the ability to provide intelligible explanations of AI model decisions to operators, customers, and regulators.
- ***Robustness***, i.e., the ability to maintain accurate, safe behavior under distribution shift, adversarial inputs, and operational conditions that differ from training environments.
- ***Fairness***, i.e., the absence of systematic *bias* in decisions that affect service quality across different customer segments, geographic areas, or demographic groups.

The SNS JU AI/ML taxonomy [53] defines these trustworthiness indicators as properties that must be designed into the MLOps lifecycle – from data collection and model training through deployment and continuous monitoring – rather than assessed retrospectively. For 6G operators, AI trustworthiness is simultaneously a regulatory obligation under the EU AI Act [51], a commercial requirement for enterprise customers whose own AI governance frameworks demand supply-chain AI assurance, and an operational necessity for agentic AI systems making autonomous decisions in safety-critical network management contexts.

National security policy has emerged as a significant structural accelerant for operator-controlled architecture adoption, reinforcing the commercial arguments of this paper with regulatory force. In Europe, the EU 5G Toolbox [60] and successive national security reviews have constrained or excluded high-risk vendors from core and management functions in multiple member states, creating a regulatory mandate for architectural disaggregation that aligns directly with the Control Compact's technical recommendations. In the United States, the CHIPS and Science Act [61] and the Secure and Trusted Communications Networks Act [62] have created fiscal incentives for Open RAN deployment explicitly motivated by supply-chain security rather than commercial economics. Japan, South Korea, and India have each adopted national industrial and security policies that favor Open RAN and domestically auditable network infrastructure. The convergence of national security policy and operator sovereignty strategy represents a structural regulatory tailwind for the operator-controlled 6G model: governments that have historically treated telecoms regulation as a consumer protection matter are increasingly treating it as a national security matter, with the architectural consequence that open, auditable, disaggregated networks are becoming the regulated norm rather than the operator's optional choice. Operators building toward the Control Compact are therefore aligned with, not merely compliant with, the direction of national security policy in the markets that represent the majority of 6G investment.

## Sustainability as Competitive Architecture

Sustainability is rapidly transitioning from a corporate responsibility commitment to a core competitive and regulatory requirement for telecommunications operators. The EU's Corporate Sustainability Reporting Directive (CSRD) [63] requires large organizations to disclose Scope 1, 2, and 3 emissions with auditable methodological rigor; enterprise procurement frameworks in most advanced economies now score sustainability performance explicitly, with some sectors requiring verified net-zero commitments as a precondition for long-term contract award; and sustainability-linked financing instruments offer preferential interest rates tied to verified energy efficiency and carbon reduction targets. For operators, this creates both a risk and an opportunity. The risk is structural: telecommunications networks account for approximately 2–3% of global electricity consumption, a figure that will grow as 6G deployment scales to denser, higher-power architectures supporting the AI and immersive use cases of Section II. The opportunity is equally structural: operators that can demonstrate verifiable, independently auditable carbon credentials can command premium pricing from enterprise customers with ambitious net-zero commitments, access sustainability-linked capital at preferential rates, and differentiate in public-sector procurement where sustainability weighting is increasingly mandatory.

The ITU-R TPR [3] formally includes energy efficiency for sustainability as one of its minimum technical performance requirements for IMT-2030, establishing 6G as the first mobile generation in which energy efficiency is a defined system-level compliance target rather than an engineering aspiration. The IOWN Global Forum's unified energy efficiency metric framework [44] provides the measurement architecture that makes this requirement auditable and comparable across operators and vendors. Three metrics are central to 6G sustainability reporting:

- The ***Telecommunications Energy Efficiency Ratio (TEER)*** measures useful network output – bits transported, connections served, SLAs delivered – per unit of energy consumed across the full network stack from radio unit to core, addressing the fundamental weakness of data-center-focused metrics such as Power Usage Effectiveness (PUE) that measure infrastructure overhead but not the efficiency of the traffic the infrastructure serves.
- The ***Network Carbon Intensity energy (NCIe)*** measures carbon dioxide equivalent emitted per bit of data transported, providing the end-to-end carbon efficiency metric that enterprise sustainability reporting requires and that regulatory frameworks are progressively mandated.
- The ***Energy Reuse Factor (ERF)*** measures the proportion of waste heat recovered and reused for other purposes [44] – as a significant contributor to lifecycle sustainability for edge compute nodes co-located with RAN infrastructure, which is a consideration that grows in importance as the AI-RAN convergence of Section V increases the thermal output of cell-site compute nodes.

As illustrated in Figure 16, the 6G sustainability KPI framework operates across three scopes that together cover the full environmental footprint of network operations. At the network layer, AI-for-RAN optimizations described in Section V deliver the primary operational efficiency gains: cell sleep control shuts down radio units during low-demand periods without service interruption, saving transmission energy proportional to the fraction of time the network operates below capacity; transmit antenna muting reduces active antenna elements during sparse-user periods; and AI-driven load balancing consolidates traffic onto fewer, more efficiently loaded nodes rather than distributing it across a maximum-coverage footprint.

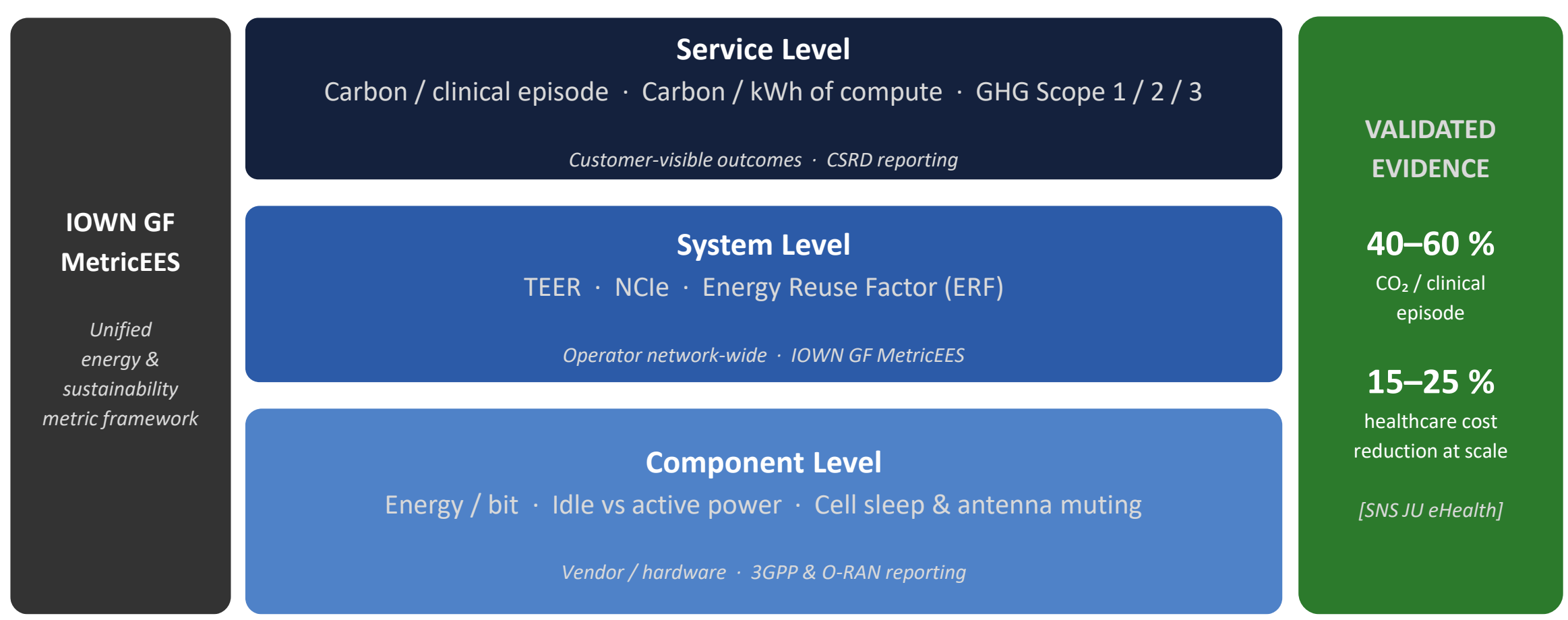

***Figure 16. The pyramid integrates component-, system- and service-level metrics; ITU-R TPR makes energy efficiency a formal compliance criterion.***

The nGRG RS02 report [17] identifies AI/ML-driven O-DU consolidation – dynamically remapping O-RU traffic to fewer active processing units during low-demand periods – as a mechanism delivering measurable energy reductions at the RAN layer that scale in proportion to the depth of the consolidation and the duration of low-demand periods. At the system layer, the TEER and Network Carbon Intensity energy (NCIe) metrics of the IOWN GF framework [44] aggregate the full stack – radio, transport, core, data center – into a single comparable efficiency indicator that operators can report to regulators, disclose in sustainability filings, and use as a benchmark for vendor selection. At the value-chain layer, the IOWN GF framework [44] identifies Scope 3 emissions – from equipment manufacture, transportation, installation, and end-of-life disposal – as the most significant gap in current sustainability reporting, and the dimension that will attract increasing regulatory scrutiny as CSRD implementation progresses. Operators that engage their supply chains on embodied carbon, design for repairability and component reuse, and implement circular economy principles for RAN hardware decommissioning will be positioned ahead of those for whom Scope 3 reporting remains an unfamiliar obligation.

The SNS JU eHealth programme [41] provides compelling evidence of the vertical sustainability impact that 6G-enabled services can deliver beyond the network itself. Across fifteen validated eHealth use cases including remote surgical assistance, AI-assisted cardiac monitoring, home-based rehabilitation, and autonomous emergency response, the programme projects carbon footprint reductions of 40 to 60 percent per clinical episode based on modelled estimates across the project portfolio [41]. These reductions are achieved primarily through the elimination of patient travel to specialist centers and specialist travel to remote sites, demonstrated in B5G/5G trials conducted under the SNS JU programme [41]; 6G is projected to extend and scale these outcomes through its combination of sub-20 millisecond end-to-end latency and enhanced positioning accuracy. The programme also documented healthcare system cost reductions of 15–25 percent at scale, driven by remote proctoring, continuous monitoring, and early intervention that prevents expensive emergency presentations. These figures demonstrate that the sustainability case for 6G is not confined to the network's own energy consumption; operators that position their networks as enablers of vertical sustainability transformations – in healthcare, logistics, manufacturing, and agriculture – can claim a proportionate share of the carbon reduction those transformations deliver, strengthening their sustainability narrative beyond what any improvement in TEER or NCIe alone could provide.

### Convergence: Trust, Sovereignty, and Sustainability as a Single Architecture

The three dimensions of this Section are architecturally convergent. The zero-trust security infrastructure that protects autonomous closed-loop operations requires the same cryptographically governed, operator-owned data layer that data sovereignty compliance demands. The data sovereignty architecture that satisfies GDPR, EHDS, and AI Act requirements is built on the same telemetry pipeline and policy enforcement layer that makes sustainability reporting auditable and manipulation resistant. The sustainability measurement framework that enables TEER and NCIe reporting is built on the same real-time observability infrastructure – eBPF-based kernel telemetry, Prometheus metrics collection, and centralized analytics – that the agentic AI operations model of Section IV requires for autonomous network management. Operators that design trust, sovereignty, and sustainability as separate compliance workstreams – each with its own team, its own data infrastructure, and its own reporting pipeline – will create redundant, inconsistent systems that satisfy none of the three requirements credibly. Operators that build a single, integrated governance architecture will find that the investment delivers returns across all three dimensions simultaneously.

The unifying principle is the one that runs throughout this paper: *operators must own the control plane, AI substrate, and data layer*. Control of the data layer makes sovereignty compliance achievable, as policy-governed data routing, localized edge processing, and cryptographically auditable access logs are only possible when the operator controls the data infrastructure, not merely the transport pipes. Control of the AI substrate makes zero-trust automation governable, because the Closed Loop Trust Management, Access Control, and Security Exposure mechanisms of ETSI ZSM 017 [27] can only be implemented by an operator that owns and can audit the models, data pipelines, and orchestration interfaces of the automation system. Control of the energy management layer makes sustainability measurable and improvable – because TEER and NCIe can only be accurately computed and acted upon by an operator that has full observability into its own network's energy consumption and traffic performance, without dependence on vendor-mediated data products. Trust, sovereignty, and sustainability are not addons to the Operator-Controlled 6G Network. They are among its most durable and commercially valuable structural properties – properties that the Control Compact makes achievable and that the vendor-dependency model structurally prevents.

## Section VII – From Vision to Deployment: A 6G Roadmap

The five reorderings proposed in this paper – Control First, Customer First, Business First, Operations First, Technology Last – are not a wish list assembled from first principles. They are the lessons of building and operating the world's first fully cloud-native, fully Open RAN mobile network. Every architectural claim in the preceding parts is grounded in capabilities that are either in production at Rakuten Mobile today or in active development in its engineering and research programs.

This final part converts the strategic framework into a concrete deployment roadmap, defines the open-source contribution and consumption disciplines that underpin the platform strategy, and articulates the calls to action that will determine whether the operator-controlled 6G compact becomes the industry default or remains the exception.

The roadmap is structured in three phases that correspond directly to the three phases of telecom evolution identified in Section I:

- ***5G-Advanced on-ramp***, in which foundational capabilities are validated in production.
- ***Early 6G commercialization phase***, in which the first outcome-guaranteed services are launched.
- ***6G-at-scale phase***, in which the Guarantee Economy, the network API economy, and the AI inference economy of Section III reach full commercial deployment.

As illustrated in Figure 17, this trajectory is not a forecast of industry averages; it is the author's deployment-grounded projection of what a specific operator – one whose cloud-native network infrastructure is already in production – can realistically achieve on a specific timeline, based on firsthand experience with the Rakuten Mobile platform [7].

## Roadmap Phase 1 (2025–2027): 5G-Advanced as the 6G On-Ramp

The most important insight from the 5G monetization shortfall is that technology readiness and operational readiness are not the same thing. 5G New Radio was standardized in 3GPP Release 15 in 2018; the cloud-native automation, programmable network exposure, and AI-driven operations that would have made 5G's differentiated capabilities commercially deliverable were not operationally mature at any operator until years later. The 6G transition must not repeat this lag. 3GPP Releases 18 and 19 – the 5G-Advanced releases [64] – deliver the critical capabilities that enable Phase 1 of the roadmap: RedCap and enhanced RedCap for cost-efficient IoT and industrial sensor connectivity; enhanced Integrated Sensing and Communication (ISAC) study items in Release 19 that lay the air-interface groundwork for the Sensing-as-a-Service revenue stream of Section III; network energy saving enhancements in Release 18 that provide the cell-sleep and antenna-muting primitives required for the AI-for-RAN energy optimization of Section V; and the CAMARA API exposure framework that begins commercializing programmable network capabilities ahead of 6G. 3GPP Release 20, whose study items are now active, initiates the technical bridge to IMT-2030 with the first 6G use-case and service requirements studies formalized in 3GPP TR 22.870 [8] and the air-interface scenarios of 3GPP TR 38.914 [33].

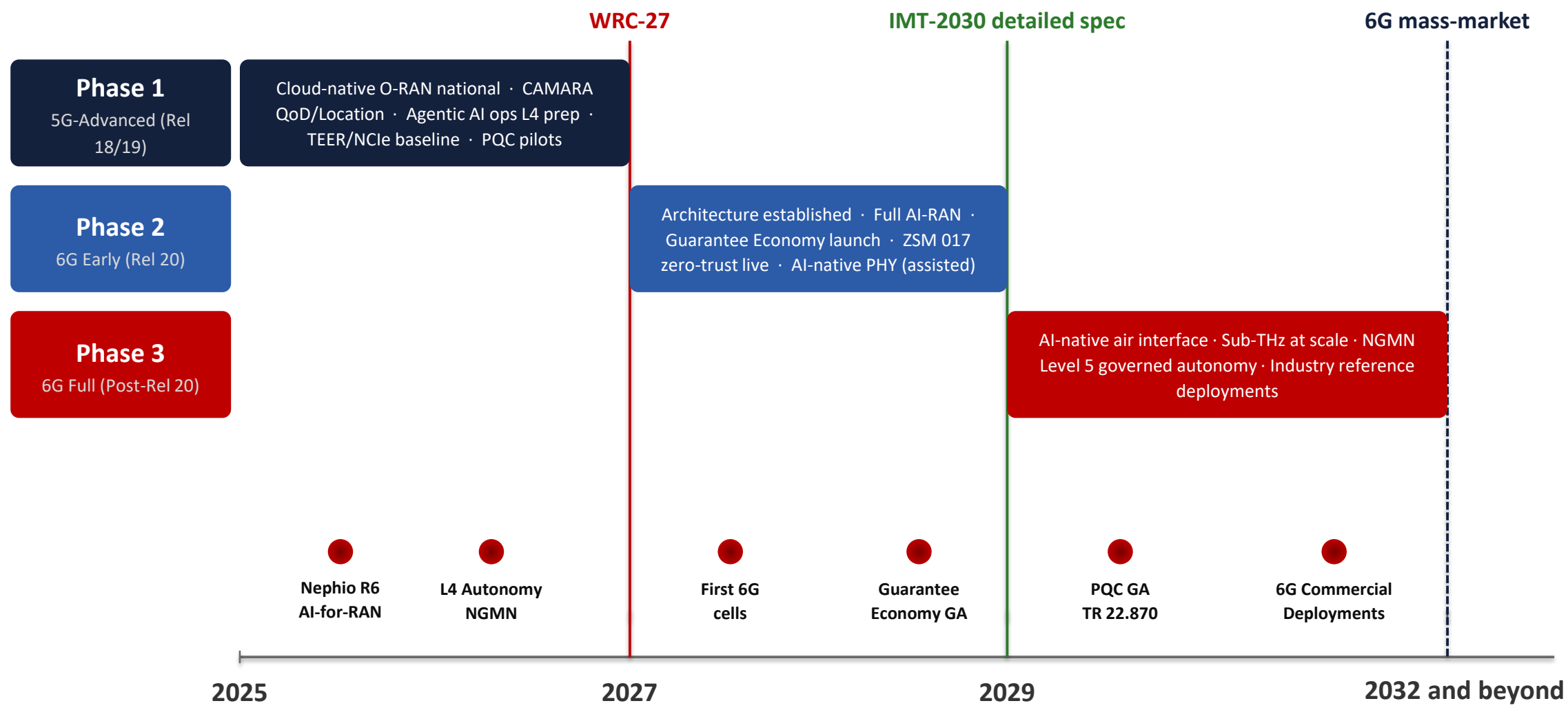

*Figure 17. Three phases align infrastructure investment, standards milestones and commercial launches; the roadmap is an example only, and not an industry average.*

Phase 1 forms a proven production baseline towards 6G. Phase 1 priorities are therefore not greenfield 6G experiments but production hardening of the capabilities that de-risk the 6G transition: full cloud-native O-RAN deployment across the national network with Kubernetes-native lifecycle management through Nephio R6; AI-for-RAN energy optimization achieving verifiable TEER and NCIe improvements reportable under the IOWN GF MetricEES framework [44]; launch of the first CAMARA-compliant Quality-on-Demand and Device Location APIs; and deployment of the agentic AI operations stack – multi-agent coordination, Network Digital Twin validation, and MLflow-governed model lifecycle management – described in Section IV.

The open-source contribution discipline that enables Phase 1 is specific and measurable. In this phase, examples of upstream projects that target that constitute the cloud-native network platform are, but it is not limited to: Nephio (intent-driven network lifecycle management) [34], Sylva (Linux Foundation Europe telco cloud stack) [35], Cilium (Tetragon) [36] and Sauron [48] (eBPF networking, observability and security), O-RAN Software Community (RIC, xApp, near-RT RIC reference implementations) [37], and kagent (the CNCF Kubernetes-native AI agent framework) [38]. Contribution is not philanthropic; it is strategic.

Operators that shape the open-source reference implementations that vendors subsequently certify against hold the architectural influence that would otherwise accrue to the vendors themselves. Consumption discipline in this phase should align production deployments with CNTi CNF conformance [32], CNCF Certified Kubernetes AI Conformance [39], and O-RAN SC J/K test specifications [37] – creating verifiable interoperability that enables multi-vendor integration without system-integrator dependency.

## Roadmap Phase 2 (2027–2029): Early 6G Commercialization

Phase 2 begins with ITU-R's finalization of the IMT-2030 detailed radio interface specifications – the technical foundation whose framework and minimum performance requirements are already established in M.2160 [2] and the TPR [3] – and with the completion of WRC-27 decisions on IMT-2030 spectrum allocations in the upper mid-band and sub-THz ranges. These international milestones open the commercial deployment window. Phase 2 is not about deploying a standards-compliant 6G air interface per se; it is about launching the service layer that 6G's performance envelope enables. The Guarantee Economy tier structure described in Section III – consumer immersive experience guarantees, enterprise determinism SLOs, AI inference QoS commitments – launches in Phase 2 on a network that combines mature 5G-Advanced infrastructure with early 6G cells at high-density urban and enterprise campus sites. The technical enablers are the AI-native air interface evolutions described in Section V – AI-assisted channel estimation, predictive beam management, and joint communication-sensing – deployed first in enterprise campus and industrial IoT scenarios where purpose-built devices can be specified with the required AI capabilities. Consumer AI-native air interface services targeting the general handset market are a Phase 3 objective, consistent with the device ecosystem analysis in Section V: AI-capable handset chipsets are not projected to reach commercial scale before 2029–2030 at the earliest. The Phase 2 timeline carries material execution risk on two dimensions: WRC-27 upper mid-band spectrum identification faces regional opposition, and 3GPP Release 20 timelines may slip by 12–18 months given current workload; operators should plan Phase 2 investment against a base-case and a delayed-standards scenario rather than treating the 2027–2029 window as firm.

A market sizing note is warranted for Phase 2 commercial planning. The Guarantee Economy revenue projections in Section III are stated across all five tiers; however, the device ecosystem analysis in Section V establishes that consumer AI-native air interface services are a Phase 3 objective – AI-capable handset chipsets at commercial scale are not available before 2029–2030. Phase 2 Guarantee Economy revenue therefore depends predominantly on enterprise, industrial IoT, and fixed-wireless use cases, with consumer premium subscription revenue (Premium Consumer Immersive tier) constrained to devices that support the required air-interface capabilities on 5G-Advanced infrastructure. Operators building Phase 2 business cases should size revenue projections against the enterprise-and-fixed-wireless addressable market in their deployment geography, rather than the full Guarantee Economy TAM, and should treat consumer AI-native service revenue as additive from Phase 3 onwards.

The AI-RAN convergence of Section V reaches commercial scale in Phase 2. Shared compute pools running AI-on-RAN inference workloads – enterprise edge AI, computer vision, natural language processing – alongside AI-for-RAN optimization functions on the same cell-site hardware generate the first AI-as-a-Service revenue at the network edge. The NGMN Agentic AI [23] framework's Level 4 autonomous operations – fully closed-loop, AI-driven management of RAN, transport, and core – is the target operational state for Phase 2, with human operators setting policy and reviewing exceptions rather than directly managing configuration changes. The network API economy matures through this phase: the CAMARA API [43] catalogue expands beyond Quality-on-Demand and Device Location to include ISAC sensing outputs, AI inference offload endpoints, and network slice self-service provisioning, creating the developer ecosystem on which the outcome marketplace of Section III depends. Zero-trust automation security, post-quantum cryptographic key management, and GDPR/EHDS-compliant data sovereignty architecture – the infrastructure of Section VI – are production-ready and certified against ETSI GR ZSM 017 [27] requirements by the time Phase 2 enterprise contracts are signed.

### Roadmap Phase 3 (2029–2032 and Beyond): 6G at Scale

Phase 3 represents the full realization of the operator-controlled 6G network described throughout this paper: nation-wide 6G coverage, mature AI-RAN infrastructure, NGMN Level 5 governed autonomy in network operations, and the Guarantee Economy as the default commercial model for enterprise and high-value consumer segments. The DOCOMO 6G White Paper's vision of a ubiquitous intelligent society [6] – in which 6G connectivity is as ambient and assumed as electrical power – is operationally achievable only at Phase 3 maturity, when the combination of pervasive coverage, sub-millisecond ultra-reliable links, and AI substrate at the network edge supports the ambient intelligence, always-on sensing, and continuous AI-assisted service delivery that consumers and enterprises will expect as the standard experience. Near-RT RIC intelligence extensions [15], dApp programmability [16], and scalable RAN architecture [17] reach their full architectural potential at this phase, enabling third-party developers to deploy real-time, sub-10ms RAN control applications at network scale through the E3 interface described in Section V.

### Industry Calls to Action

The operator-controlled 6G network is not a unilateral operator achievement. It requires coordinated action across a seven-stakeholder ecosystem: *operators*, *vendors*, *hyperscalers*, *regulators*, *standards and technical specifications bodies*, *academia*, and *investors*. As illustrated in Figure 18, each stakeholder group holds a specific set of enabling actions without which the vision described in this paper cannot be realized on the timeline the competitive and regulatory environment demands. The following calls to action are specific and accountable, not general exhortations to 'collaborate' or 'innovate'.

For ***operators***, the call to action is architectural: make the decision to own the AI substrate and the data layer before the vendor ecosystem crystallizes around proprietary AI platforms that are as difficult to disaggregate as the RAN vendor stacks of the 4G era. The window in which operator-led, open-source AI infrastructure can be established as the industry default is open now, with the CNCF Kubernetes AI Conformance program, the O-RAN SC AI/ML framework, and the OCUDU Ecosystem Foundation providing the organizational infrastructure for collective action. Operators that defer this decision will cede the AI control plane to vendors and hyperscalers, recreating the dependency cycle that this paper documents as the root cause of 5G's monetization shortfall. Concretely: adopt CNTi CNF conformance and CNCF AI conformance as procurement requirements in all 6G and 5G-Advanced vendor contracts signed from 2026 forward; join and contribute to the CAMARA project to shape the API exposure layer before it is vendor-mediated; and deploy the agentic AI operations framework described in Section IV on your own data and your own infrastructure, not as a managed service from a vendor whose interests in perpetuating operational complexity are structural. The critical precondition for this strategy is engineering capability: owning the AI substrate requires the AI/ML and cloud-native talent to make that ownership meaningful rather than nominal. NGMN [5] documents the skills gap as a primary barrier to cloud-native adoption; closing it through hiring, retraining, and engineering culture change is necessary but not sufficient: the global supply of engineers with both deep telecom protocol knowledge and cloud-native AI/ML expertise is structurally limited, and hyperscalers and AI companies compete for the same talent pool at substantially higher compensation. The three mitigations available – MLOps-as-a-Service to reduce specialized engineering burden, operator consortia to pool talent, and university partnership pipelines – are individually insufficient to close the gap within the Phase 1–2 horizon for most operators; the realistic near-term outcome for many is reliance on vendor-managed AI platforms as a tactical compromise, which is acceptable provided operators retain model audit rights, explainability requirements, and interface portability commitments in vendor contracts. Structural mitigations include platform-as-a-service adoption for specific MLOps functions that reduces the minimum talent floor; operator consortia – coordinated through NGMN [5] – that share AI/ML engineering infrastructure and training investment; and targeted university partnerships around 6G AI that begin building the pipeline over the Phase 1–2 horizon. This transition will also require upfront investment – in talent, platform re-engineering, and the operational cost of parallel running during migration – that will appear in capital and operating expenditure before it appears in revenue growth. Incumbent operators should model this investment as foundational infrastructure for the 6G architectural generation and should communicate this thesis explicitly to investors and boards who may otherwise apply near-term margin pressure to decisions whose returns are measured over a five-to-seven-year horizon.

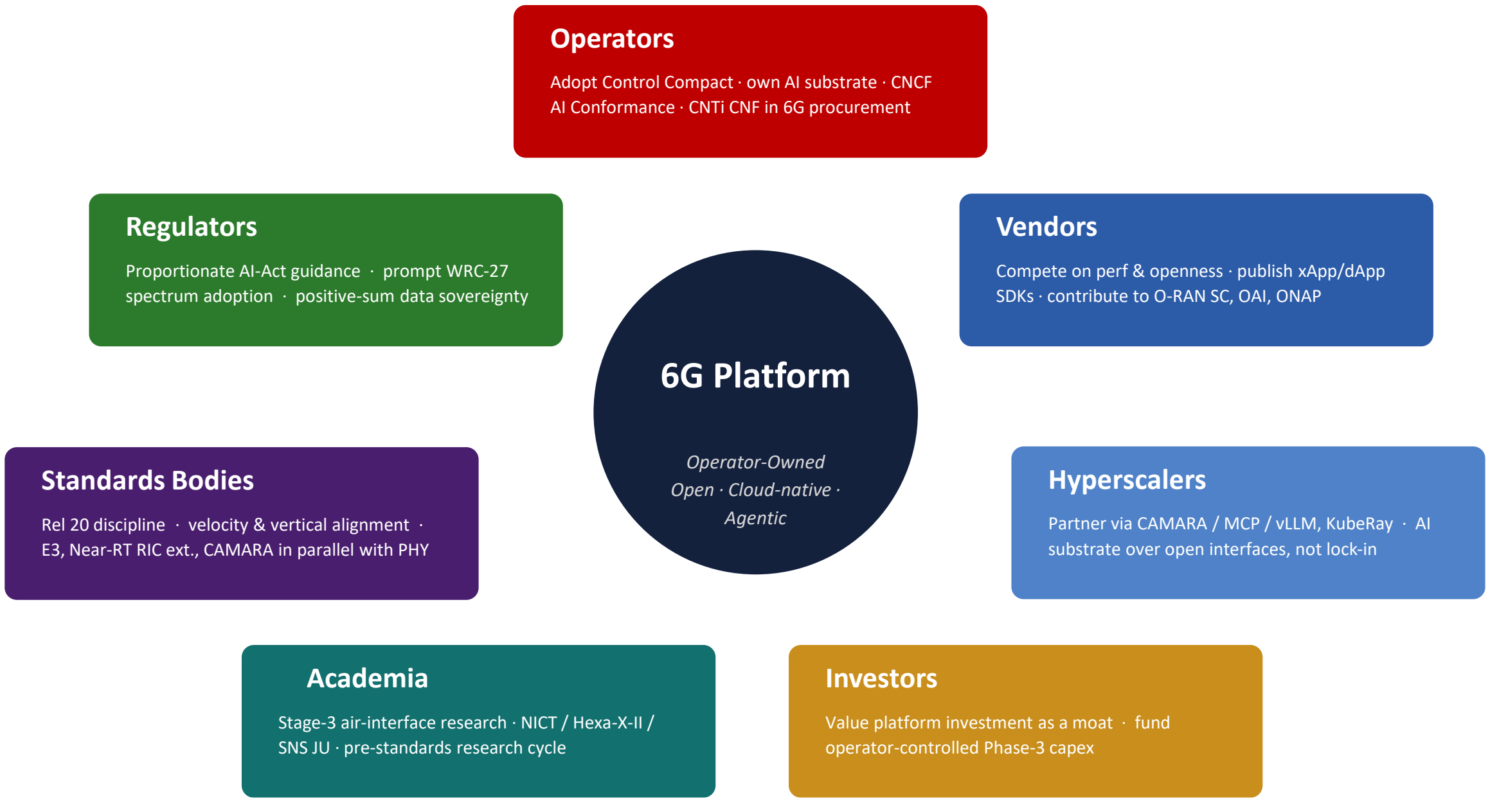

*Figure 18. Hub-and-spoke: 6G platform is the integration substrate; each stakeholder holds a specific, time-bound call to action.*

For ***vendors*** – RAN, core, and OSS/BSS suppliers – the call to action is to compete on performance, interoperability, and developer experience rather than on lock-in. The operators that will grow through the 6G era are those that control their own platforms; the vendors that will grow with them are those that provide genuinely interoperable, open-API, cloud-native functions that perform better than the alternatives rather than those that create switching costs. This means adopting O-RAN specifications as the primary integration interface, not as a compliance minimum; publishing full xApp and dApp development documentation that enables third-party developers to build on vendor RAN without requiring vendor professional services; and contributing network function implementations to open-source projects – O-RAN SC [37], OpenAirInterface [65], ONAP [66] – as evidence of genuine commitment to interoperability rather than as a marketing posture. Vendors that position their AI capabilities as open platforms that operators govern – rather than as managed AI services that operators subscribe to – will earn the trusted-partner role that the Control Compact demands.

For ***hyperscalers***, the call to action is competitive-economics discipline. AWS Private 5G, Azure Private MEC, and Google Distributed Cloud Edge each position the hyperscaler as the primary AI and orchestration layer, reducing the operator to a spectrum license holder and physical transport provider – a substitution strategy that hyperscaler business development teams are actively pursuing. The case for a complementary rather than substitution positioning is not that hyperscalers should be generous; it is that the competitive economics favor complementarity: the 6G operator controlling a licensed AI substrate at the network edge, with carrier-grade SLA accountability and regulatory relationships that no hyperscaler can replicate from a central cloud region, provides compute and inference infrastructure that extends hyperscaler AI reach rather than duplicating it. The hyperscaler-operator relationship that creates the most value is one in which hyperscaler foundation models, data platforms, and developer ecosystems are composed with operator-controlled edge infrastructure and network APIs through open interfaces – CAMARA, MCP, CNCF-standardized AI serving platforms – rather than through proprietary integration arrangements that recreate the vendor lock-in dynamic in a new technology layer. The competitive asymmetry must be acknowledged directly: hyperscalers command larger AI engineering teams, deeper enterprise AI platform integrations, and broader developer ecosystems than any MNO. Microsoft Azure OpenAI Service, Google Gemini Enterprise, and AWS Bedrock are already the dominant platforms through which enterprises build AI applications, and these platforms are integrated with hyperscaler networking products in ways that make operator edge inference a narrower competitive moat than the framing of this paper implies. The operator's defensible differentiation is licensed spectrum, physical edge proximity below 5 ms, regulatory accountability, and data sovereignty guarantees that no hyperscaler can replicate from a central cloud region. Hyperscalers should invest in CAMARA API consumption and contribution, in open-source AI serving frameworks (vLLM, KubeRay) that operators can deploy on their own edge infrastructure, and in joint go-to-market models that share revenue without requiring enterprise data to transit hyperscaler infrastructure.

For ***regulators***, the call to action is enabling policy that accelerates the transition to operator-controlled, AI-driven network management without creating compliance architectures that inadvertently entrench incumbent vendors. Spectrum policy for IMT-2030 should prioritize making upper mid-band spectrum – the 7–24 GHz range where 6G's capacity and latency targets are most achievable – available for mobile broadband use with appropriate coexistence frameworks. Upper mid-band allocation at WRC-27 faces significant regional opposition – from administrations in China, Russia, and across Sub-Saharan Africa and South-East Asia that favor fixed-satellite, fixed-service, and radio astronomy co-existence over new mobile identification in the 7–24 GHz range. Operators and national administrations must engage CEPT, APT, and ATU preparatory meetings now to build the coalition for mobile identification; the outcome is not guaranteed, and the 6G capacity roadmap requires contingency planning for partial or delayed WRC-27 spectrum decisions. National administrations should adopt confirmed WRC-27 outcomes promptly rather than waiting for multilateral consensus to progress at the pace of the 5G spectrum cycle. AI Act implementation guidance for telecom should recognize the operational necessity of autonomous closed-loop network management and provide proportionate compliance pathways for Level 4 AI operations that do not require per-action human approval for routine network management decisions – pathways that maintain rigorous human oversight of policy, exception handling, and model governance without requiring a human in the loop for every cell-sleep event or load-balancing optimization. Data sovereignty frameworks should create positive-sum architectures in which operators can demonstrate GDPR and EHDS compliance through verifiable technical means rather than through compliance documentation that adds administrative overhead without improving actual data protection outcomes. Neutral host and shared active RAN arrangements – in which multiple operators share radio infrastructure while maintaining independent spectrum licenses, control planes, and SLA enforcement mechanisms – should be affirmed as compatible with operator sovereignty in regulatory frameworks, not treated as dilutions of it; the Control Compact architecture is deployable in multi-tenant shared infrastructure scenarios where each operator retains its own AI substrate and data layer above the shared radio layer.

For ***technical specification and standards bodies*** – 3GPP, ITU-R, ETSI, O-RAN Alliance, GSMA – the call to action is velocity and vertical alignment. The 6G standardization timeline is tight: ITU-R M.2160 was finalized in November 2023, the TPR in February 2026, and the first IMT-2030 radio interface specifications must be complete in time to enable commercial deployment in the 2028–2030 window. 3GPP Release 20 study items must therefore be conducted with the discipline and urgency that Release 15 NR demonstrated, without the feature accumulation that inflated Releases 16 and 17. Vertical alignment means standardizing the service-layer APIs and management interfaces – CAMARA, ZSM, NGMN Level 4/5 agentic operations [23] – in parallel with the radio access layer, so that when 6G air interfaces are commercially deployable the service monetization and operations automation infrastructure is production-ready, not eighteen months behind. O-RAN Alliance should prioritize the dApp and E3 interface specifications [16] and the near-RT RIC extensions [15] that enable third-party application development on RAN infrastructure – the programmability layer that converts the 6G radio platform into the network

API economy of Section III. Operators and vendors must also plan for the possibility that Release 20 timelines slip by 12–18 months – as Releases 16 and 17 each took approximately two years from specification freeze to commercial equipment availability – and should develop Phase 2 contingency scenarios that accelerate Release 19 capabilities rather than waiting exclusively for Release 20 features.

For ***academia and research institutes***, the call to action is applied research that closes the gap between IMT-2030 framework targets [2][3] and the engineering realities of achieving them at commercial scale. The AI-native air interface evolution described in Section V – particularly Stage 3, end-to-end AI-generated waveforms trained on channel conditions and service requirements simultaneously – requires fundamental advances in learning-based communications that the telecoms research community is uniquely positioned to deliver. Collaboration with field trial programmes, with NICT's 6G testbed infrastructure, and with the Hexa-X-II and SNS JU research ecosystems provides the empirical grounding that prevents theoretical research from diverging from operational constraints. Research outputs that feed directly into 3GPP study items – through the pre-standardization research cycle that has always been the most effective path from academic insight to commercial deployment – have the highest leverage for accelerating the 6G transition on the timeline this paper describes.

For ***investors***, the call to action is capital allocation discipline that rewards operator-controlled platform investment over managed-service dependency. The operators that will generate sustainable ARPU growth in the 6G era are those that own the AI substrate, the network API exposure layer, and the data pipeline that enables guaranteed service delivery and outcome-based pricing – capabilities that require sustained capital investment in cloud-native infrastructure, MLOps, and open-source engineering talent. Investment frameworks that penalize this investment as cost rather than valuing it as the construction of a durable competitive moat will inadvertently favor the vendor-dependency model that has suppressed operator returns throughout the 4G and 5G eras. The case for operator platform investment is grounded in Rakuten Mobile's disclosed operational evidence: a fully cloud-native, disaggregated network costs less to operate per bit than its traditional-architecture counterparts, generates higher ARPU through ecosystem monetization, and provides the platform flexibility to launch new service categories without proportional capital expenditure – the fundamental operating leverage that the software industry has demonstrated and that the telecoms industry is only now beginning to realize.

## Conclusion: The Operator-Controlled, Customer-First 6G Compact

The argument of this paper can be stated simply: *6G will not deliver structural revenue growth for operators unless it is designed from the start as an operator-controlled platform – one in which the operator owns the operating model, the AI substrate, the data layer, and the customer relationship, and uses that ownership to deliver verified outcomes rather than undifferentiated connectivity*. The precedent for this transition already exists in production at Rakuten Mobile: a national network run as software, operated autonomously, and generating ecosystem revenue that demonstrates the ARPU uplift that the *Guarantee Economy* promises.

The five reorderings – *Control First*, *Customer First*, *Business First*, *Operations First*, *Technology Last* – are not a rhetorical device; they are a sequencing of investments and architectural decisions that any operator can follow with existing technology and existing standards, calibrated to the scale and market context of their own deployment.

The *Control Compact* is not an abstract principle. It is the practical conclusion that follows from twenty years of evidence that operators who cede system design, AI strategy, and operational control to vendors and hyperscalers do not recover structural revenue growth – and from Rakuten Mobile's operational evidence that operators who reclaim that control can build networks that grow with their customers' ambitions rather than constraining them. 6G is the last near-term opportunity to establish the operator-controlled model as the industry default before the AI platform consolidation that is now under way in the hyperscaler ecosystem extends into the telecoms stack.

The open research questions that will determine the pace and scope of this transition include: the economics of brownfield network transformation at national scale; the formal verification requirements for LLM-based agents (and/or their evolution) operating autonomously in safety-critical network control loops; the device ecosystem timeline for AI-native physical layer deployment in consumer handsets beyond enterprise and IoT; and the market-acceptance and regulatory conditions under which the Guarantee Economy scales from enterprise pilots to mass-market adoption. Resolving these questions – through joint operator experimentation, open standards, and cross-industry evidence sharing – is the productive research and innovation agenda (RIA) that the industry should now pursue.

# Annexes

## Annex A – Glossary of Acronyms

| Acronym | Expansion | Context / Key document |
|---|---|---|
| 3GPP | 3rd Generation Partnership Project | Standards body for mobile network specifications (NR, 5GC, 6G) |
| 5GC | 5G Core | Service-based architecture core network for 5G |
| 5G-A | 5G-Advanced | 3GPP Releases 18–19; evolutionary capabilities bridging 5G to 6G |
| 6G | Sixth-Generation Mobile Networks | IMT-2030 systems; ITU-R M.2160 framework and TPR. |
| AI | Artificial Intelligence | Machine learning, deep learning, and agentic AI systems |
| AI-RAN | AI-Radio Access Network | Dual function: AI-for-RAN (optimization) and AI-on-RAN (edge AI hosting) |
| APPI | Act on the Protection of Personal Information | Japan's primary data protection legislation |

| AR | Augmented Reality | Overlay of digital information on physical world; 6G immersive use case |
|---|---|---|
| ARPU | Average Revenue Per User | Operator revenue metric; target for Guarantee Economy uplift |
| B5G | Beyond 5G | Pre-standardization term for IMT-2030 / 6G research phase |
| CAMARA | Cloud Artefact for Mobile Access to Resources through APIs | Linux Foundation project; operator API exposure framework |
| CAPEX | Capital Expenditure | Network infrastructure investment; contrasted with OPEX |
| Cilium | Cilium (eBPF networking) | CNCF project; kernel-level eBPF-powered networking and security for Kubernetes |
| CNCF | Cloud Native Computing Foundation | Hosts Kubernetes, Cilium, Tetragon, kagent, OpenTelemetry |
| dApp | Distributed Application (O-RAN) | Sub-10 ms real-time RAN control application; E3 interface; O-RAN nGRG |
| DL | Downlink | Base station to device transmission direction |
| E3 | E3 Interface (O-RAN) | Open interface between O-RAN controller and dApps; sub-10 ms control loop |
| eBPF | Extended Berkeley Packet Filter | Kernel-level programmable packet processing and security monitoring |
| EHDS | European Health Data Space | EU framework for health data interoperability; HL7 FHIR-based |
| ERF | Energy Reuse Factor | IOWN GF MetricEES metric; proportion of waste heat recovered and reused |
| ETSI | European Telecommunications Standards Institute | ZSM, NFV, MEC standards body |
| EU | European Union | GDPR, EU AI Act, EHDS, CSRD, NIS2 regulatory jurisdiction |
| FHIR | Fast Healthcare Interoperability Resources | HL7 standard for health data exchange; required by EHDS |
| GDPR | General Data Protection Regulation | EU data privacy law; applies to all 6G networks processing EU personal data |
| GPU | Graphics Processing Unit | Primary compute unit for AI inference and baseband processing in AI-RAN |
| GSMA | Global System for Mobile Communications Association | Industry body; Open Gateway, CAMARA, NESAS/SCAS programmes |
| IMT | International Mobile Telecommunications | ITU-R framework; IMT-2020 = 5G; IMT-2030 = 6G |
| IoT | Internet of Things | Massive machine-type connectivity; $10^6$ devices/km$^2$ at 6G |
| ISAC | Integrated Sensing and Communication | Combined radio sensing and communication; 6G use-case family |
| ITU-R | ITU Radiocommunication Sector | IMT-2030 framework, TPR, spectrum decisions (WRC-27) |

| KPI | Key Performance Indicator | Quantified network or service metric |
|---|---|---|
| LLM | Large Language Model | Foundation AI model; basis for agentic network operations (Sections IV–V) |
| MCP | Model Context Protocol | Standardized agent-to-tool interface for LLM-based network automation agents |
| ML | Machine Learning | Subset of AI; includes supervised, unsupervised, reinforcement learning |
| MLflow | MLflow | Open-source ML lifecycle management; experiment tracking, model registry |
| MLOps | Machine Learning Operations | Engineering discipline for production ML model lifecycle management |
| MNO | Mobile Network Operator | Licensed mobile network operator: Rakuten Mobile is an MNO |
| MR | Mixed Reality | Combination of AR and VR; 6G immersive use case |
| NaaS | Network-as-a-Service | On-demand programmable network capacity; ETSI ZSM 019 framework |
| NCIe | Network Carbon Intensity Energy | IOWN GF metric: $CO_2e$ per bit transported; sustainability reporting |
| Nephio | Nephio (Linux Foundation) | Intent-driven cloud-native network lifecycle management; Kubernetes-native |
| NICT | National Institute of Information and Communications Technology | Japan's 6G testbed and research institute |
| NIST | National Institute of Standards and Technology (USA) | AI RMF, PQC algorithm standardisation (2024), Zero-Trust SP 800-207 |
| NR | New Radio | 5G radio access technology (3GPP Rel-15 and beyond) |
| NTN | Non-Terrestrial Networks | Satellite and HAPS components of 6G coverage architecture |
| O-DU | O-RAN Distributed Unit | Baseband processing; AI/ML-driven consolidation for energy efficiency |
| O-RAN | Open Radio Access Network | Disaggregated, open-interface RAN architecture; O-RAN Alliance specifications |
| O-RU | O-RAN Radio Unit | RF front-end; separated from baseband by Open Fronthaul (eCPRI) |
| OPEX | Operational Expenditure | Network running costs; target for autonomous operations reduction |
| OSS | Operations Support Systems | Network management software layer; cloud-native OSS is Section IV's foundation |
| PQC | Post-Quantum Cryptography | Quantum-resistant algorithms; NIST 2024 finalization; required by 3GPP TR 22.870 [8] |
| PUE | Power Usage Effectiveness | Data-center energy efficiency metric; IOWN GF MetricEES context |
| QoD | Quality-on-Demand | CAMARA API for guaranteed bandwidth/latency; Guarantee Economy primitive |
| QoS | Quality of Service | 5G/6G traffic differentiation and priority management |
| RAG | Retrieval-Augmented Generation | LLM architecture: augments inference with dynamic knowledge base retrieval |
| RAN | Radio Access Network | Base station and radio infrastructure; O-RAN, AI-RAN in 6G context |
| RIC | RAN Intelligent Controller | O-RAN platform for xApp and dApp execution; near-RT RIC in 6G |
| SLA | Service-Level Agreement | Contractual commitment to service performance parameters |
| SLM | Small Language Model | Compact LLM for on-device or edge inference with low latency |
| SLO | Service-Level Objective | Specific measurable target within an SLA; Guarantee Economy billing unit |
| SNS JU | Smart Networks and Services Joint Undertaking | EU research programme; eHealth and AI/ML Landscape references |
| Sauron | Sauron eBPF Platform | Rakuten Mobile in-house developed solutions for advanced networking, observability and security tooling |
| Sylva | Sylva (Linux Foundation Europe) | Carrier-grade open-source cloud platform specification for telecom |
| TEER | Telecommunications Energy Efficiency Ratio | IOWN GF metric: useful network output per unit energy consumed |
| Tetragon | Tetragon (CNCF) | eBPF-based runtime security; kernel-level process and network visibility |
| UL | Uplink | Device to base station transmission direction |
| URLLC | Ultra-Reliable Low-Latency Communication | 5G service class; evolved in 6G for sub-0.1 ms, 99.9999 % reliability |
| VR | Virtual Reality | Immersive digital environment; 6G use-case driver |
| xApp | O-RAN Extended Application | Application on near-RT RIC; 10 ms–1 s control loop |
| XR | Extended Reality | Umbrella term for AR, VR, and MR |
| ZSM | Zero-touch Network and Service Management | ETSI ISG framework; five-level autonomy scale |

## Annex B – Mapping of Customer Outcomes to 3GPP Use-Case Families

The six outcome families of Section II map to the 3GPP TR 22.870 IMT-2030 use-case categories and the ITU-R TPR minimum performance requirements as follows.

| **Outcome Family (Section II)** | **TR 22.870 Use-Case Family** | **Key ITU-R TPR KPIs** | **Guarantee Economy tier** |
|---|---|---|---|

| Immersive Consumer Experience | Enhanced Mobile Broadband Plus (eMBB+); Immersive Communication | DL 36 Gbit/s; UL 18 Gbit/s; E2E latency ≤20 ms; connection density $10^6$/km$^2$ | Premium Consumer (immersive guarantee) |
|---|---|---|---|
| Mission-Critical Determinism | Ultra-Reliable Low-Latency Communication (URLLC+); Critical and Safety Communication | User-plane latency 1 ms; reliability ≥99.999 % (HRLLC baseline), ≥99.9999 % (safety-critical, TR 22.870); jitter <1 µs | Enterprise Determinism SLO |
| Pervasive AI as a Network Service | AI and Machine Learning as a Network Service; Distributed Intelligence | AI inference latency <10 ms (edge); model update <100 ms; AI-RAN compute density | AI-as-a-Service billing |
| Sensing as a Network Service | Integrated Sensing and Communication (ISAC); High-Precision Positioning | ≤0.75 m indoor / ≤6 m urban positioning (ITU-R TPR); sensing update rate <1 s; spatial resolution 0.1 m | Sensing-as-a-Service API |
| Sustainable Connectivity | Energy Efficiency for Sustainability; Green Network Operations | TEER improvement vs. 5G; NCIe reduction target; Scope 1+2 carbon reduction | Sustainability-certified tier |
| Sovereign Trust | Secure and Trustworthy Networks; Data Governance and Sovereignty | PQC-secured control plane; zero-trust closed loops; GDPR/EHDS compliance attestation | Sovereign enterprise contracts |

## Annex C – The Three Phases of Telecom: Capability and Revenue Map

| **Dimension** | **Stage 1: Connectivity (1G–4G)** | **Stage 2: Programmability (5G / 5G-A)** | **Stage 3: Monetization (6G)** |
|---|---|---|---|
| Primary product | Voice minutes; data megabytes; geographic reach | Programmable slices; differentiated QoS; API exposure (CAMARA) | Verified outcomes; guaranteed SLOs; AI inference; sensing |
| Revenue model | Per-minute / per-MB tariffs; flat-rate consumer subscriptions | Enterprise SLA contracts; early API revenue; IoT connectivity | Outcome-based pricing; Guarantee Economy; API marketplace; AI-as-a-Service |
| Technology paradigm | Circuit-switched → IP; OFDM; MIMO; carrier aggregation | Cloud-native NFs; network slicing; MEC; O-RAN disaggregation | AI-native air interface; AI-RAN; agentic operations; ISAC; dApps; sub-THz |

| Operator strategy | Infrastructure build-out; coverage and capacity leadership | Virtualization; vendor diversification; API platform launch | Operator-controlled AI substrate; data sovereignty; ecosystem multiplier |
|---|---|---|---|
| AI role | None (rule-based SON) | ML-assisted optimization; AI recommendations to human operators | Agentic AI (Level 4–5); closed-loop autonomous management; AI substrate as revenue layer |
| Edge compute | None | Multi-access Edge Computing (MEC); low-density deployment | AI-RAN converged compute; dense edge inference; real-time dApp hosting |
| Customer relationship | Consumer: mass market; Enterprise: static connectivity contracts | Enterprise: SLA-differentiated; Developer: early API access | Consumer: experience guarantees; Enterprise: outcome SLOs; Developer: outcome marketplace |
| Competitive dynamic | Coverage race; spectrum licensing; equipment vendor dependency | Slicing and API differentiation; O-RAN disaggregation begins | AI substrate and data sovereignty as moat; operator-controlled platform vs. hyperscaler dependency |

## Annex D – The Control Compact: Operator-Owned, Federated, and Commodity

The Control Compact (Section I) defines three layers of the 6G stack by strategic ownership. Operators that misclassify capabilities – consuming as commodity what should be owned or building what should be federated – recreate the vendor-dependency cycle that suppressed 4G and 5G monetization.

| **Operator must OWN** | **Operator should FEDERATE** | **Operator may CONSUME as commodity** |
|---|---|---|
| End-to-end operating model and system architecture | Foundation AI models (LLMs, domain-adapted) | Silicon (SoCs, GPUs, FPGAs) |
| AI strategy: model selection, training data, MLOps governance | Shared spectrum capacity (neutral host, national roaming) | Optical transport components |
| Customer relationship, identity, and trust layer | Network API exposure marketplaces (CAMARA, Open Gateway) | Generic cloud-native services (undifferentiated compute) |
| Data layer: telemetry pipelines, model training data, inference context | Sustainability accounting and carbon reporting platforms | Standard undifferentiated network functions (e.g. generic NAT) |

| Security governance: zero-trust policies, PQC key management | Partner settlement and onboarding infrastructure | Off-the-shelf COTS hardware (commodity servers, switches) |
|---|---|---|
| Network observability: eBPF telemetry, real-time analytics | Spectrum-sharing and dynamic allocation frameworks (WRC-27) | Standardized conformance testing tools |
| Open-source platform contributions and consumption discipline | Multi-operator federation for roaming and API interoperability | Public cloud regions for non-latency-sensitive workloads |
| Agentic AI operations: closed-loop automation, digital twin | Cross-operator research programmes (Hexa-X-II, SNS JU, Next G) | Generic DevOps tooling (CI/CD pipelines, version control) |

## Annex E – The Guarantee Economy: Service-Tier Catalogue and SLO/KPI Library

The five commercial tiers of the Guarantee Economy (Section III) are defined by their guaranteed SLOs, target customer segments, and billing models. SLO values are derived from ITU-R TPR and 3GPP TR 22.870.

| Tier | Target segment | Guaranteed SLOs | Breach consequence | Billing model |
|---|---|---|---|---|
| Premium Consumer Immersive | Consumer: gaming, XR, live-event immersive | DL ≥500 Mbit/s; latency ≤20 ms; availability 99.99 % | Bill credit; automatic QoS re-routing | Monthly premium subscription; experience-tier add-on |
| Enterprise Determinism | Manufacturing, logistics, surgery, robotics, critical comms | Latency ≤1 ms; reliability 99.9999 %; jitter <1 µs; positioning ≤0.75 m indoor / ≤6 m urban (ISAC-enhanced; specialized indoor deployments) | SLA penalty; incident report; automated remediation SLA | Per-SLO outcome contract; annual framework + usage |
| AI Inference Edge | Enterprise AI apps, real-time computer vision, NLP at edge | Inference latency <10 ms; throughput ≥[agreed tokens/s]; model availability 99.99 % | Fallback to cloud; bill credit | Per-inference billing; committed capacity reservation |
| Sensing-as-a-Service | Smart city, autonomous logistics, | Positioning accuracy ≤0.75 m indoor / ≤6 m urban; | API SLA credit; fallback sensing source | Per-API-call or per-area subscription model |

| | environmental monitoring | sensing update ≤1 s; sensor coverage SLA | | |
|---|---|---|---|---|
| Network-as-a-Service (NaaS) | Hyperscalers, enterprise private network, cloud-native B2B | Slice provisioned <5 min; guaranteed bandwidth and latency; isolation guarantee | Provisioning SLA credit; automated slice re-instantiation | Usage + committed capacity; outcome-based enterprise contract |

**Annex F – Mapping of Capabilities to TMF Requirements and ETSI Framework**

| Capability | TM Forum Autonomous Networks level | ETSI ZSM closed-loop stage | Key reference |
|---|---|---|---|
| eBPF / Tetragon kernel telemetry and OpenTelemetry metrics collection | Level 3 (Conditional Autonomy) | Monitor | Section IV, Section VI |
| AI-assisted anomaly detection and root-cause analysis (LLM-RAG) | Level 3–4 | Analyze | Section IV |
| Multi-agent AI decision engine (NGMN Level 4 closed loop) | Level 4 (Fully Autonomous) | Decide | Section IV |
| Nephio intent-driven RAN and core lifecycle management | Level 3–4 | Execute | Section IV, Section V |
| Network Digital Twin (pre-execution validation) | Level 4–5 | Decide → Execute (safety gate) | Section IV |
| AI-for-RAN: cell sleep, beam management, load balancing | Level 4 | Decide + Execute | Section V |
| AI-on-RAN: edge AI inference hosting on RAN compute | Level 4 | Execute (resource allocation) | Section V |
| Symworld API hub: CAMARA QoD, Location, Sensing exposure | Level 3 (service exposure) | Execute (API fulfilment) | Section III |
| ETSI ZSM 017 Closed Loop Trust Management and Access Control | Level 4–5 (security governance) | All stages (cross-cutting) | Section VI |
| MLflow model versioning, audit trail, EU AI Act compliance | Level 5 (Governed Autonomy) | All stages (governance) | Section IV, Section VI |

### Annex G – Sustainability KPI Catalogue

KPI definitions aligned with IOWN GF MetricEES v1.2, ITU-T L.1470 and L.1480, and ETSI ES 203 228. All network-layer KPIs are reported at full-stack scope (radio + transport + core + data center) unless noted.

| KPI | Definition | 6G target / benchmark | Standard alignment | GHG scope |
|---|---|---|---|---|
| TEER (Telecom Energy Efficiency Ratio) | Useful network output (bits, connections, SLOs delivered) per unit of energy consumed across the full network stack | >10× improvement vs. 5G baseline (ITU-R TPR requirement) | IOWN GF MetricEES v1.2; ITU-T L.1480 | Scope 1+2 |
| NCIe (Network Carbon Intensity Energy) | $CO_2$ equivalent emitted per bit of data transported end-to-end across the full network stack | Year-on-year reduction target; aligned with net-zero trajectory | IOWN GF MetricEES v1.2; GHG Protocol | Scope 1+2 |
| ERF (Energy Reuse Factor) | Fraction of total thermal output recovered and reused for other purposes (e.g. district heating, on-site heating) | >20 % at edge compute nodes co-located with RAN | IOWN GF MetricEES v1.2; ETSI ES 203 228 | Scope 1 |
| PUE (Power Usage Effectiveness) | Total facility power divided by IT equipment power; measures data-center infrastructure overhead | <1.3 at edge nodes; <1.2 at centralized sites | IOWN GF MetricEES v1.2; ISO/IEC 30134-2 | Scope 1 |
| Cell sleep rate | Percentage of total RU-time in low-power sleep mode without service-affecting interruption | >40% during off-peak periods (AI-for-RAN optimization) | 3GPP Release 18 energy saving; O-RAN nGRG RS02 | Scope 1 |
| O-DU consolidation ratio | Fraction of O-DU processing capacity active vs. provisioned during low-demand periods (AI/ML-driven) | >30 % average consolidation across 24-hour traffic cycle | O-RAN nGRG RS02 | Scope 1 |
| Scope 1 (direct) | Direct GHG emissions from operator-owned or controlled sources (generators, fuel use, refrigerant leakage) | Net-zero by 2040; intermediate 50 % reduction by 2030 | GHG Protocol; CSRD | Scope 1 |

| Scope 2 (energy) | Indirect emissions from purchased electricity, steam, heat | 100 % renewable electricity by 2030 | GHG Protocol; CSRD; ITU-T L.1470 | Scope 2 |
|---|---|---|---|---|
| Scope 3 (value chain) | All other indirect emissions: equipment manufacture, logistics, installation, end-of-life disposal | 50 % reduction vs. 2025 baseline by 2035; CSRD reporting from 2026 | GHG Protocol; CSRD; IOWN GF MetricEES | Scope 3 |
| Vertical carbon saving (eHealth) | Carbon reduction per clinical episode enabled by 6G-connected remote care vs. in-facility equivalent | 40–60 % per episode (SNS JU eHealth programme) | SNS JU eHealth; NHS / EU health carbon frameworks | Scope 3 (value-chain) |

## Annex H – CAMARA / Open Gateway 6G API Candidate List

APIs listed are either currently in the CAMARA project specification pipeline or identified as 6G-scope candidates based on TR 22.870 use-case requirements and nGRG technical reports.

| API name | Primary use case | Key parameter / SLO | TR 22.870 family | Phase |
|---|---|---|---|---|
| Quality-on-Demand (QoD) | Guaranteed bandwidth and latency for enterprise applications | Guaranteed bitrate; latency profile; priority class | URLLC+; eMBB+ | 1 (5G-A) |
| Device Location | Asset tracking, geofencing, logistics optimization | Accuracy: <5 m (5G-A); ≤0.75 m indoor / ≤6 m urban (6G ISAC) | ISAC; High-precision positioning | 1 (5G-A) → 2 (6G) |
| Number Verification | SIM-based authentication; fraud prevention | Real-time verification; API response <100 ms | Sovereign Trust | 1 (5G-A) |
| SIM Swap | Account takeover fraud prevention; identity assurance | Swap event notification; recency timestamp | Sovereign Trust | 1 (5G-A) |
| Device Status | Roaming state, connectivity status for IoT management | Status update latency; reachability accuracy | Massive IoT | 1 (5G-A) |
| Edge Discovery | Route AI inference to nearest MEC node | Node latency <10 ms; compute availability signal | AI/ML as a network service | 2 (6G) |

| AI Inference Offload | On-demand LLM / CV / NLP inference at network edge | Inference latency <10 ms; throughput SLO; model availability | Distributed intelligence; AI NaaS | 2 (6G) |
|---|---|---|---|---|
| ISAC Sensing | Environmental sensing, occupancy detection, velocity measurement | Resolution <10 cm; update rate <1 s; coverage SLA | ISAC | 2 (6G) |
| Network Slice Self-Service | Automated enterprise private network provisioning | Provisioning time <5 min; guaranteed BW + latency isolation | NaaS; Enterprise determinism | 2 (6G) |
| AI-RAN Exposure (xApp/dApp lifecycle) | Third-party real-time RAN control applications | E3 sub-10 ms control loop; xApp <1 s; dApp <10 ms | Programmable RAN | 2–3 (6G) |
| Energy Transparency | Carbon intensity reporting per session / per API call | NCIe per session; real-time energy attribution | Green network; CSRD | 2 (6G) |
| Sovereign Attestation | Verifiable proof of data residency and processing jurisdiction | GDPR/EHDS compliance certificate; cryptographic proof | Sovereign Trust | 2–3 (6G) |

## Annex I – Regulatory Matrix: Key Jurisdictions

A simplified overview of the regulatory landscape affecting 6G deployment across primary markets and the European context. Operators should conduct full legal analysis for each deployment jurisdiction.

| **Jurisdiction** | **Spectrum / IMT-2030** | **Data sovereignty** | **AI governance** | **Security / resilience** |
|---|---|---|---|---|
| Japan (JP) | Sub-6 GHz + mmW licensed; NICT 6G testbed active; WRC-27 preparation ongoing | Act on Protection of Personal Information (APPI, amended 2022); Cross-border transfer restrictions | AI Strategy 2022; NICT AI governance; 3GPP TR 22.870 alignment | Cyber Security Basic Act; Radio Act resilience; Critical infrastructure designation |
| European Union (EU) | RSPG upper mid-band guidance; WRC-23 outcomes implemented; 6–17 GHz under study | GDPR (mandatory); European Health Data Space (EHDS); Data Act 2025 | EU AI Act (high-risk systems for critical infrastructure; prohibited uses) | NIS2 Directive (mandatory); European Cybersecurity Act; CSRD (sustainability reporting) |

| Australia (AU) | ACMA 5G/6G spectrum roadmap; mmW 26 GHz licensed; 6G study items 2026+ | Privacy Act 2024 amendments; CDR (Consumer Data Right); Data sovereignty Bill pending | AI Safety Standard (DISR 2024); NIST AI RMF alignment; Essential 8 (cyber) | SOCI Act (critical infrastructure); ISM / Essential 8; Defense Industry Security |
|---|---|---|---|---|
| Singapore (SG) | IMDA 5G nationwide; 6G advisory panel 2025; WRC-27 APAC coordination | Personal Data Protection Act (PDPA); Model AI Governance Framework; Digital Economy Agreements | IMDA AI Governance Framework v2 (2020+); AI Verify toolkit | Cybersecurity Act 2018; Critical Information Infrastructure; CSA cyber hygiene |
| South Korea (KR) | 28 GHz + sub-6 GHz 5G; 6G R&D programme (ETRI); WRC-27 preparation | Personal Information Protection Act (PIPA, 2023 amendment); Data cross-border rules | National AI Act (in progress); KISA AI security guidelines; ETRI 6G AI framework | NIS Act; Critical infrastructure law; KICS security standards |
| India (IN) | Bharat 6G Vision (2023); DOT 6G task force; 700 MHz + 3.5 GHz 5G | Digital Personal Data Protection Act (DPDP, 2023); Data localization provisions | NITI Aayog AI strategy; MEITY AI governance framework; Responsible AI guidelines | IT Act; CERT-In; Critical information infra protection orders |
| United States (US) | FCC mid-band (3.45, CBRS, 6 GHz); NGSO spectrum; WRC-27 via ITU-R WP5D | State-level (CCPA/CPRA +); Federal sectoral laws (HIPAA, GLBA); No federal omnibus | NIST AI RMF 1.0; Executive Order 14110 (AI Safety); Biden → Trump transition continuity | NIST Cybersecurity Framework; NIST PQC standards (2024); CISA critical infrastructure |

# Disclosure

The author is Senior Vice President, Advanced Research and Innovation, at Rakuten Mobile Inc., Tokyo, Japan. *The assumptions and views reported herein are solely those of the author and do not necessarily represent those of Rakuten or its affiliates.* This work received no external funding; the author was supported by Rakuten Mobile Inc. in his employed capacity.

## Acknowledgements

This paper draws on the author’s over 25 years of work experience[3] and contributions to the information and communication technology sector, as well as the work of the most respected industry and academic programs and organizations. Their published outputs are cited throughout, including those from the ITU-R, 3GPP, NGMN Alliance, O-RAN Alliance, ETSI, AI-RAN Alliance working groups, IOWN Global Forum, and the Linux Foundation.

The author thanks the Rakuten Mobile Inc. engineering and operations teams, whose daily work operating the world's first fully cloud-native, fully Open RAN mobile network at commercial scale provides the deployment evidence upon which every claim in this paper rests. The author also acknowledges Sharad Sriwastawa of Rakuten Mobile and Rakuten Symphony for the opportunity to work on this subject and for his invaluable support and contribution.

## References

[1] ITU-R, “Future Technology Trends of Terrestrial IMT towards 2030 and Beyond,” ITU-R Report M.2516-0, International Telecommunication Union, Geneva, Nov. 2022.
[2] ITU-R, “Framework and Overall Objectives of the Future Development of IMT for 2030 and Beyond,” ITU-R Recommendation M.2160-0, International Telecommunication Union, Geneva, Nov. 2023.
[3] ITU-R WP5D, "6G Technical Performance Requirements (TPR)," ITU-R Working Party 5D, International Telecommunication Union, Feb. 2026.
[4] Ericsson, “Network for AI Experiences,” Ericsson AB, Stockholm, 2025.
[5] NGMN Alliance, “Framework for Network Simplification – An Operator View,” v1.0, NGMN, Frankfurt, Feb. 2026.
[6] NTT DOCOMO, “5G Evolution and 6G White Paper,” Version 5.0, NTT DOCOMO, Tokyo, Jan. 2023.
[7] Rakuten Group, Inc., “FY2025 Full Year Financial Results Presentation,” Rakuten Group, Inc., Tokyo, Feb. 2026.
[8] 3GPP, “Study on 6G Use Cases and Service Requirements – Stage 1,” 3GPP TR 22.870 v2.0.0, 3rd Generation Partnership Project, 2026.
[9] C. Chatzieleftheriou and E. Liotou, "A Survey on AI for 6G: Challenges and Opportunities," IEEE Open Journal of the Communications Society, vol. 7, pp. 3189–, 2026. DOI: 10.1109/OJCOMS.2026.3677293
[10]X. Zheng, H. Xiao, S. Jin, Z. Wang, W. Tian, W. Liu, J. Cao, J. Shen, Z. Shi, Z. Zhang, and N. Yang, "AI-Native 6G Physical Layer with Cross-Module Optimization and Cooperative Control Agents," arXiv:2601.02827v2 [eess.SP], Jan. 2026. Available at: https://arxiv.org/abs/2601.02827
[11]AI-RAN Alliance, “AI-RAN Architecture,” Version 1.2, AI-RAN Alliance, 2024.

[3] https://www.linkedin.com/in/dr-david-soldani/

[12]AI-RAN Alliance, “State-of-the-Art AI/ML Applications for Improvements in RAN Performance,” WG1 AI-for-RAN Technical Report, AI-RAN Alliance, 2024.

[13]AI-RAN Alliance, “AI on RAN: Enabling Monetizable Differentiated Connectivity for AI,” Working Group 3 White Paper, AI-RAN Alliance, 2024.

[14]S. Salmi, M. A. Ouameur, M. Bagaa, G. C. Alexandropoulos, A. Tahenni, D. Massicotte, and A. Ksentini, "AI-Native O-RAN Architectures for 6G: Towards Real-Time Adaptation, Conflict Resolution, and Efficient Resource Management," TechRxiv (preprint), posted Sep. 2025. DOI: 10.36227/techrxiv.175825547.74922399/v1

[15]O-RAN Alliance nGRG, “Evolution of Near-Real-Time RAN Intelligent Controller Towards 6G,” Technical Report nGRG-RR-2025-04, O-RAN Alliance, Jan. 2026.

[16]O-RAN Alliance nGRG, “dApps: Distributed Applications for Open RAN,” Technical Report nGRG-RR-2025-05 v2.0, O-RAN Alliance, Jan. 2026.

[17]O-RAN Alliance nGRG, “Scalable and User-Centric RAN Architecture: Service Requirements and Design Considerations,” Technical Report nGRG-RS02, O-RAN Alliance, Jan. 2026.

[18]H. Harkous, A. Kak, A. Urie, H. Straulino, H. Wu, M. Laitila, H.-T. Thieu, N. Choi, T. Van de Velde, and M. Kanerva, "Flat UP: Toward RAN-Core Convergence for the 6G User Plane," IEEE Communications Magazine, vol. 63, pp. 62–68, May 2025. DOI: 10.1109/MCOM.003.2400403

[19]ETSI, “Zero-touch Network and Service Management (ZSM); Reference Architecture,” ETSI GS ZSM 002 v1.1.1, ETSI, Sophia Antipolis, Aug. 2019.

[20]ETSI, “Zero-touch Network and Service Management (ZSM); ZSM Framework for Network-as-a-Service (NaaS),” ETSI GR ZSM 019 v1.1.1, ETSI, Sophia Antipolis, Jan. 2026.

[21]ETSI, “Zero-touch Network and Service Management (ZSM); Study on the Utilization of Agents in Autonomous Networks,” ETSI GR ZSM 020 v1.1.1, ETSI, Sophia Antipolis, Jan. 2026.

[22]TM Forum, “Autonomous Networks Levels Evaluation Methodology,” IG1252 v3.0.0, TM Forum, May 2025. Available at: https://www.tmforum.org/resources/standard/ig1252-autonomous-networks-levels-evaluation-methodology/

[23]NGMN Alliance, “Agentic AI-Based Operating Models,” V1.0, NGMN, Frankfurt, Mar. 2026.

[24]H. Sun, Y. Liu, A. Al-Tahmeesschi, A. Nag, M. Soleimanpour, B. Canberk, H. Arslan, and H. Ahmadi, "Advancing 6G: Survey for Explainable AI on Communications and Network Slicing," IEEE Open Journal of the Communications Society, vol. 6, 2025. DOI: 10.1109/OJCOMS.2025.3534626

[25]C. Provvedi, L. Seidenari, B. Picano, and R. Fantacci, "Intent-LLM: A Framework for Automated Network Configuration Through Code Generation," IEEE Transactions on Cognitive Communications and Networking, vol. 12, pp. 7246–7248, 2026. DOI: 10.1109/TCCN.2026.3683230

[26]M. Souppaya, K. Scarfone, and D. Yaga, “Zero Trust Architecture,” NIST Special Publication 800-207, National Institute of Standards and Technology, Aug. 2020. Available at: https://doi.org/10.6028/NIST.SP.800-207

[27]ETSI, “Zero-touch Network and Service Management (ZSM); Closed-Loop Automation Security Aspects,” ETSI GR ZSM 017 v1.1.1, ETSI, Sophia Antipolis, Jan. 2026.

[28]E. Choudhary, A. K. Yadav, R. Kumar, and M. Liyanage, "A Practical Approach to Transitioning 5G-AKA Toward Fully and Hybrid Post-Quantum Security for 5G and Beyond Communication," IEEE Open Journal of the Communications Society, vol. 7, pp. 2980–, 2026. DOI: 10.1109/OJCOMS.2026.3677962

[29]M. Altintaş, S. N. Karhan, Y. E. Tok, A. G. Toprak, and Ö. B. Mercan, "When Beneficial Intelligence Turns Hostile: A Survey of Adversarial Threats in AI-Native 6G," IEEE Open Journal of the Communications Society, vol. 7, pp. 3468–, 2026. DOI: 10.1109/OJCOMS.2026.3678511

[30]NGMN Alliance, “AI Surge and Its Implications for 6G,” v1.0, NGMN, Frankfurt, Feb. 2026.

[31]GSMA Open Gateway, “A global framework of common network APIs that simplifies access to mobile operator networks and helps expose core capabilities of the open network to developers and cloud providers.” Available at: https://www.gsma.com/solutions-and-impact/gsma-open-gateway/

[32]Cloud Native Telco Initiative (CNTi), “Driving Cloud-Native Adoption in Telecom,” The Linux Foundation Project. Available at: https://lfnetworking.org/cloud-native-telco-initiative-cnti-driving-cloud-native-adoption-in-telecom/

[33]3GPP, “Study on 6G Scenarios and Requirements,” 3GPP TR 38.914 v0.4.0, 3rd Generation Partnership Project, Mar. 2026.

[34]Nephio, “Cloud Native Network Automation,” The Linux Foundation Project. Available at: https://nephio.org/

[35]Sylva, “A fundamental step to Telco Cloud & Edge homogenization and sustainability,” The Linux Foundation Project. Available at: https://sylvaproject.org/

[36]Cilium, “eBPF-based Networking, Observability, Security,” The Linux Foundation Project. Available at: https://cilium.io/

[37]O-RAN SC, “O-RAN Software Community (SC),” The Linux Foundation Project. Available at: https://o-ran-sc.org/

[38]kagent, “Bringing Agentic AI to Cloud Native,” The Linux Foundation Project. Available at: https://kagent.dev/

[39]CNCF, “Cloud Native Computing Foundation,” The Linux Foundation Initiative. Available at: https://www.cncf.io/

[40]O-RAN Alliance, “Transforming Radio Access Networks Towards Open, Intelligent, Virtualized and Fully Interoperable RAN.” Available at: https://www.o-ran.org/

[41]SNS JU Technology Board, "Towards 6G-Enabled eHealth," Smart Networks and Services Joint Undertaking, Brussels, Feb. 2026.

[42]Government of India, Department of Telecommunications, Ministry of Communications, "Bharat 6G Vision," DoT, New Delhi, India, Mar. 2023. Available at: https://bharat6galliance.com/bharat6G/Home/content/Bharat-6G-Vision-Document/Bharat-6G-Vision-Document

[43]CAMARA, “APIs enabling seamless access to Telco network capabilities,” The Linux Foundation. Available at: https://camaraproject.org/

[44]IOWN Global Forum, “Toward a Unified Metric for Energy Efficiency and Sustainability in Next-Generation Networks,” IOWN GF MetricEES v1.2, IOWN Global Forum, Jan. 2026.

[45]D. Soldani and A. Manzalini, “Horizon 2020 and Beyond: On the 5G Operating System for a True Digital Society,” IEEE Vehicular Technology Magazine, Vol. 10, Issue 1, March 2015, ISSN [20]Information: DOI: 10.1109/MVT.2014.2380581.
[46]TM Forum, "REST API Design Guidelines," TMF630, TM Forum. Available at: https://www.tmforum.org/resources/standard/tmf630-rest-api-design-guidelines/
[47]TM Forum, "Service Ordering Management API," TMF641, TM Forum. Available at: https://www.tmforum.org/resources/standard/tmf641-service-ordering-management-api/
[48]D. Soldani, et al., “eBPF: A New Approach to Cloud-Native Observability, Networking and Security for Current (5G) and Future Mobile Networks (6G and Beyond),” IEEE Access, May 2023. Available at: https://ieeexplore.ieee.org/document/10138542
[49]D. Brown, “Rakuten Mobile Adopts eBPF to Strengthen Anomaly Detection and Security in Cloud-Native Telecom Networks,” August 12, 2025. Available at: https://ebpf.foundation/rakuten-mobile-adopts-ebpf-to-strengthen-anomaly-detection-and-security-in-cloud-native-telecom-networks/
[50]OCUDU Ecosystem Foundation, “A collaborative program hosted by the Linux Foundation dedicated to accelerating open, secure, and interoperable Open RAN CU/DU implementations,” The Linux Foundation. Available at: https://ocudu.org/
[51]Regulation (EU) 2024/1689, “Artificial Intelligence Act.” Available at: https://eur-lex.europa.eu/eli/reg/2024/1689/oj/eng
[52]O-RAN Alliance nGRG, “Generative AI in Network,” Technical Report nGRG-RR-2025-02, O-RAN Alliance, 2025.
[53]SNS JU Smart Networks Working Group, “The AI/ML Landscape for Smart Networks and Services: Taxonomy, Standards and Innovation Pathways,” SNS JU Reliable Software Networks Working Group, Dec. 2025.
[54]Cloud Native Telco Day EU 2026. Available at: https://colocatedeventseu2026.sched.com/overview/area/Cloud+Native+Telco+Day
[55]NIST, “Module-Lattice-Based Key-Encapsulation Mechanism Standard (ML-KEM),” FIPS 203, National Institute of Standards and Technology, Aug. 2024. Available at: https://doi.org/10.6028/NIST.FIPS.203
[56]NIST, “Module-Lattice-Based Digital Signature Standard (ML-DSA),” FIPS 204, National Institute of Standards and Technology, Aug. 2024. Available at: https://doi.org/10.6028/NIST.FIPS.204
[57]NIST, “Stateless Hash-Based Digital Signature Standard (SLH-DSA),” FIPS 205, National Institute of Standards and Technology, Aug. 2024. Available at: https://doi.org/10.6028/NIST.FIPS.205
[58]EU General Data Protection Regulation (GDPR). Available at: https://gdpr-info.eu/
[59]European Health Data Space Regulation (EHDS). Available at: https://health.ec.europa.eu/ehealth-digital-health-and-care/european-health-data-space-regulation-ehds_en
[60]European Commission, “EU Toolbox for 5G Security,” NIS Cooperation Group, Brussels, Jan. 2020. Available at: https://digital-strategy.ec.europa.eu/en/library/eu-toolbox-5g-security

[61]United States, “CHIPS and Science Act,” Public Law 117-167, Available at: https://www.congress.gov/bill/117th-congress/house-bill/4346
[62]United States, “Secure and Trusted Communications Networks Act,” Public Law 116-124, Available at: https://www.congress.gov/bill/116th-congress/house-bill/4998
[63]European Union, “Corporate Sustainability Reporting Directive (CSRD), Directive (EU) 2022/2464,” Official Journal of the European Union, 2022. Available at: https://finance.ec.europa.eu/financial-markets/company-reporting-and-auditing/company-reporting/corporate-sustainability-reporting_en
[64]5G Americas, “5G Advanced Overview,” White Paper, July 2025. Available at: https://www.5gamericas.org/wp-content/uploads/2025/07/5G-Advanced-Overview.pdf
[65]OpenAirInterface, “The fastest-growing open-source community in cellular wireless.” Available at: https://openairinterface.org/
[66]Open Network Automation Platform (ONAP). Available at: https://onap.org/.

## Author

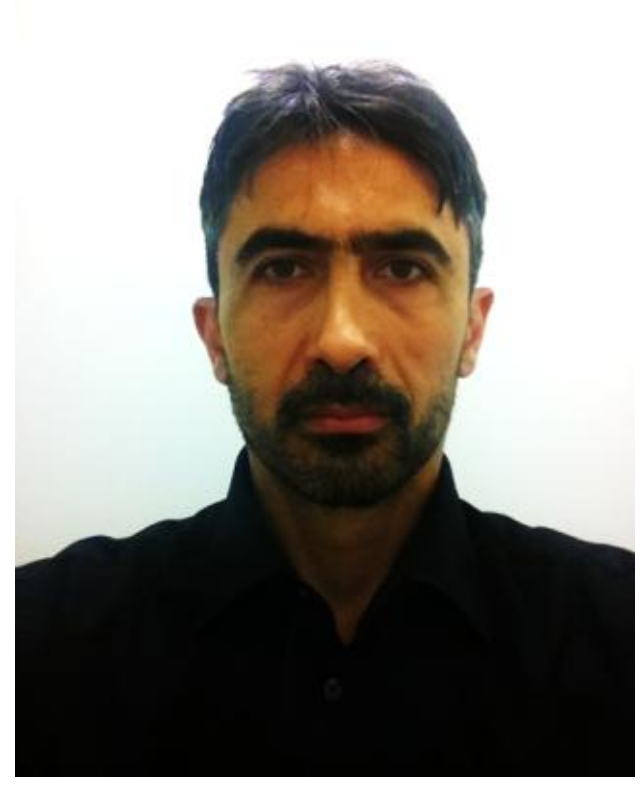

**DAVID SOLDANI (Senior Member, IEEE)** received the M.Sc. degree (magna cum laude) in engineering from the University of Florence, Italy, in 1994, and the D.Sc. degree in technology (Hons.) from the Helsinki University of Technology, Finland, in 2006. Throughout his career, he has held prestigious academic positions, including a Visiting Professor with the University of Surrey, U.K., in 2014, an Industry Professor with The University of Technology Sydney (UTS), Australia, in 2016, and an Adjunct Professor with the University of New South Wales (UNSW), in 2018. In his professional roles, he has served in significant leadership positions. He has held the positions of Chief Information and Security Officer (CISO), e2e, Global, as well as SVP Innovation and Advanced Research with Rakuten. Previously, he held the position of Chief Technology and Cyber Security Officer (CTSO) with the ASIA Pacific Region, Huawei. He was the Head of 5G Technology, e2e, Global, with Nokia, the Head of the Central Research Institute (CRI), and the VP of strategic research and innovation in Europe with Huawei European Research Center. He is currently with Rakuten Mobile Inc.

He can be reached at: https://www.linkedin.com/in/dr-david-soldani/